\documentclass[preprint,aip,cha]{revtex4-1}

\usepackage[english]{babel}
\usepackage[utf8]{inputenc}
\usepackage{float}
\usepackage{graphicx}
\usepackage{dcolumn}
\usepackage{bm}
\usepackage{amsmath}
\usepackage{amssymb}
\usepackage{datetime}
\usepackage{hyperref}
\usepackage{bigints}
\usepackage{siunitx}
\usepackage[caption=false]{subfig}
\usepackage{todonotes}

\begin{document}

\title{Fluid and hybrid simulations of the ionization instabilities in Hall thruster}
\author{\firstname{O.}~\surname{Chapurin}}
\email{alex.chapurin@usask.ca}
\affiliation{Department of Physics and Engineering Physics, University of Saskatchewan, Saskatoon SK S7N 5E2, Canada}
\author{\firstname{A. I.}~\surname{Smolyakov}}
\affiliation{Department of Physics and Engineering Physics, University of Saskatchewan, Saskatoon SK S7N 5E2, Canada}
\author{\firstname{G.}~\surname{Hagelaar}}
\author{\firstname{J.B.}~\surname{Boeuf}}
\affiliation{LAPLACE, Université de Toulouse, CNRS, INPT, UPS, 118 Route de Narbonne, 31062 Toulouse, France}
\author{\firstname{Y.}~\surname{Raitses}}
\affiliation{Princeton Plasma Physics Laboratory, Princeton, New Jersey 08540, USA}

\begin{abstract}
Low-frequency axial oscillations (5-50 kHz) stand out as a pervasive feature observed in many types of Hall thrusters. While it is widely recognized that the  ionization effects play the central role in this mode, as manifested via the large scale oscillations of neutral and plasma density, the exact mechanism(s) of the instabilities remain unclear. To gain further insights into the physics of the breathing mode and evaluate the role of kinetic effects, a one-dimensional time-dependent full nonlinear low-frequency model  describing neutral atoms, ions, and electrons, is developed in full fluid formulation and compared to the hybrid model in which the ions and neutrals are kinetic. Both models are quasineutral and share the same electron fluid equations that include the electron diffusion, mobility across the magnetic field, and the electron energy evolution. The ionization models are also similar in both approaches. The predictions of fluid and hybrid simulations are compared for different test cases. Two main regimes are identified in both models: one with pure low-frequency behaviour and the other one, where the  low-frequency oscillations coexist with higher frequency oscillations (with the characteristic time scale of the ion channel flyby time, 100-200 kHz).  The other test case demonstrate the effect of a finite  temperature of injected atoms which is shown to have a substantial effect on the oscillation amplitude.

\end{abstract}

\maketitle

\section{Introduction}

Hall thrusters are successfully used for electric propulsion in space, e.g.\ for satellite orbit keeping, and becoming an enabling technology of choice for long-term missions, such as trips to Mars. Despite the relatively long history of practical use (since 1972\cite{morozov2000fundamentals}), the crucial physical aspects of their operation are poorly understood. In the absence of predictive modeling capabilities, scaling of these devices for large (e.g., for long-term missions) and for low (for microsatellites) power is very difficult and expensive. The quantitative understanding of the physics of these devices remains an important task.

Plasma discharge in Hall thrusters (HT) is supported by the electrons drifting in the closed (periodic) azimuthal $\mathrm{E}\times\mathrm{B}$ direction, while the thrust is created by ions (effectively unmagnetized due to large gyroradius) accelerated by the electric field in the axial direction. One of the characteristics of Hall thrusters is the presence of turbulence and structures (azimuthal and axial) that affect their operation. Studies of nonlinear phenomena in these plasmas are not only of great practical importance but also address fundamental problems of plasma physics and plasma turbulence. In particular, the turbulent electron transport in such devices is orders of magnitude larger than the classical collisional transport (across the magnetic field) predicts. Inhomogeneous plasmas with $\mathrm{E}\times\mathrm{B}$ electron drift typically prone to various drift instabilities, both due to the fluid \cite{smolyakov2016fluid} and kinetic mechanisms \cite{lafleur2016theory, janhunen2018nonlinear, janhunen2018evolution, villafana2021rz, jimenez2021twod}, which drive high cross-field electron currents (for more details see Ref.~\onlinecite{kaganovich2020physics}).

Among the plethora of wave phenomena in a Hall thruster device, low-frequency oscillations propagating in axial direction stand as one of the most common and observed in most types of Hall thrusters \cite{morozov2000fundamentals}. They appear as the axial discharge current oscillations with frequencies of 5-50 kHz\cite{esipchuk1974plasma}. 
A strong periodic depletion of atoms in the ionization region is observed during the oscillations, suggesting the ionization nature. In the literature they are known as breathing modes (due to slow periodic plasma bursts out of the channel exhaust). Analytical studies of these phenomena are difficult due to the importance of nonlinear effects and the global nature of solutions, thus numerical methods have to be used. Qualitatively, the oscillation period of breathing modes depends on the travel time of neutral particles to the ionization region, e.g.\ for the characteristic \SI{1}{cm} and atom velocity $\SI{150}{m/s}$ gives \SI{15}{kHz}. 
Overall, we understand the phenomenology of the oscillations but cannot accurately predict their existence and amplitude. Generally accepted phenomenological description of these oscillations is described as the following sequence: decrease of the discharge current $\,\to\,$ decrease of ionization $\,\to\,$ increase of the neutral density in the exhaust region $\,\to\,$ increase of the electron conductivity in that region $\,\to\,$ increase of the current and ionization $\,\to\,$ neutral depletion $\,\to\,$ decrease of the current and so on. According to this picture, the oscillation frequency is related to the time necessary for the neutrals to refill the ionization region. 

The 0D predator-prey model proposed earlier \cite{fife1997numerical,barral2008origin,hara2014perturbation} is appealing because of its simplicity but fails to identify the conditions for the instability. Moreover, more accurate treatments show that the basic two-component (plasma-neutral) system with uniform ion and neutral velocities is stable \cite{hara2014perturbation,chapurin2021mechanism,LafleurJAP2021}. A simple model was proposed that the ion back-flow  region  which occurs near the anode as a result of large contribution of the electron diffusion current (due to the density gradient) and quasineutrality constrain provides a critical excitation mechanism for the breathing mode \cite{chapurin2021mechanism}. Linear resistively unstable modes \cite{chable2005numerical} and fluctuations of electron temperature
and power absorption \cite{hara2014perturbation,LafleurJAP2021} were also investigated as possible triggers of the breathing modes.

In general, several  physical mechanisms affect the breathing mode excitation and characteristics: electron momentum and energy losses to the wall, anomalous cross-field transport and heating, the ion backflow, and recombination at the anode. These mechanisms are inter-related, depend in a complex way on the magnetic field configuration, and are not easily quantifiable.
Numerical models that include many of these effects were proposed \cite{makowski2001review,hagelaar2004modelling,boeuf1998low,barral2001fluid,barral2008origin,barral2009low}. However, some calibration and adjustment of the parameters are required to satisfactorily reproduce the breathing modes characteristics observed experimentally \cite{giannetti2021numerical}. Therefore further insights on key physical processes are required to expand the predicting powers of such models, especially to new parameters range and new operational regimes.

While many numerical models for breathing modes based on fully fluid formulations, time-dependent, the hybrid modeling was also undertaken using the kinetic description for ions and neutrals \cite{boeuf1998low,MorozovPPR2000h,chable2005numerical,barral2001fluid,ShashkovPoP2017,gavrikov2021hybrid}. 
The extent to which the ion and neutrals kinetic effects influence the breathing mode excitation and characteristics remain a mute point of many studies. The goal  of this paper is to analyze the role of ions and neutrals kinetic effects under the same physics of the electron dynamics which is treated with the fluid theory.    We use the axial one-dimensional  full fluid and hybrid models  and compare their results.  

The basic fluid model describes ion and atoms with first two fluid moments (conservation of mass and momentum), and electrons are considered in drift-diffusion approximation with a full electron energy balance. In the hybrid model heavy species (ions, atoms) are kinetic (via particle-in-cell method) and electrons are fluid (modelled in exactly the same way in both approaches). Both models include plasma recombination at the anode and neutral dynamics with ionization due to electron-neutral impact. Plasma discharge is supported by the ionization process driven by the axial current due to the  applied potential across the domain. For the fluid simulations BOUT++ computational framework\cite{dudson2009bout++} is used. The hybrid code was developed in the LAPLACE laboratory, France \cite{hagelaar2002two, hagelaar2003role, hagelaar2004modelling}. The parameters of the simulations are chosen according to the ``Fluid/Hybrid'' test case in the LANDMARK  (Low temperAture magNetizeD plasMA  benchmaRKs) benchmarking project\cite{landmark}.
Previously, a comparison between fluid and hybrid models for the axial direction of Hall thruster configuration was presented in Ref.~\onlinecite{barral2001fluid}, however, this model did not include electron pressure gradients, thus omitting the effects of  electron diffusion resulting in the formation of the presheath region near the anode and ions transition through the ion-sound barrier. Neither  it   included the full electron energy balance.

One of the important finding of the present paper is the identification of  two distinct regimes of breathing oscillations; the result which was confirmed with both fluid and hybrid models during this benchmark. We show that the regime with higher electron energy losses exhibits the low-frequency mode (${\sim}\SI{14}{kHz}$) that coexists with the higher frequency ion ``transient-time'' oscillations (${\sim} \SI{150}{kHz}$)\cite{esipchuk1974plasma}. In the second regime, with low electron energy losses, purely breathing oscillations are observed, the so-called solo regime. We believe that different mechanisms are involved in these regimes. 

For the first regime, we identify the higher frequency oscillations as the excitation of the resistive modes (convective instability with the characteristic ion fly-by frequency)\cite{litvak2001resistive, chable2005numerical,fernandez2008growth,koshkarov2018current}. Such resistive-type modes appear in simple models without ionization or electron diffusion. The main feature of resistive modes is a strong dependence of growth rate and frequency on the electron mobility (resistivity)\cite{fernandez2008growth}. Similar features are shown in this work, while that the low frequency mode  (breathing mode) has weak or no dependency on the electron mobility (see Appendix~\ref{apdx_resistive}).

In fact, the frequency of resistive modes can vary significantly, ranging 0.1-10 MHz, and may become close to the breathing modes at the lower end of its spectrum. Some axial thruster models (with ionization but without electron diffusion)\cite{morozov2000fundamentals,MorozovPPR2000f} claimed that these modes might be responsible for the breathing modes observed in Hall thrusters. 
For clarity, here we will call breathing modes only those associated with atom depletion, whose frequencies scale accordingly to atom fly-by time and ionization processes.

In our earlier work, we have proposed a reduced model (only ion and atom dynamics included) for the second, solo regime of the breathing mode. In this regime, the instability is triggered by the ion backflow (negative ion velocity) in the near-anode presheath region \cite{chapurin2021mechanism}. It was demonstrated that such configuration is prone to low-frequency oscillations, where the ion backflow region is necessary. Recently, it has been shown in Ref.~\onlinecite{gavrikov2021some} with a more rigorous formal analysis that the sign-alternating ion velocity profile with a positive slope (i.e.\ negative ion velocity near anode and positive towards the exit) indeed is a necessary condition to excite the oscillations. It is also pointed out that the problem cannot be reduced to the 0D predator-prey model. 

Finally, we demonstrate the effect of atom temperature in the solo regime. We find that a finite energy spread   of injected atoms strongly suppress the oscillations amplitude compared to the cases of the injection with the same velocity (called below as monokinetic injection). It is found that even a small spread in atom temperature for the solo regime notably lowers the amplitude of breathing mode. For the first regime, with the presence of resistive modes, the atom temperature effect is negligible.
We also show the role of ion heating (due to the resistive modes) and selective ionization of neutrals.

The paper is organized as follows. In Section~\ref{sec_model} the detailed description of both fluid and hybrid models is given. Section~\ref{sec_results} defines the main features of two distinct regimes of low-frequency oscillations and presents results for three test cases with a detailed comparison between models.

\section{Fluid and hybrid models of low-frequency dynamics}\label{sec_model}

Detailed description of the full fluid model and the hybrid model is presented in this section. Two models share the same electron fluid equations (drift-diffusion approximation and energy evolution). The models are considered in the electrostatic and  quasineutral approximation, with the three species: neutral atoms, ions, and electrons. The ionization effects included via the electron-atom collisions, serving as a mechanism for supporting plasma discharge. Atom losses are only due to ionization effects, radial atom losses were not included. The models also includes the self-consistent electric field, the anode plasma recombination, the electron pressure effects, and the electron heat flux across the magnetic field.

\begin{figure}[ht]
\centering
\includegraphics[width=0.5\textwidth]{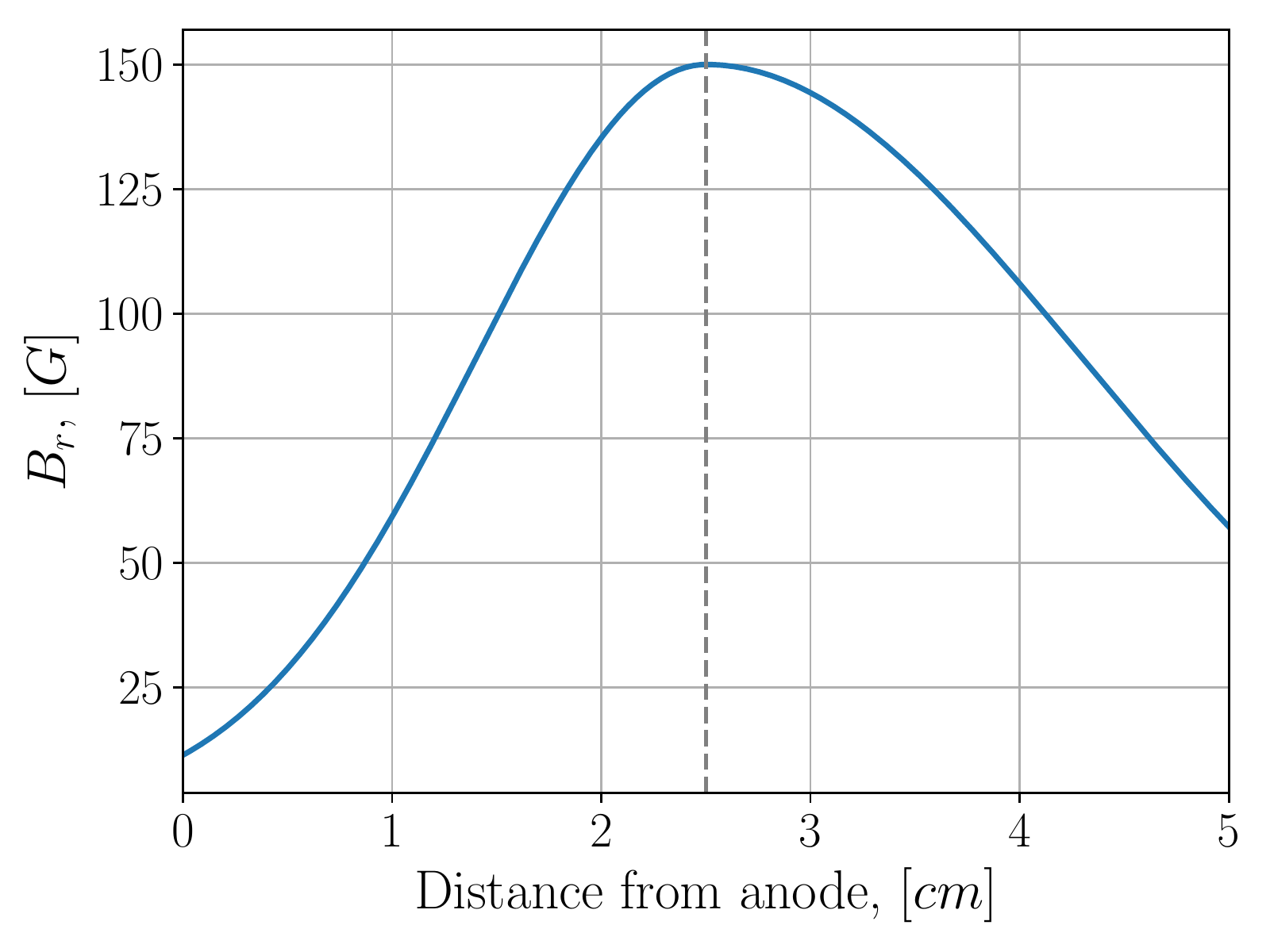}
\caption{The magnetic field profile used in simulations, with the channel exit located \SI{2.5}{cm} from anode (dashed line).}
\label{b-profile}
\end{figure}

The simulated length of $\SI{5}{cm}$ is assumed in the axial direction of a Hall thruster ($x$-direction), with the channel exit in the middle where the radial magnetic field has its maximum, Fig.~\ref{b-profile}. The profile of the magnetic field amplitude given by $B = B_0 \exp{\left[-\left(x-x_0\right)^2/2\delta_B^2\right]}$, 
where $x_0 = \SI{2.5}{cm}$ is the channel exit location, and $\delta_B$ is the characteristic width coefficient for the magnetic field profile, which are set  $\delta_{B,\text{in}} = \SI{1.1}{cm}$, $\delta_{B,\text{out}} = \SI{1.8}{cm}$, respectively for the inner and outer regions\cite{landmark}.

\subsection{Fluid model}\label{fluid_model}

First, a short description of each species dynamics is given and then the full system of time-dependent equations is formulated. For the neutral atoms, the constant flow velocity $V_a$ along the channel is considered, and the continuity equation with the source term is used to describe their dynamics:
\begin{equation} \label{na}
\frac{\partial n_a}{\partial t} + V_a \frac{\partial n_a}{\partial x} = -\beta n_a n_e,
\end{equation}
where $n_a$ is the atom number density, $n_e$ is the electron number density, $\beta(\varepsilon)$ is the ionization rate coefficient that depends on the electron energy $\varepsilon = (3/2) T_e$, where $T_e$ is the electron temperature. The ionization due to electron-atom impact produces a pair of ion and electron with a loss of neutral atom, hence the sink term $-\beta n_i n_e$ (same and opposite sign source terms are included in the ion and electron continuity equations). The ionization rate $\beta(\varepsilon) = \langle \sigma(v) v \rangle$ is obtained via BOLSIG\cite{hagelaar2005solving} by averaging over Maxwellian EEDF with the cross-sections from the SIGLO database\cite{siglo}. 

The ion species are unmagnetized (gyroradius is much larger than the thruster dimensions for a typical magnitude of a magnetic field in the thruster) and described with the conservation of number density and momentum equations:
\begin{eqnarray}
&& \frac{\partial n_i}{\partial t} + \frac{\partial}{\partial x}\left(n_i V_i\right) = \beta n_a n_e, \label{ic} \\
&& \frac{\partial V_i}{\partial t} + V_i\frac{\partial V_i}{\partial x} = \frac{e}{m_i}E + \beta n_a \left( V_a - V_i \right), \label{im}
\end{eqnarray}
where $n_i$ is the ion density, $V_i$ is the ion flow velocity, $E$ is the axial  electric field, $e$ is the elementary charge, $m_i$ is the ion mass (Xenon, \SI{131.293}{amu}). The ion pressure term and generalized viscosity tensor are neglected in this model (ions are ballistic with temperatures much lower than that of electron component). 

The magnetized electron species are described with the first three fluid moment equations: (electron inertia is neglected)
\begin{eqnarray}
&& \frac{\partial n_e}{\partial t} + \frac{\partial}{\partial x}\left(n_e V_{ex}\right) = \beta n_n n_e, \label{ec} \\
&& 0 = -\frac{e}{m_e} \mathbf{E} - \frac{e B}{m_e} \left( \mathbf{V}_{e \perp} \times \mathbf{\hat{z}}\right) - \frac{1}{n_e m_e} \frac{\partial \left(n_e T_e \right)}{\partial x} - \nu_{m} \mathbf{V}_{e \perp}, \label{em} \\
&& \frac32 \frac{\partial }{\partial t} \left(nT_e\right) + \frac52 \frac{\partial }{\partial x}\left(n_e V_{ex} T_e \right) + \frac{\partial q_e}{\partial x} = -n_e V_{ex}\frac{\partial \phi}{\partial x} - n_e n_a \mathrm{K} - n \mathrm{W}, \label{ee}
\end{eqnarray}
where $n_e$ is the electron density, $\mathbf{V}_{e \perp} = (V_{ex}, V_{e \theta})$ is the electron flow velocity perpendicular to the magnetic field ($x$ and $\theta$ are axial and azimuthal coordinates), $m_e$ is electron mass, $B$ is the external magnetic field, $\nu_m$ is the total electron momentum exchange frequency, $\mathrm{W}$ is the anomalous energy loss coefficient, $\mathrm{K}$ is the collisional energy loss coefficient, generated by BOLSIG\cite{hagelaar2005solving} as a table-valued function, and $q_e$ is the electron heat flux. Phenomenological  anomalous electron energy loss coefficient (e.g., described the sheath radial energy losses) $\mathrm{W}$ is introduced \cite{boeuf1998low} as
\begin{equation}\label{el_an_loss}
\mathrm{W} = \nu_{\varepsilon} \varepsilon \exp{\left(-U/\varepsilon\right)},
\end{equation}
where $\varepsilon = 3 T_e /2$, $U = \SI{20}{eV}$, $\nu_{\varepsilon}$ is the anomalous losses frequency coefficient. The heat flux across the magnetic field is
\begin{equation}
q_e = - \frac{5}{2} \mu_e n T_e \frac{\partial T_e}{\partial x}.
\end{equation}

The electron momentum conservation equation~(\ref{em}) is simplified assuming no pressure gradients nor equilibrium electric fields in other than axial direction, then the axial electron velocity (denoted further as $V_e$) can be expressed as
\begin{equation}\label{ve}
V_{e} = -\mu_e E - \frac{\mu_e}{n_e} \frac{\partial (nT_e)}{\partial x},
\end{equation}
where the electron mobility $\mu_e$ is the well-known classical electron mobility across the magnetic field:
\begin{equation}\label{muex}
\mu_e = \frac{e}{m_e \nu_m} \frac{1}{1+\omega_{ce}^2/\nu_m^2},
\end{equation}
where $\omega_{ce}=eB/m_e$ is the electron cyclotron frequency. Eq.~(\ref{ve}) commonly called the drift-diffusion equation. The model of electron transport as based on the assumption of the following total electron momentum exchange collision frequency:
\begin{equation}\label{nue}
\nu_{m} = \nu_{en} + \nu_{walls} + \nu_{B},
\end{equation}
where the electron-neutral collision frequency $\nu_{en}$, electron-wall collision frequency $\nu_{walls}$, and anomalous Bohm frequency $\nu_{B}$ are given with
\begin{eqnarray}
\nu_{en} = k_m n_a, \\
\nu_{walls} = \alpha \SI{e7}{[s^{-1}]}, \\
\nu_{B} = \left(\beta_a/16\right) eB/m_e.
\end{eqnarray}
where $k_m = \SI{2.5e-13}{m^{-3} s^{-1}}$, $\alpha$ and $\beta_a$ are free parameters. 
For the electron mobility model different parameters are used inside and outside the channel (denoted additionally as \textit{in}, \textit{out}): the near wall conductivity contribution $\alpha_{\text{in}} = 1, \ \alpha_{\text{out}} = 0$, the anomalous contribution is set to $\beta_{a,\text{in}} = 0.1$, $\beta_{a,\text{out}} = 1$.

Here we seek the low-frequency and bulk plasma modes, thus electron inertia is neglected (as shown above) and further the full plasma quasineutrality is assumed. One can quantify\cite{makowski2001review} the quasineutral approximation using the Poisson equation $\varepsilon_0 \partial E / \partial x = e(n_i - n_e)$ ($\varepsilon_0$ is the permittivity of free space). With the typical values of electric field $E \approx \SI{e4}{V/m}$ and the size of acceleration zone $\SI{1}{cm}$, average plasma density $n_0 = \SI{e17}{m^{-3}}$, the difference $(n_i - n_e)/n_0 \approx \num{5e-4}$.
Instead of the Poisson equation, the electric field is found from the electron momentum equation as shown below. Note that while the quasineutrality neglects a potential drop on the Debye sheath near the anode, it still allows the presheath region to form if the electron pressure is included\cite{CohenZurPoP2002,AhedoPoP2005,dorf2003anode, dorf2005experimental}, as in our electron model. The presheath is the region where the electric field is induced to accelerate ions towards a plasma boundary to compensate the electron current due to the pressure gradient. 

Thus, the full system of time-dependent fluid equations to be solved include Eqs.~(\ref{na},\ref{ic},\ref{im},\ref{ee}), along with the drift-diffusion form for the electron velocity Eq.~(\ref{ve}). Full quasineutrality $n = n_i = n_e$ is enforced and the self-consistent electric field is found via the electron drift-diffusion equation~(\ref{ve}), given by
\begin{equation}\label{ef}
E = \frac{J_T}{en\mu_e} - \frac{V_i}{\mu_e} - \frac{1}{n} \frac{\partial n T_e}{\partial x},
\end{equation}
where the total current density $J_T = en\left(V_i - V_e\right)$. Here $J_T$ is constant in space (divergenceless current), which can be seen by combining continuity equations for ions and electrons with quasineutral assumption. We will use the integral approach, consisting of the evaluation the total current density $J_T$ via the constraint $\int_0^L E dx = U_0$ (then it is substituted to Eq.~(\ref{ef}) to evaluate the electric field), which yields to
\begin{equation} \label{jt-nodiff}
J_T = \dfrac{U_0 + \bigintss_0^L \left( \dfrac{V_i}{\mu_e}  + \dfrac{1}{n} \dfrac{\partial p_e}{\partial x} \right) dx} {\bigintss_0^L \dfrac{dx}{en\mu_e}},
\end{equation}
where $L$ is the system length, $U_0$ is the applied voltage (without sheath voltage).

The fluid model is solved with the following boundary conditions. A constant mass flow rate $\dot{m}$ and the full recombination of plasma that flows to the anode determines the value of $n_a$ at the anode boundary,
\begin{equation}\label{nn_backflow}
n_a(0) = \frac{\dot{m}}{m_i A V_a} - \frac{nV_i(0)}{V_a},
\end{equation}
where $A$ is the anode surface area. The value of the ion velocity is  imposed at the anode $V_i(0) = -b_v \sqrt{T_e/m_i}$ with the parameter $b_v=0\text{--}1$, the Bohm velocity factor that can be varied. Both anode and cathode electron temperature are fixed with $T_e(0) = T_e(L) = \SI{2}{eV}$. 

As noted in Ref.~\onlinecite{makowski2001review}, plasma acceleration in the configuration of the axial direction of a Hall thruster shows similarities to the flow in de Laval nozzle. Indeed, the whole acceleration region can be split into subsonic $V_i < c_s$ and supersonic $V_i > c_s$ regions, where $c_s = \sqrt{T_e/m_i}$ is the ion sound speed. While in de Laval nozzle the transition through sonic point happens at the region with the smallest cross-section of the channel (due to extrema condition and regularity requirement), for the bounded plasma configuration the position is determined via the nonlinear relationship between plasma parameters (and their first derivative) at the sonic point and a value of the total current\cite{smolyakov2019stationary}. Note that the total current $J_T$, given by Eq.~(\ref{jt-nodiff}), is a function of $U_0$ and the integral dependence on all main plasma parameters, thus the problem is inherently nonlocal, which has no analogy with the standard Laval nozzle. Another difference is that in the axial direction of Hall thrusters the presheath region can induce the backward ion flow in a large portion of the thruster channel (see Case 2 below).

\subsection{Hybrid model}\label{hybrid_model}

The hybrid model has the same electron equations as in the fluid model, while ions and neutrals are modeled via particle-in-cell (PIC) method \cite{hagelaar2002two, hagelaar2003role, hagelaar2004modelling}. The plasma recombination effect is also included via the relationship~(\ref{nn_backflow}), but the ion velocity at the anode is not forced to satisfy the Bohm velocity. The ionization is included via the electron-atom impact with the Monte Carlo simulations (via the null collision method\cite{birdsall1991particle, hagelaar2008modelling}) with the known ionization rate $\beta(\varepsilon)$, obtained in the same way as described above the fluid model.
Neutral atoms are injected with the constant flow rate $\dot{m}$ either with the constant velocity $V_a$ (monokinetic), thus $f(v_x) = \delta(v_x-V_a)$, or with the half-Maxwellian velocity distribution function at the left wall (anode),
\begin{equation}\label{halfMaxw}
    f(v_x) = \frac{2v_x}{v_{Ta}^2} \exp ({-\frac{v_x^2}{v_{Ta}^2}}), \ \ v_x > 0,
\end{equation}
where $v_{Ta}^2 = 2 k_B T_a/m_a$ is the atom thermal speed ($m_a = m_i$), $T_a$ is the atom temperature (in \si{K}). For the half-Maxwellian injection~(\ref{halfMaxw}) the average flow velocity is $v_{Ta}/\sqrt{\pi}$.
Ions are assumed only singly charged and unmagnetized, thus they are only accelerated by an electric field. Ions are produced accordingly to the ionization rate coefficient, i.e.\ self-consistently with the electron temperature evolution and local atom density. Both atoms and ions are lost at boundaries. In this model the ion velocity is not forced to the Bohm velocity, like in the fluid ion model. With the quasineutral approach ($n_e = n_i$ is forced at every time step), plasma density is evaluated from the ion particle distribution and thereafter used for the electron temperature~(\ref{ee}) and the electric field~(\ref{ve}) calculations. Formally, the evolution of the distribution function for ions $f_i(x,v_x,t)$ and atoms $f_a(x,v_x,t)$ is described with the Boltzmann equation for each specie:
\begin{eqnarray}
    \frac{\partial f_i}{\partial t} + v_{ix}\frac{\partial f_i}{\partial x} + \frac{e}{m_i}E\frac{\partial f_i}{\partial v} = S(x,v_x), \label{boltz_i}\\
   \frac{\partial f_a}{\partial t} + v_{ax}\frac{\partial f_a}{\partial x} =- S(x,v_x), \label{boltz_a}
\end{eqnarray}
where $S(x,v_x)$ is the collisional source term due to the ionization. In the case of the monokinetic target species (atoms), the ionization leads to the ion creation with the atoms velocity $V_a$ (constant) and the source term can be expressed\cite{boeuf1998low} as $S(x,v_x) = \beta n_e n_a \delta(v_x-V_a)$. For the simulations with finite atom temperature, newly created ions assigned velocities by sampling from the isotropic Maxwellian distribution with the temperature $T_a$ with the standard sampling techniques\cite{cartwright2000loading}. Solutions to Eqs.~(\ref{boltz_i},\ref{boltz_a}) effectively obtained by solving the motion equations for the corresponding specie (the method of characteristics via PIC method). 
Eqs.~(\ref{boltz_i},\ref{boltz_a}) and the electron fluid equations~(\ref{ee}, \ref{ve}) form the complete set of equations solved in the hybrid model.

\section{Simulation results and comparison}\label{sec_results}
\begin{figure}[ht]
\centering
\subfloat[]{\includegraphics[width=0.49\textwidth]{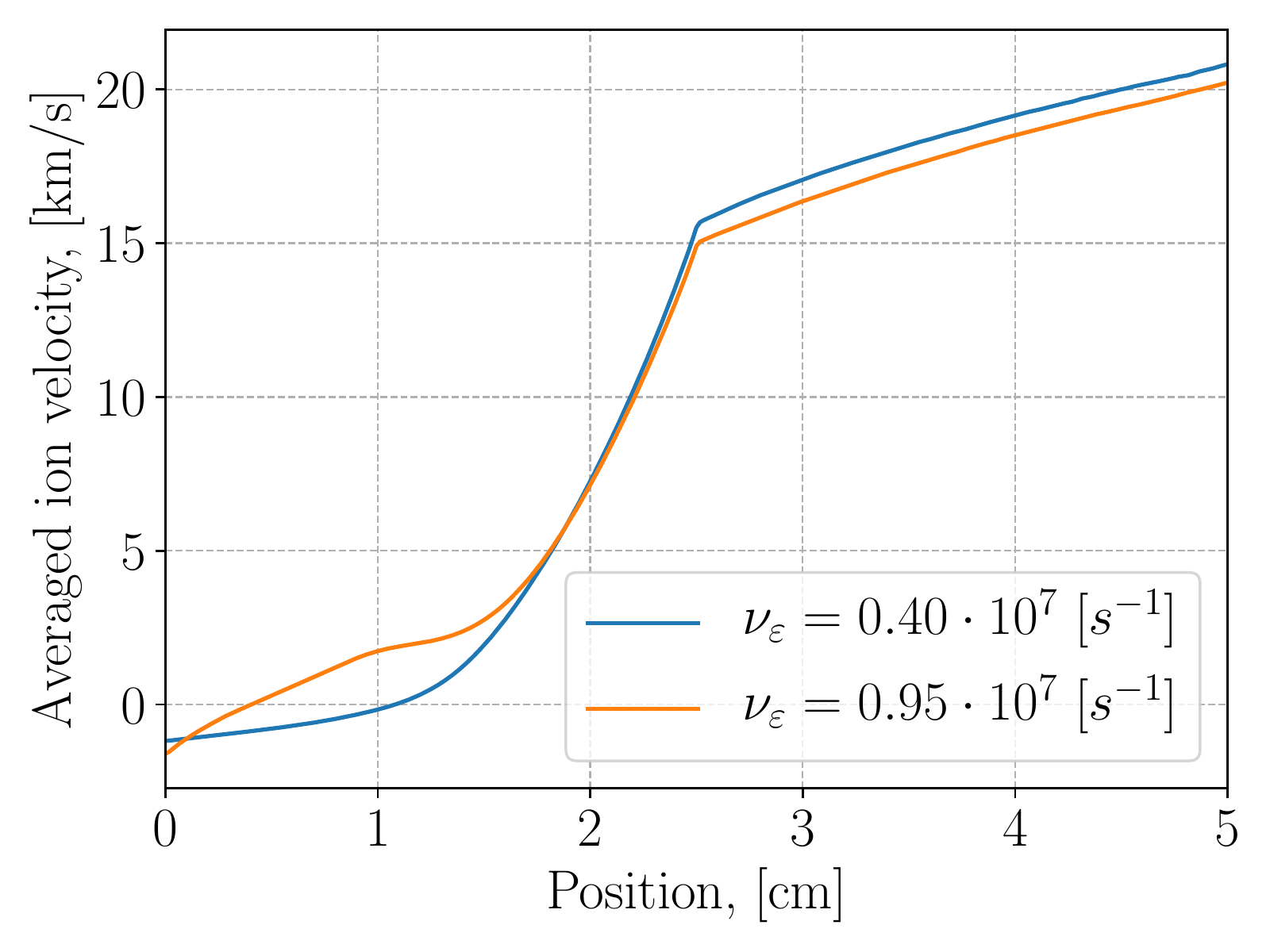}\label{vi_pros_nueps}}
\subfloat[]{\includegraphics[width=0.49\textwidth]{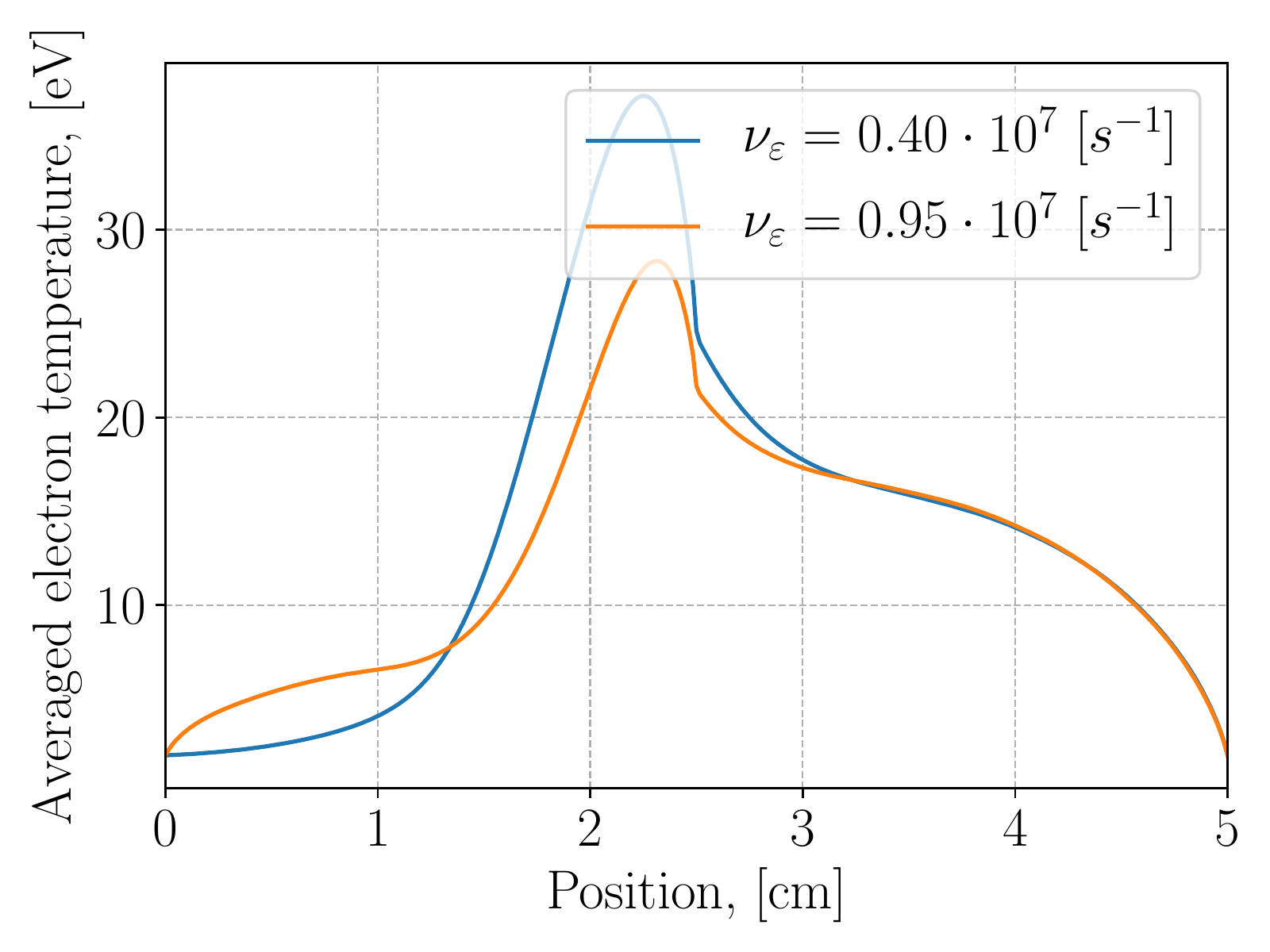}\label{te_pros_nueps}}
\caption{Averaged in time profiles of ion velocity (a) and electron temperature (b) for Case 1 ($\nu_{\varepsilon} = \SI{0.4e7}{s^{-1}}$) and Case 2 ($\nu_{\varepsilon} = \SI{0.95e7}{s^{-1}}$). The result obtained with the hybrid model.}
\end{figure}

\begin{figure}[H]
\centering
\includegraphics[width=0.65\textwidth]{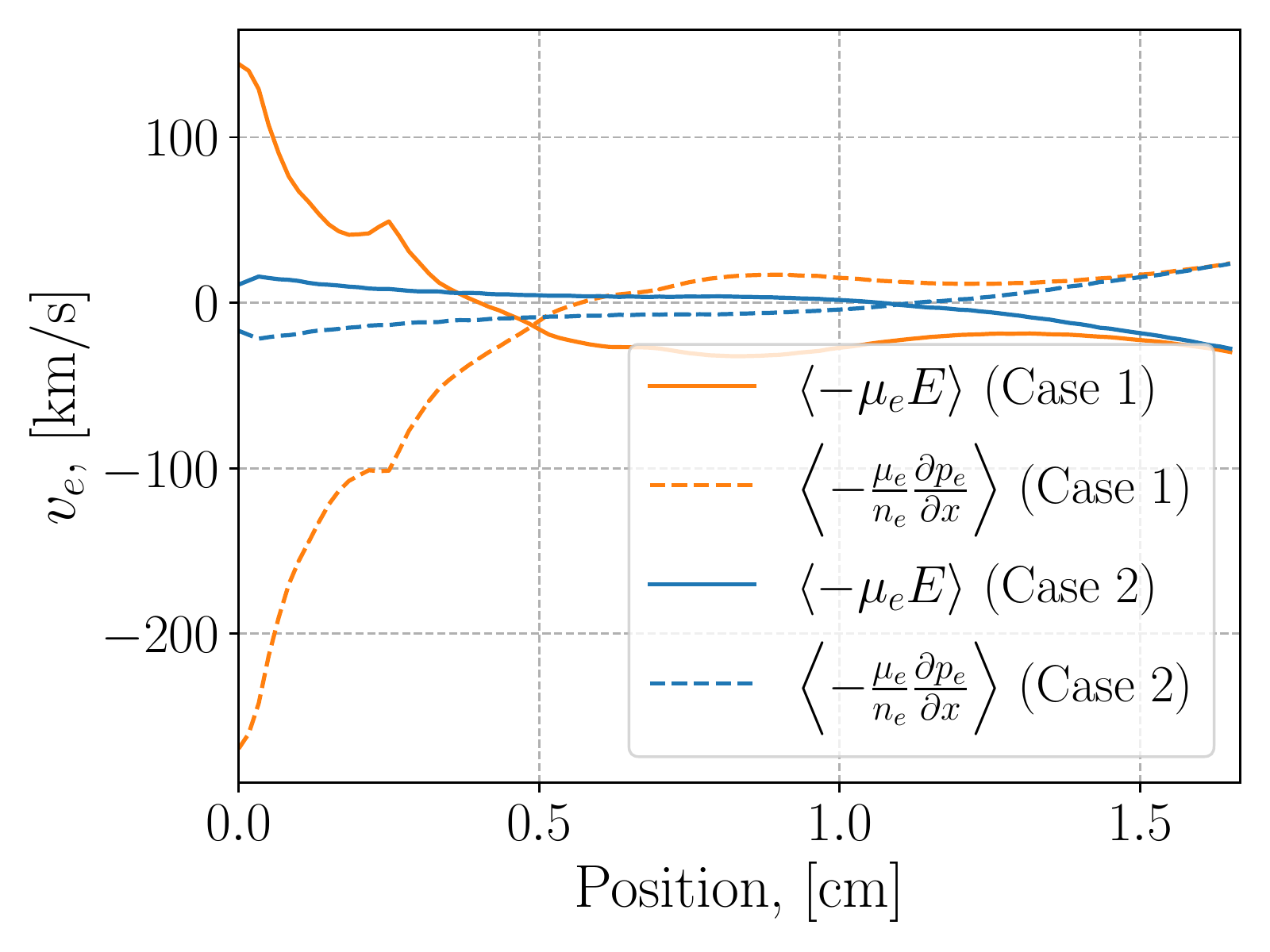}
\caption{Electron flow velocity components, response to the electric field and due to due for Case 1 and 2 (result obtained with the hybrid model).}
\label{ve_chann}
\end{figure}

\begin{figure}[H]
\centering
\includegraphics[width=0.65\textwidth]{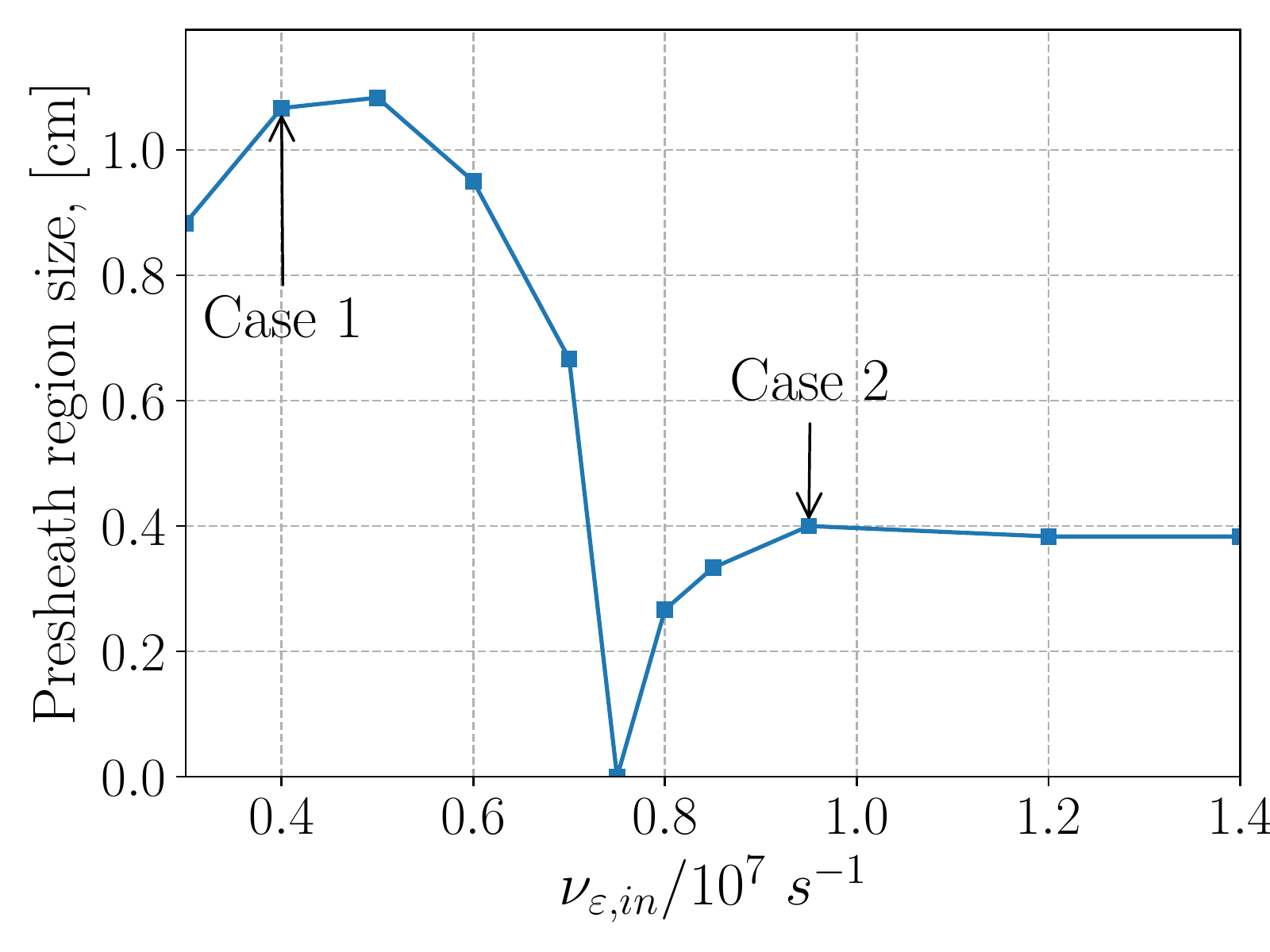}
\caption{Extent of the presheath zone, defined as the region with negative electric field near the anode, as a function of anomalous electron energy losses coefficient $\nu_{\varepsilon}$ (result obtained with the hybrid model).}
\label{ps_size}
\end{figure}

The fluid and the hybrid models described above were studied for the three test cases (denoted as Cases 1-3). Cases 1-2 will demonstrate two distinct regimes of low-frequency oscillations, and Case 3 shows the effect of atom temperature. Note that both Cases 1-2 use monokinetic atoms with the velocity $v_a = \SI{150}{m/s}$. Cases 1-2 are chosen with the following observation: the larger values of $\nu_{\varepsilon,\text{in}}$ allow the high frequency (of ion fly-by time) modes to appear and coexist with the low-frequency modes, represented in Case 1. The simulations with lower value of $\nu_{\varepsilon,\text{in}}$ reveal only the large amplitude low-frequency oscillations, we call it the solo regime. In Case 1 the anomalous energy loss coefficient $\nu_{\varepsilon,\text{in}} = \SI{0.95e7}{s^{-1}}$, and in Case 2 $\nu_{\varepsilon,\text{in}} = \SI{0.4e7}{s^{-1}}$. This is the only parameter distinguishing Cases 1 and 2. Note that for all cases reported in this paper we keep $\nu_{\varepsilon,\text{out}} = \SI{e7}{s^{-1}}$. Besides the different time-dependent behaviour, these regimes show a notable difference in the time-averaged axial profiles of the ion velocity and the electron energy, see Figs.~\ref{vi_pros_nueps},\ref{te_pros_nueps}. 
For both Cases the ion velocity profile is similar near the exit and beyond, but in the near-anode region the ion backflow region (where ions are moving towards the anode) is much shorter for Case 1 (Fig.~\ref{vi_pros_nueps}). The ion backflow region is associated with the presheath formation near the anode (with negative electric field). The size of the presheath region as a function of $\nu_{\varepsilon,\text{in}}$ is shown in Fig.~\ref{ps_size}, where the transition between regimes with large and short backflow region happens near the value $\nu_{\varepsilon,\text{in}} = \SI{0.75e7}{s^{-1}}$. 
The electron temperature for Case 2 mostly concentrated near the channel exit (Fig.~\ref{te_pros_nueps}) with very low values in the near-anode region, and for Case 1 its spread more uniformly and to the near-anode region, which might be due to the higher electron flow velocity near the anode, see Fig.~\ref{ve_chann} for the time-averaged electron velocity components from Eq.~(\ref{ve}). Note substantially higher electron current due to the pressure gradient in Case 1 for $x < \SI{0.4}{cm}$, which also results in a larger total electron flow velocity. Indeed, the average gradient parameter $L_{n}^{-1} = \partial_x n_i / n_i$ is about five times larger for Case 1. Recall, that the presheath region in formed due to diffusive electron current by inducing ion current (total current is conserved), generating negative electric in this region. All features presented above clearly distinguish Case 1 and Case 2.

Finally, Case 3 demonstrates the effect of finite atom temperature, where atoms injected to the system with the half-Maxwellian distribution, Eq.~(\ref{halfMaxw}). The main effect of atom temperature is found to be a significant reduction of breathing mode amplitude to those observed in Case 2. Besides finite atom temperature, all parameters for Case 3 are exactly the same as in Case 2, and the atom temperature is set to $T_a = \SI{500}{K}$ (average injected flow velocity \SI{142}{m/s}, close to monokinetic \SI{142}{m/s} used in Cases 1-2).


\subsection{Case 1: Low electron energy losses; the co-existence of  low and high frequency  modes}

\begin{figure}[H]
\centering
\subfloat[]{\includegraphics[width=0.49\textwidth]{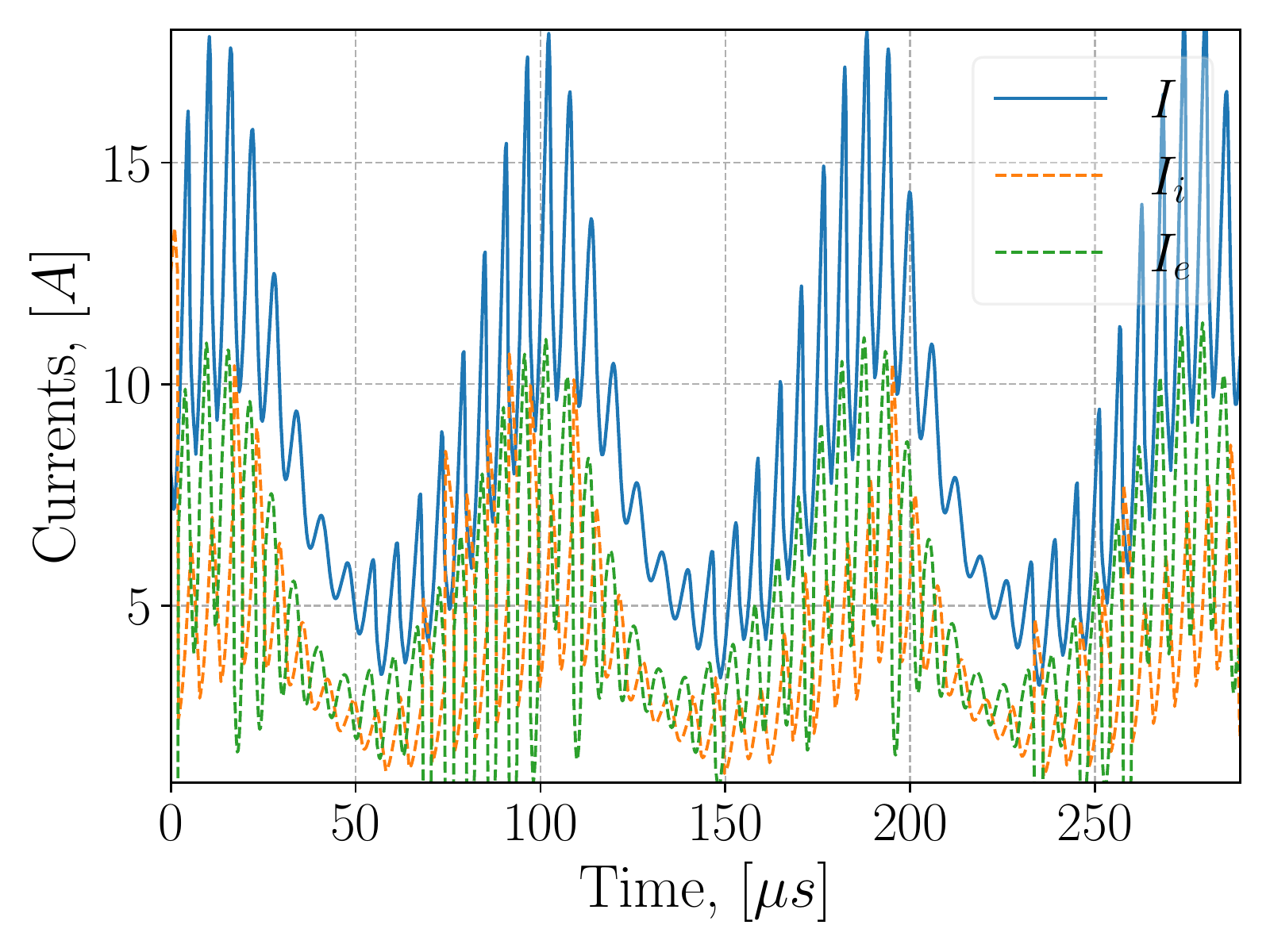}\label{case1_fluid_cur}}
\subfloat[]{\includegraphics[width=0.49\textwidth]{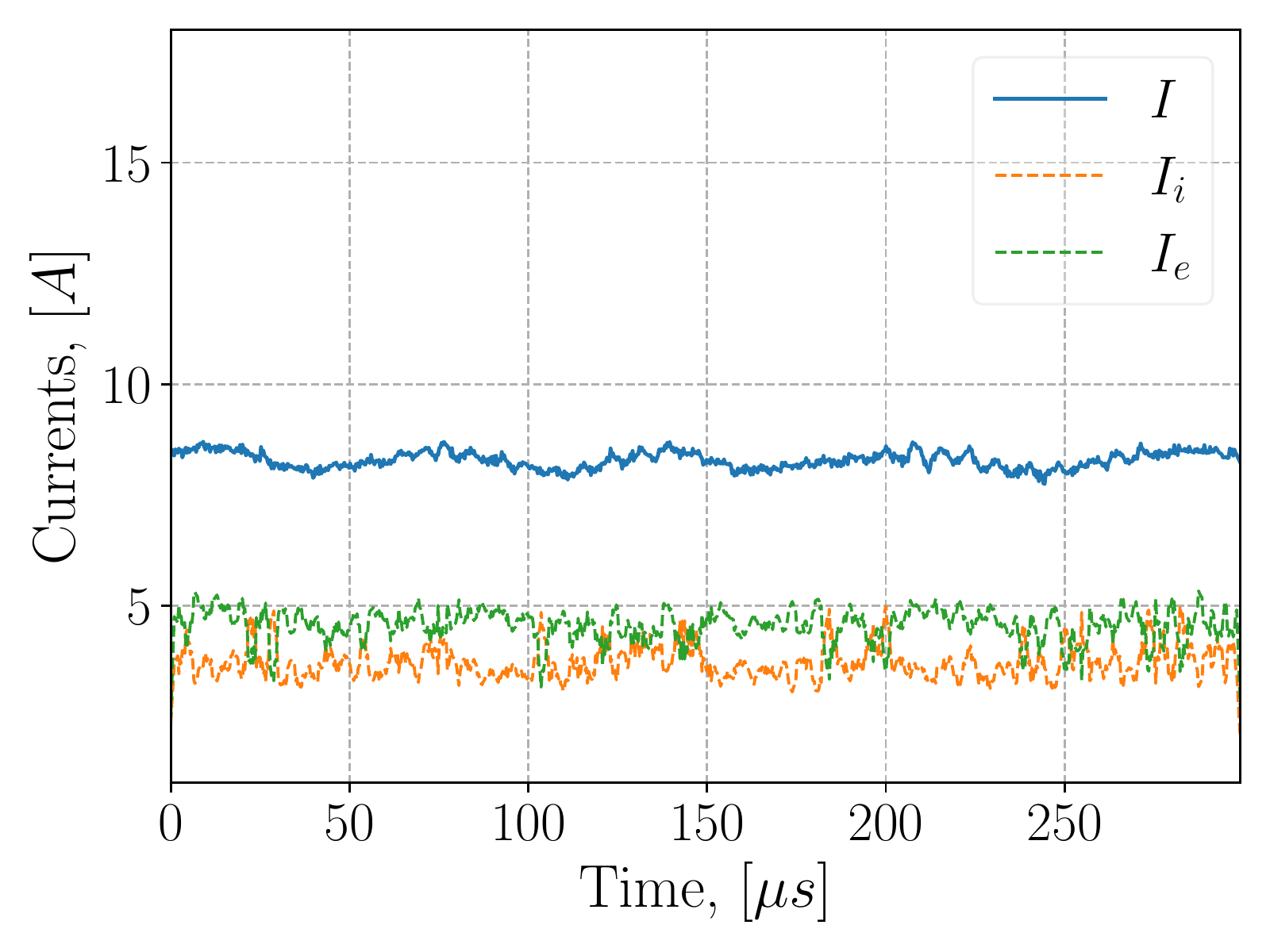}\label{case1_hybrid_cur}}
\caption{Amplitudes of the total, ion, and electron currents in fluid model (a) and hybrid model (b). Ion and electron currents are evaluated at $x = \SI{5}{cm}$. Small averaging window of length \SI{2.3}{\mu s} is applied to the ion and electron currents (to filter out high-frequency noise).}
\end{figure}

This case exhibits both low and higher frequency oscillations in the fluid and the hybrid models. The hybrid model shows a smaller total current amplitude, see Figs.~\ref{case1_fluid_cur}, \ref{case1_hybrid_cur}. The time-averaged total currents are close, \SI{8.2}{A} in the hybrid model, and \SI{8.3}{A} in the fluid model. The time-averaged ratio of the ion current (at the plume exit, $x = \SI{5}{cm}$) to the total current is 45\% in the hybrid model and 48\% in the fluid.

\begin{figure}[H]
\centering
\subfloat[]{\includegraphics[width=0.49\textwidth]{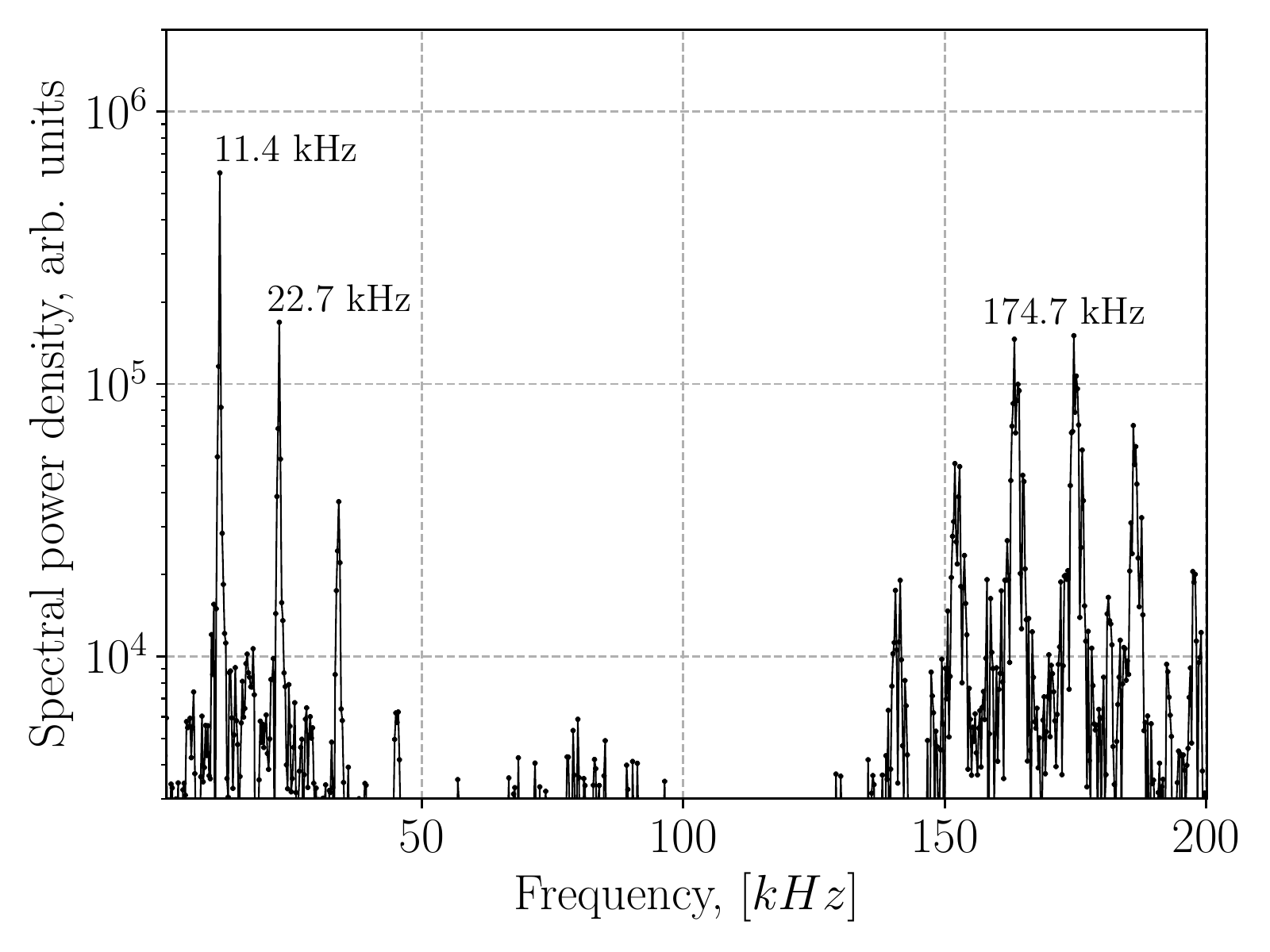}\label{case1_fluid_ps}}
\subfloat[]{\includegraphics[width=0.49\textwidth]{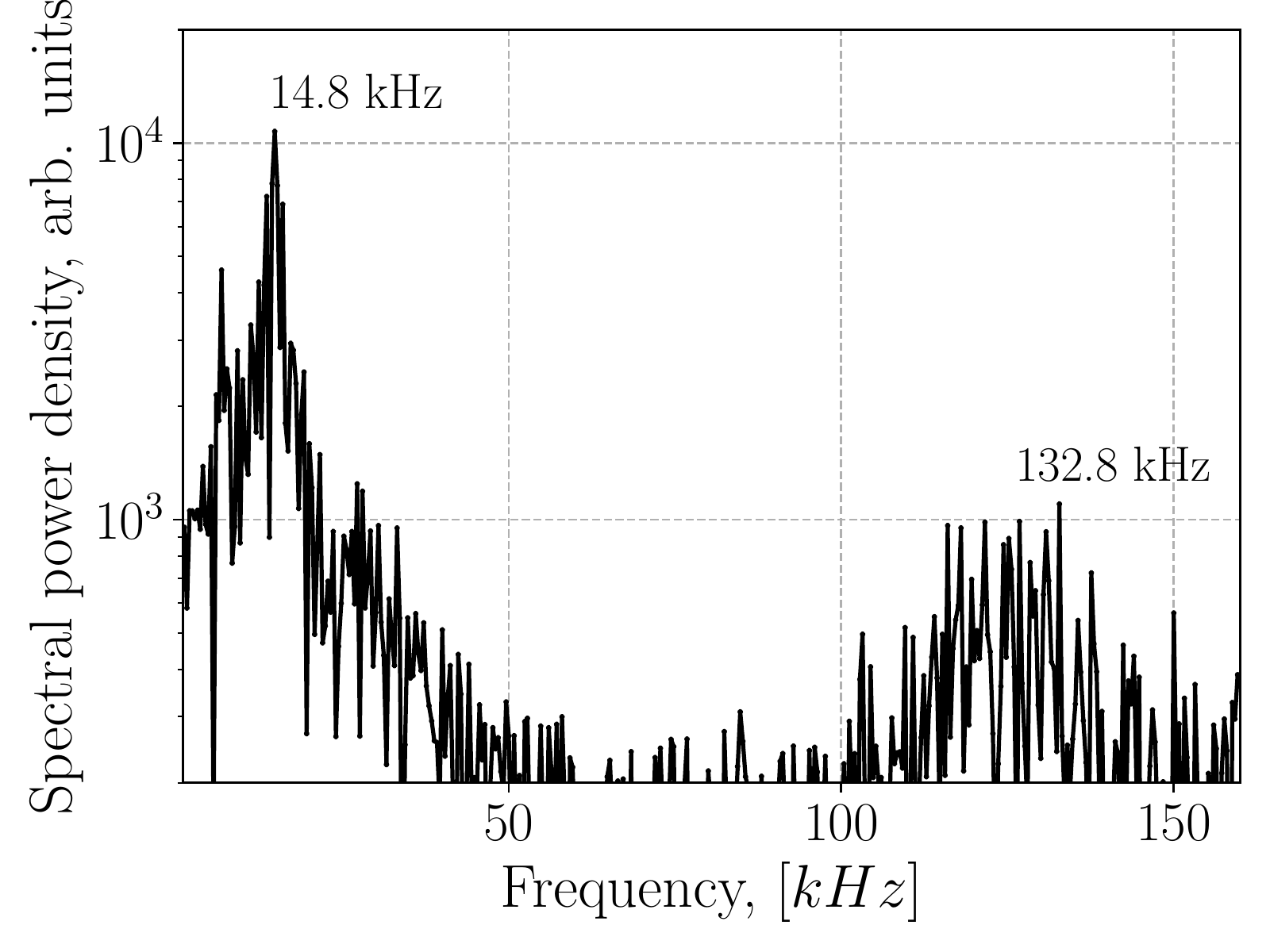}\label{case1_hybrid_ps}}
\caption{Spectral density of the total current yield in fluid model (a) and hybrid model (b).}
\end{figure}

Spectral power of the total current  also shows some differences in both low- and high-frequency range, Figs.~\ref{case1_fluid_ps},\ref{case1_hybrid_ps}. The main low-frequency mode in the fluid model is \SI{11.4}{kHz}, while it is \SI{14.4}{kHz} in the hybrid model. The total current signal in the hybrid model contains more noise (statistical noise due to the use of macroparticles), but the high-frequency component is clearly seen at around \SI{125}{kHz}. In the fluid model the high-frequency mode is larger, centered at about \SI{175}{kHz}.

\begin{figure}[H]
\centering
\subfloat[]{\includegraphics[width=0.49\textwidth]{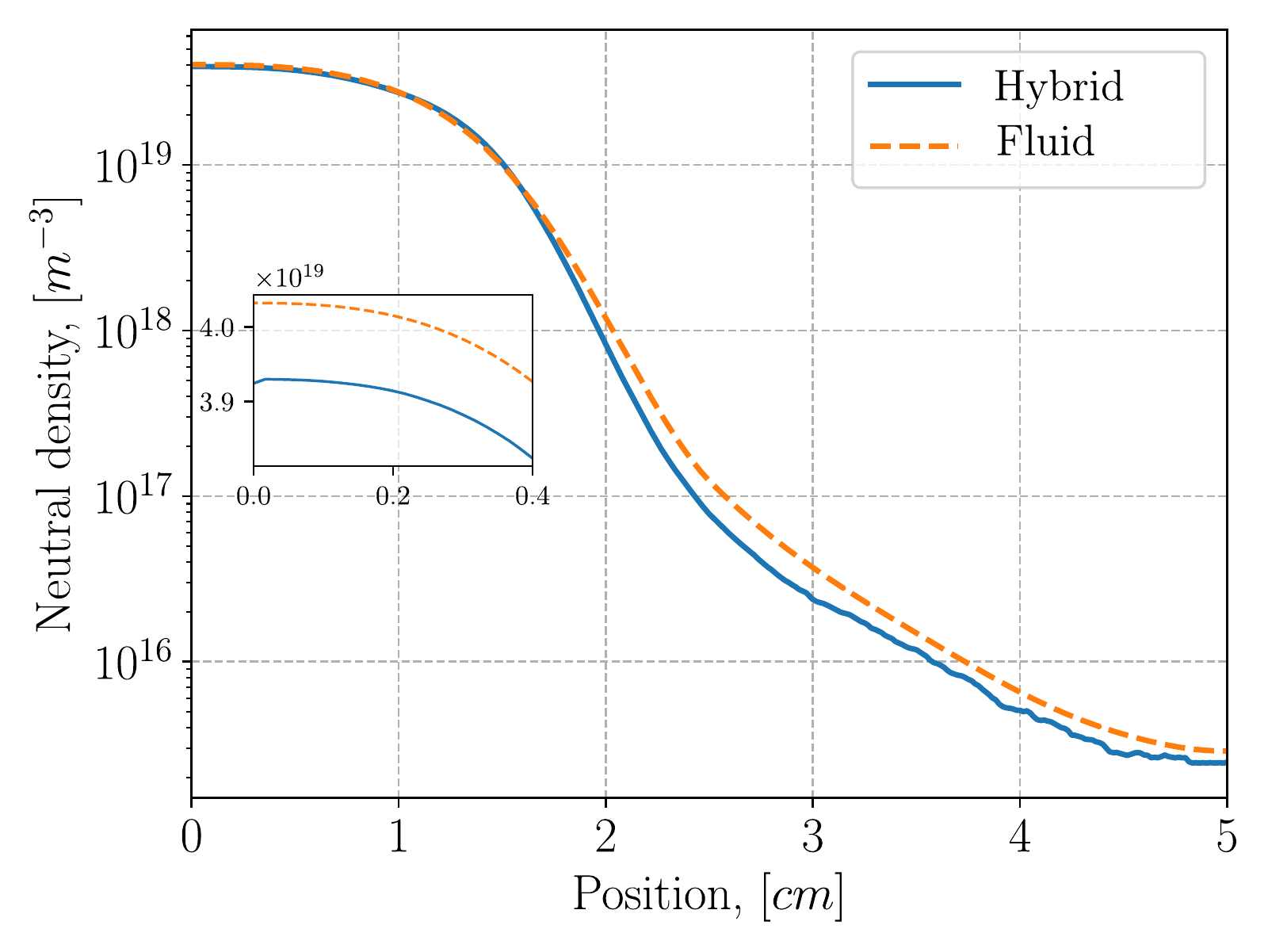}\label{case1_nn_cmpr1}}
\subfloat[]{\includegraphics[width=0.49\textwidth]{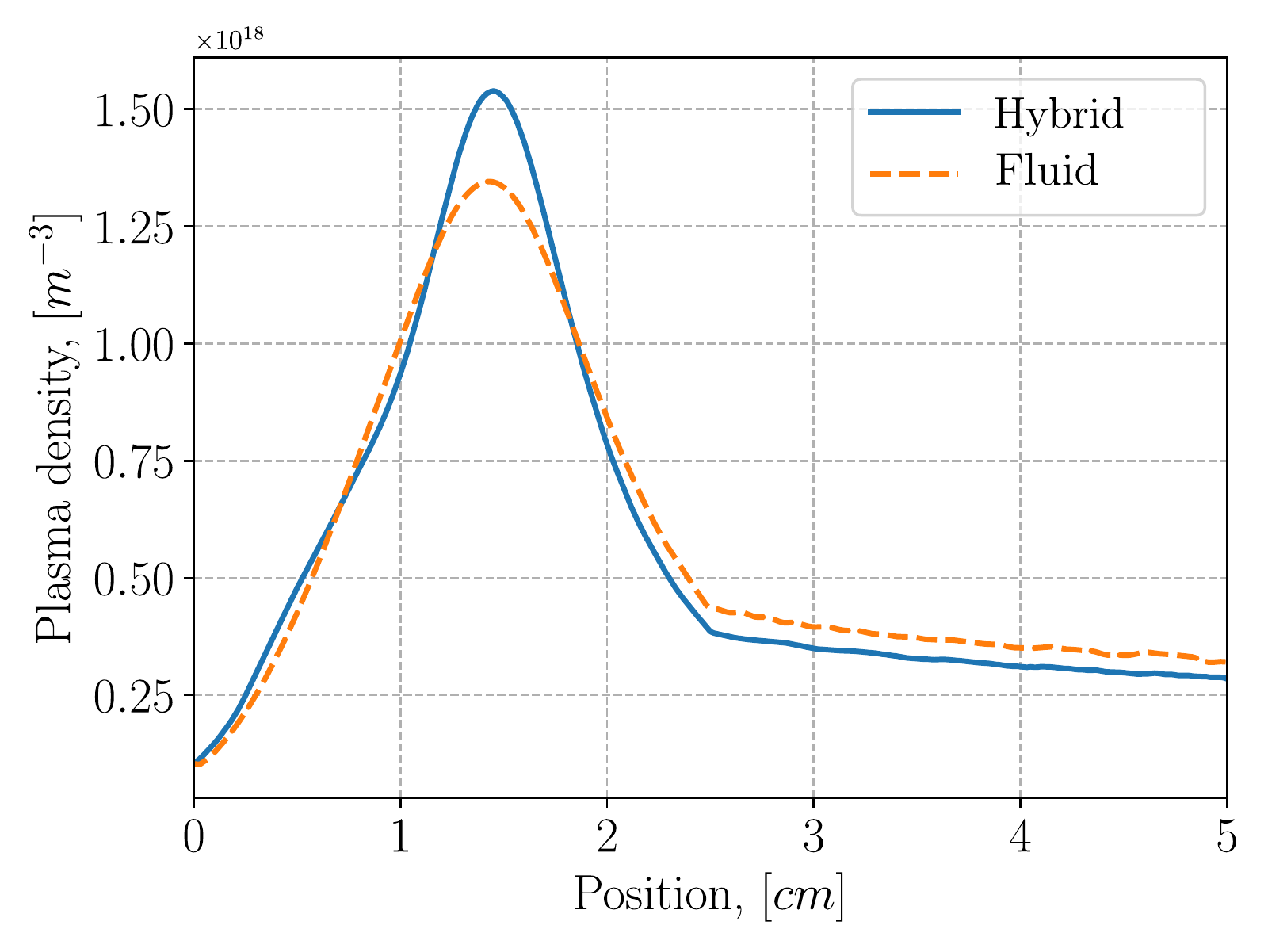}\label{case1_ni_cmpr1}} \\
\subfloat[]{\includegraphics[width=0.49\textwidth]{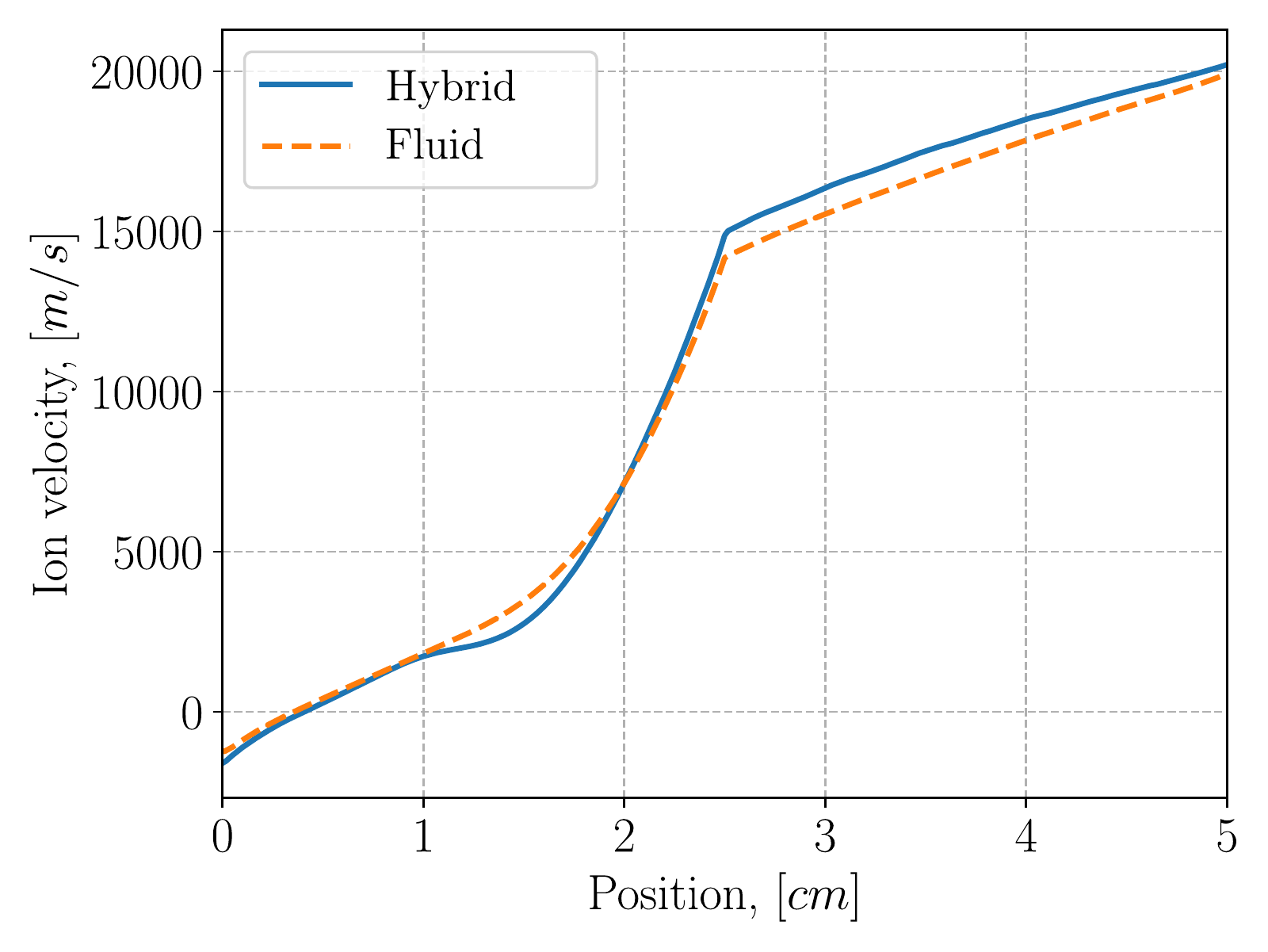}\label{case1_vi_cmpr1}}
\subfloat[]{\includegraphics[width=0.49\textwidth]{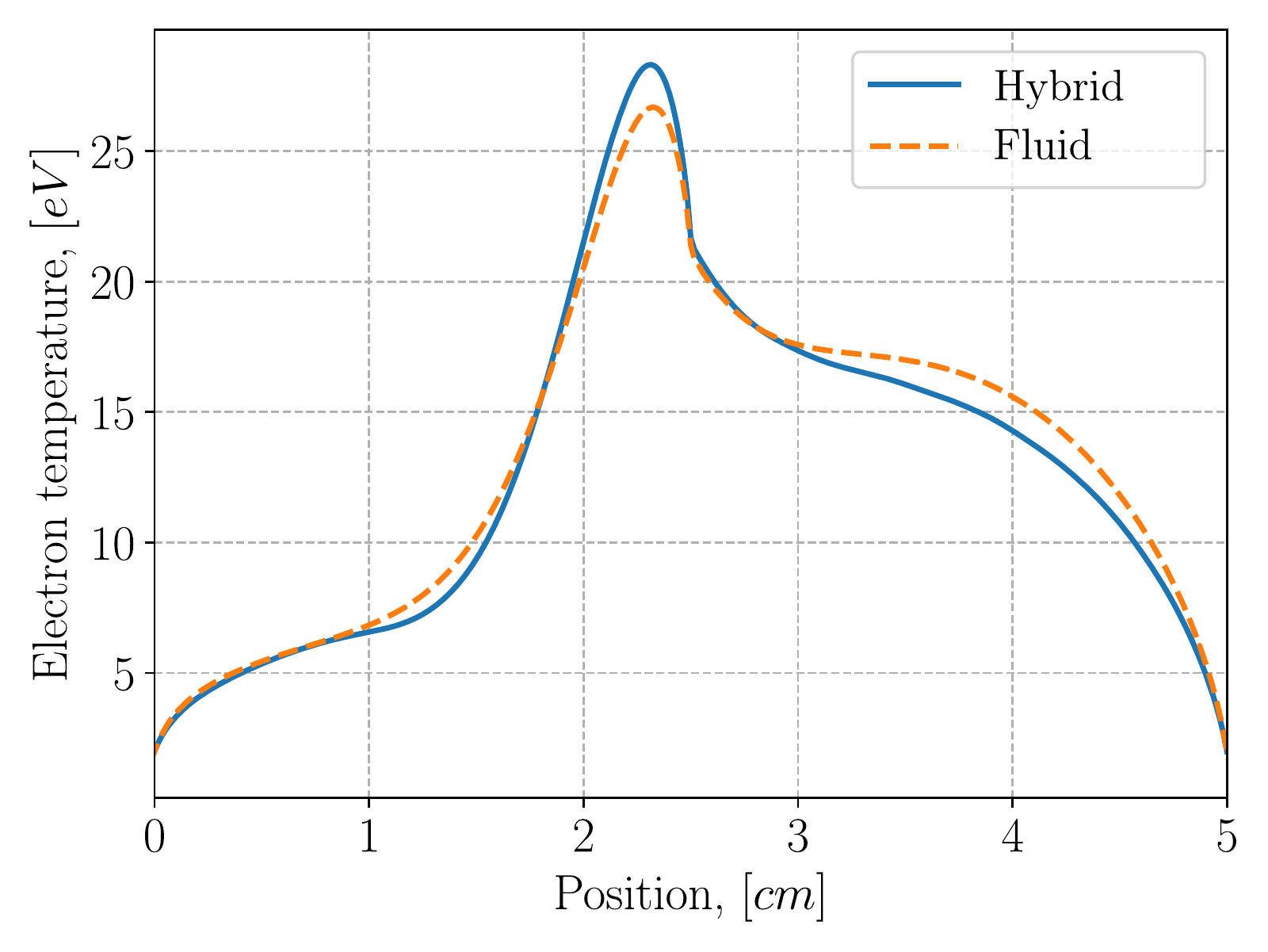}\label{case1_te_cmpr1}}
\caption{Comparison of time-averaged axial macroscopic profiles resulted from fluid and hybrid models of neutral density (a), plasma density (b), ion flow velocity (c), and electron temperature (d).}
\end{figure}

Besides the currents, a more rigorous comparison between the two models is shown in terms of time-averaged axial profiles of various plasma quantities. Due to the oscillatory nature of these solutions, the averaging time window was chosen as the ten periods of the corresponding main low-frequency mode in each simulation. The profiles depicted in Figs.~\ref{case1_nn_cmpr1}-\ref{case1_te_cmpr1}. The main discrepancy lies in the peak plasma density in the ionization (source) region at about \SI{1.3}{cm} from the anode (higher value in the hybrid model). Also, the ion velocity in the plume ($x>\SI{2.5}{cm}$) is slightly higher in the hybrid model. 

\begin{figure}[H]
\centering
\subfloat[]{\includegraphics[width=0.5\textwidth]{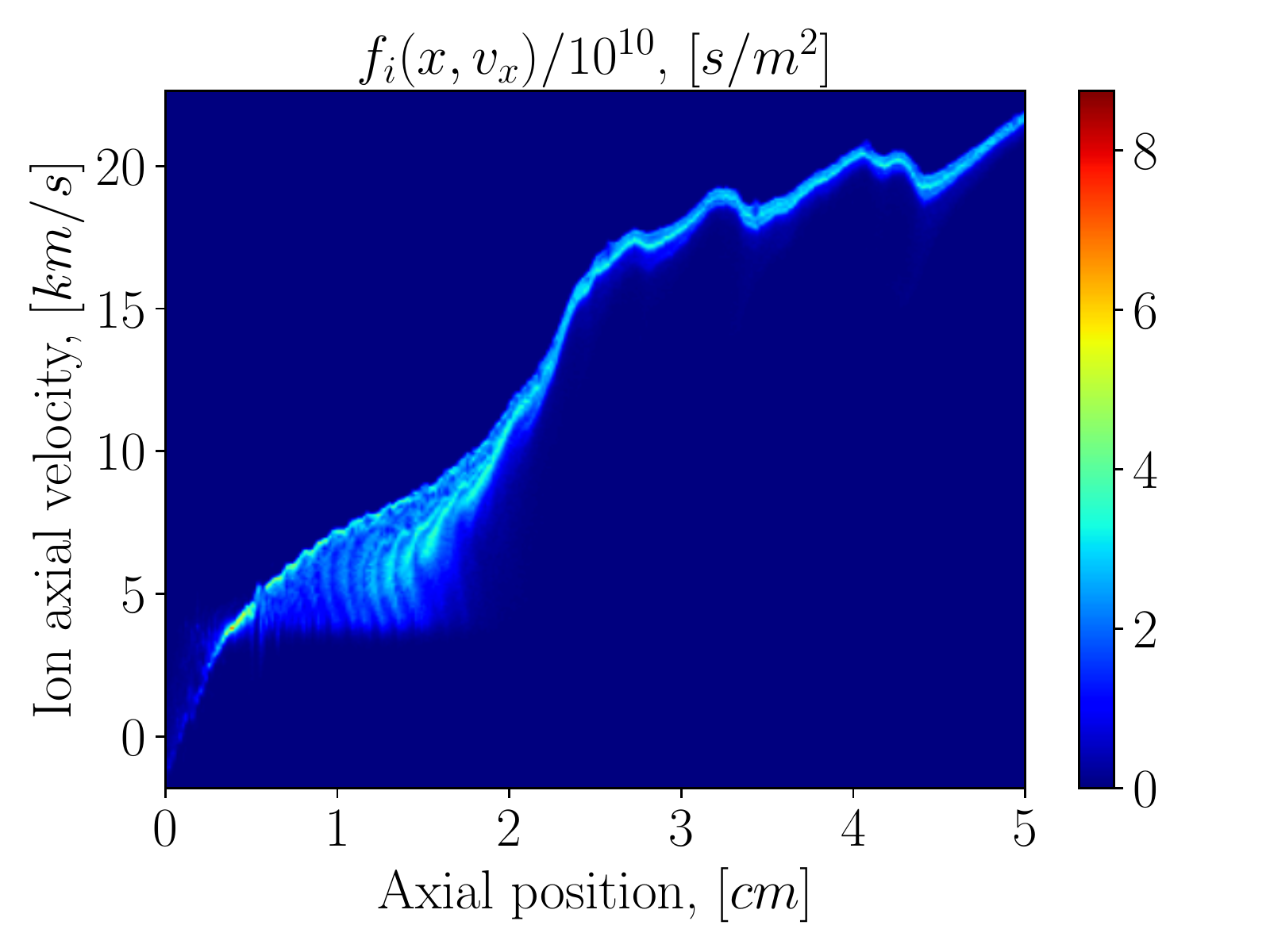}\label{ivdf_case1}}
\subfloat[]{\includegraphics[width=0.5\textwidth]{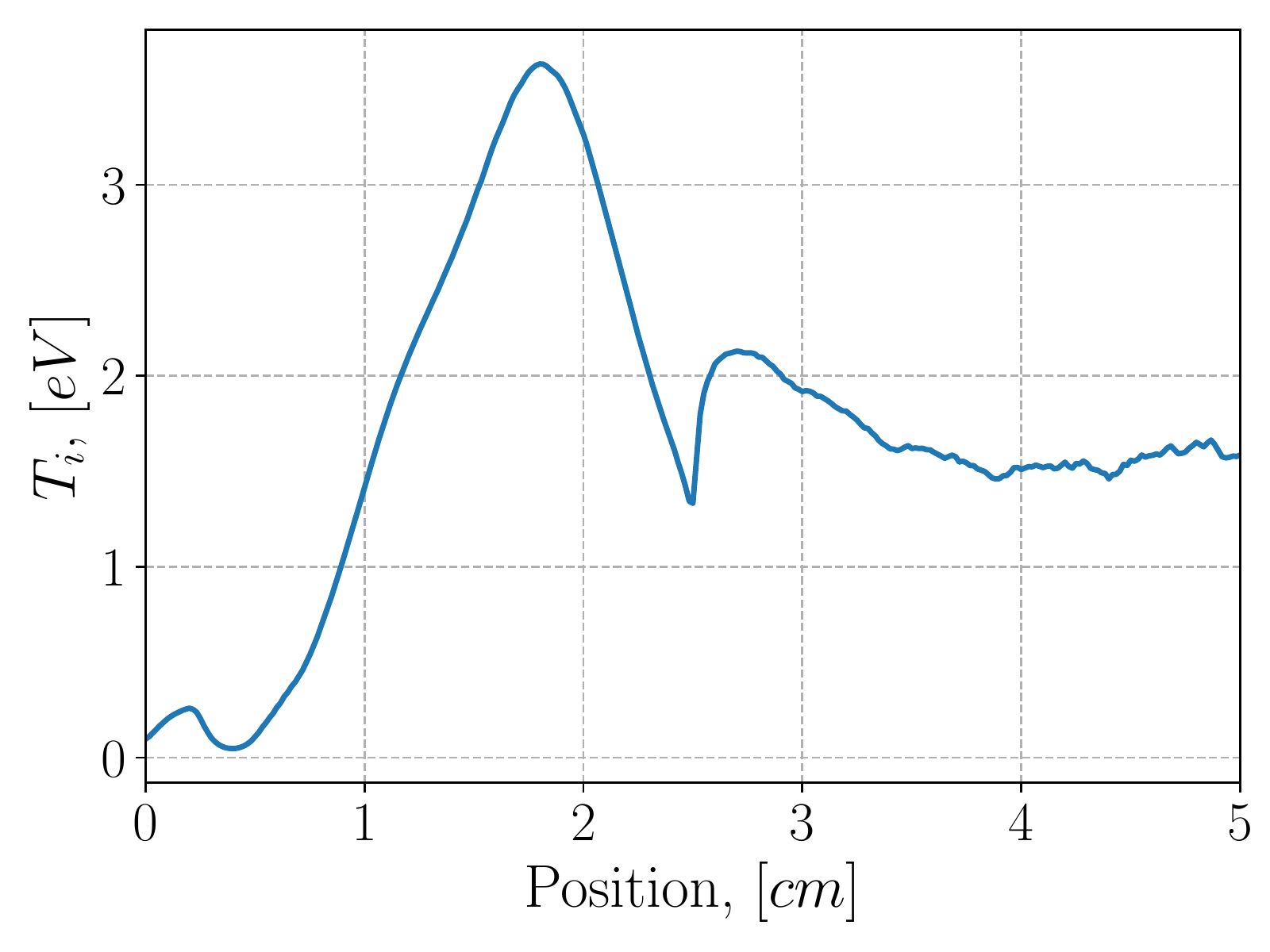}\label{tix_case1}}
\caption{Instantaneous image of ion distribution function (in space of axial coordinate and axial velocity) in the hybrid model (a). Ion temperature spatial profile (time-averaged) evaluated from ion kinetic representation in the hybrid model (b).}
\end{figure}

The ion phase space (hybrid model) is shown in Fig.~\ref{ivdf_case1}. The ion velocity distribution function (IVDF) is highly inhomogeneous inside the channel, suggesting that the higher fluid moments may play a role (recall, that the closure in the ion fluid model is given by nullified pressure term). Typically, ion pressure effects are neglected due to a low ion temperature; ion formation is due to the ionization process as they carry the low atom temperature even accelerating by the axial electric field.

\begin{figure}[H]
\centering
\subfloat[]{\includegraphics[width=0.49\textwidth]{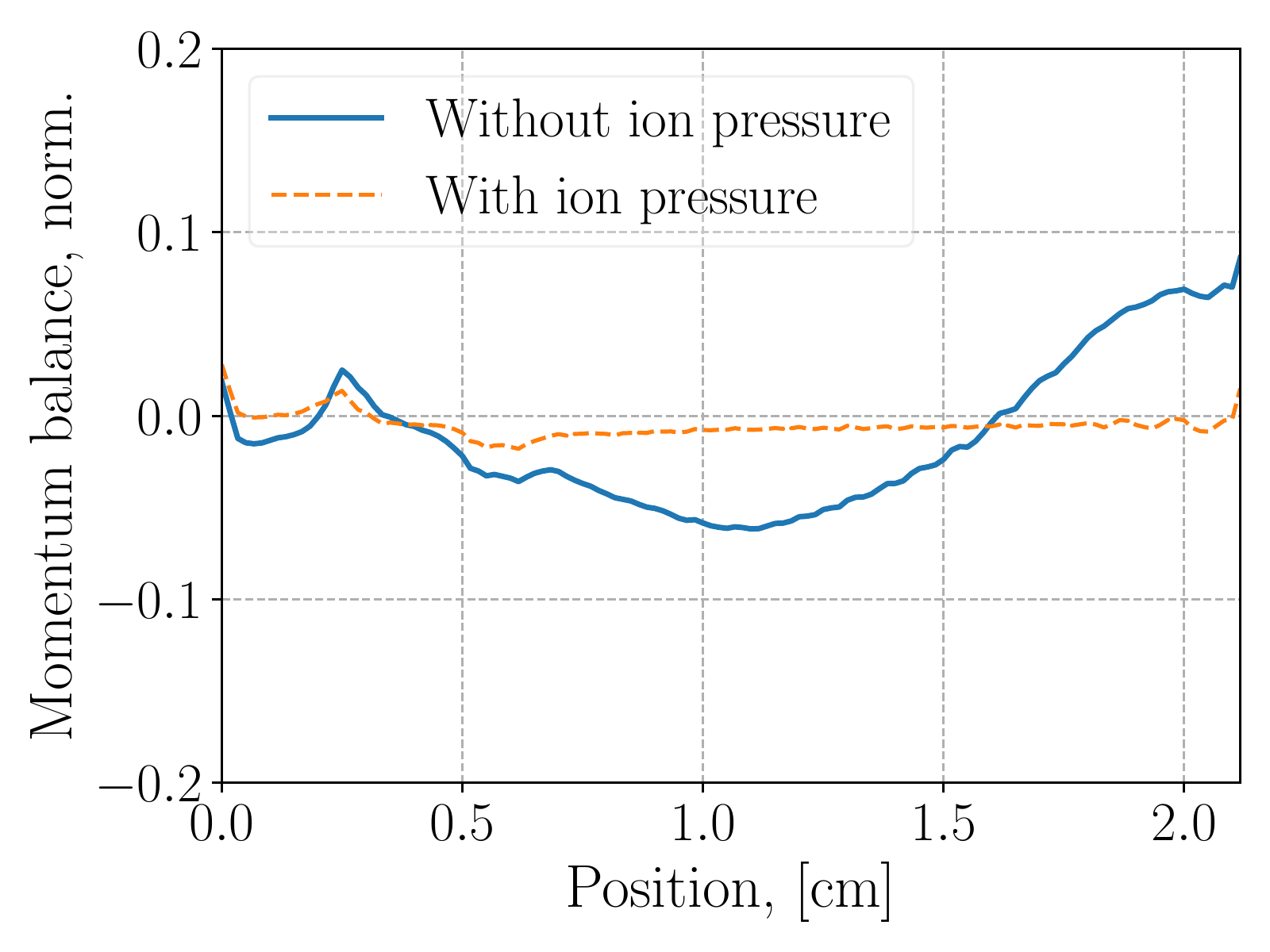}\label{ion_mom_bal_case1}}
\subfloat[]{\includegraphics[width=0.49\textwidth]{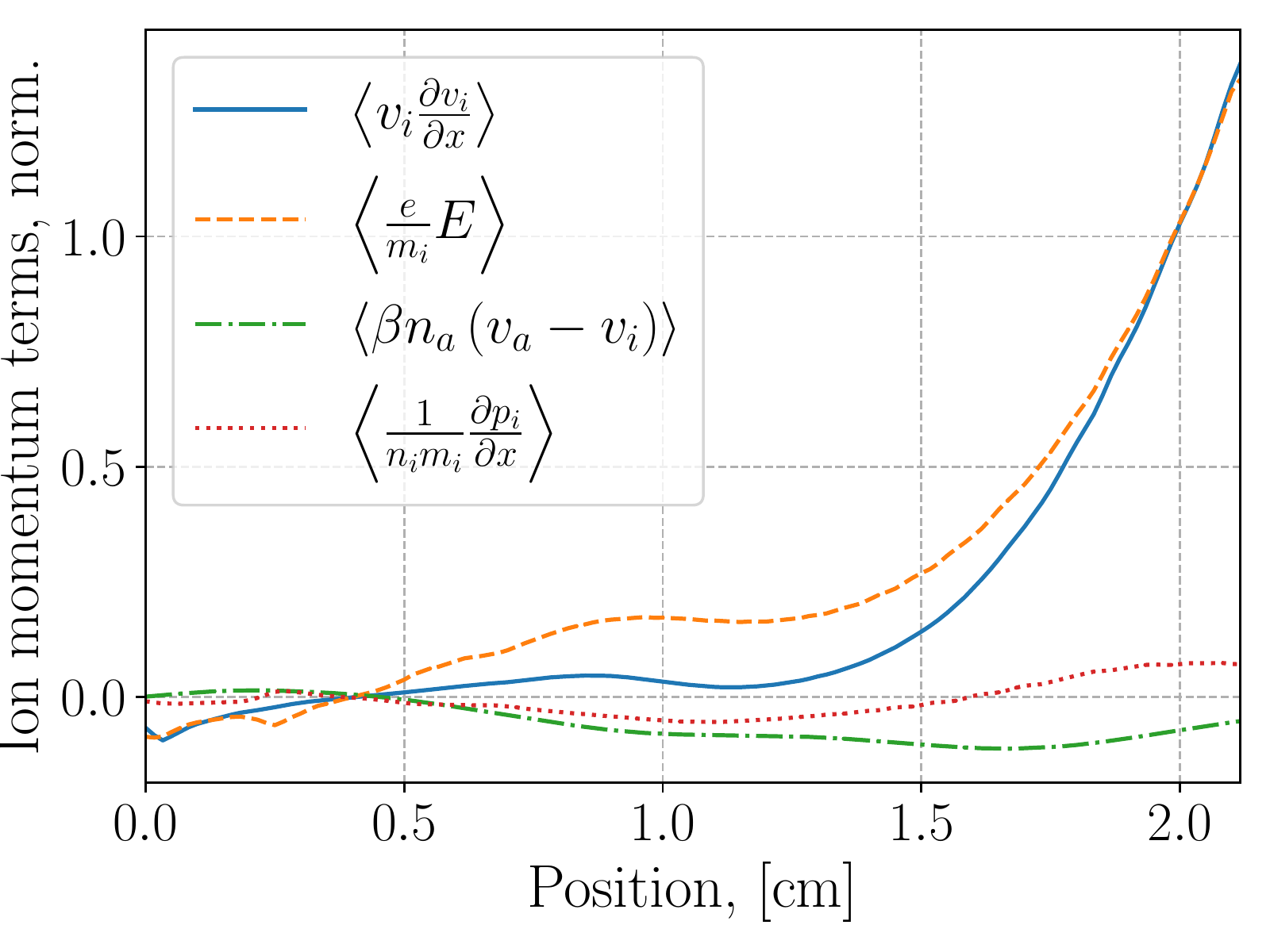}\label{ion_mom_terms_case1}}
\caption{Difference between left and right hands sides of the ion momentum balance equation~(ion-mom) evaluated from the kinetic description of the hybrid model (a). Separate terms of the same equation averaged in time (denoted with angle brackets) over few periods of the low-frequency component (b).}
\end{figure}

To check the validity of this assumption for this case the fluid moments were calculated from the kinetic particle representation in the hybrid model. It allows to test the ion momentum balance equation (to identify the role of the ion pressure term) in a more complete form
\begin{equation}\label{im_with_pi}
\frac{\partial V_i}{\partial t} + V_i\frac{\partial V_i}{\partial x} = \frac{e}{m_i}E - \frac{1}{n_i}\frac{\partial p_i}{\partial x} + \beta n_a \left( V_a - V_i \right), 
\end{equation}
where the fluid moments, such as ion density $n_i$, ion flow velocity $V_i$, and ion pressure $p_i$ are evaluated from the ion distribution function in the following way:
\begin{eqnarray}\label{ion-mom}
    && n_i = \int f_{i} dv_{ix}, \\
    && V_i = \frac{1}{n_i} m_i \int v_x f_i dv_{ix}, \\
    && p_{i} = m_i \int v_{i}^{\prime} v_{i}^{\prime} f_i dv_{ix},
\end{eqnarray}
$v_{i}^{\prime} = v_{i} - V_{i}$ is the random component of particle velocity, and $V_{i}$ is the average (flow) velocity. The time-averaged profile of the ion temperature $T_i = p_i/n_i$ is shown in Fig.~\ref{tix_case1}, revealing higher than typical values of the ion temperature, with the average over the whole domain of \SI{1.7}{eV}. For the momentum balance test, each term in Eq.~(\ref{im_with_pi}) was evaluated as a function of time and space and then averaged in time over a few periods of the main low-frequency mode. Fig.~\ref{ion_mom_bal_case1} shows the difference of left- and right-hand sides of equation~(\ref{im_with_pi}) that ideally must add up to zero. The plotted terms were normalized to the value $V^2/L$ where the ballistic ion velocity is $V^2 = 2 e U_0/m_i$ with the potential difference $U_0 = \SI{300}{V}$ over the system length $L$. It is seen that the ion pressure term notably improves the overall ion momentum balance, suggesting the fluid model for this configuration should not ignore the ion pressure.

At this moment, it is interesting to further inspect each term in the Eq.~(\ref{im_with_pi}).
The unsteady term $\partial V_i/\partial t$ is negligible due to time averaging, while the rest four terms have comparable values inside the channel, see Fig~\ref{ion_mom_terms_case1}. In the near-anode region, the ion pressure is negligible ($T_i \approx \SI{0.1}{eV}$), and no ionization, so the ions accelerate towards the anode in a weak negative electric field ballistically. Then in the ionization (source) region, we see all terms are comparable. The ion pressure and the collisional drag compensate the ballistic acceleration, so the ion inertia remains low. Finally, to the right in the acceleration zone ($x > \SI{1.5}{cm}$) the inertial and ballistic terms start to dominate (which continues outside the channel). It can be seen that the ion pressure term changes the sign at $x \approx \SI{1.6}{cm}$ (due to ion density profile, Fig.~\ref{case1_ni_cmpr1}) and contributes to the ion acceleration. At the same time, the collisional drag continues to slow down ions.

\begin{figure}[H]
\centering
\subfloat[]{\includegraphics[width=0.49\textwidth]{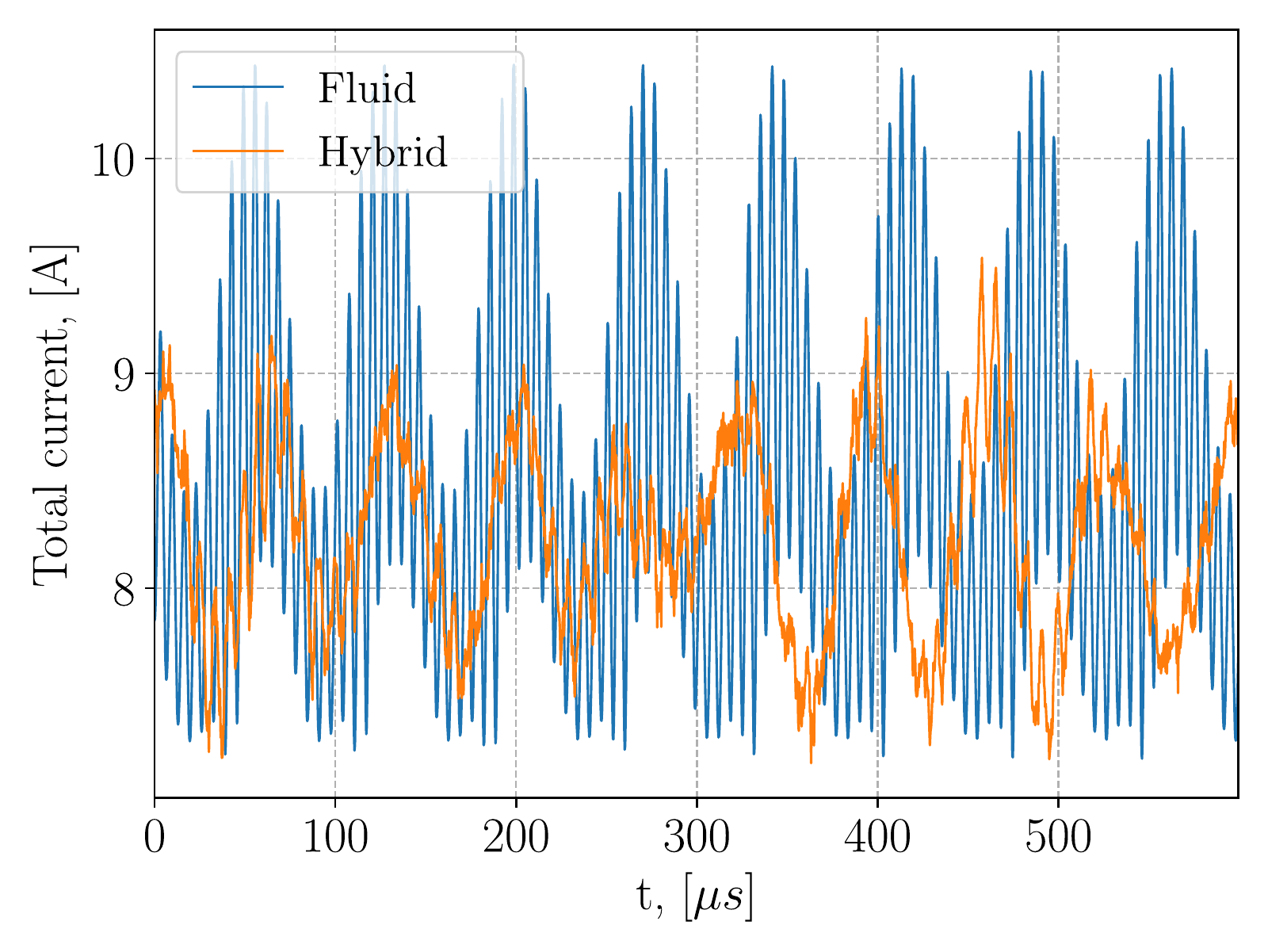}\label{cur_cmpr_case1fixed}}
\subfloat[]{\includegraphics[width=0.49\textwidth]{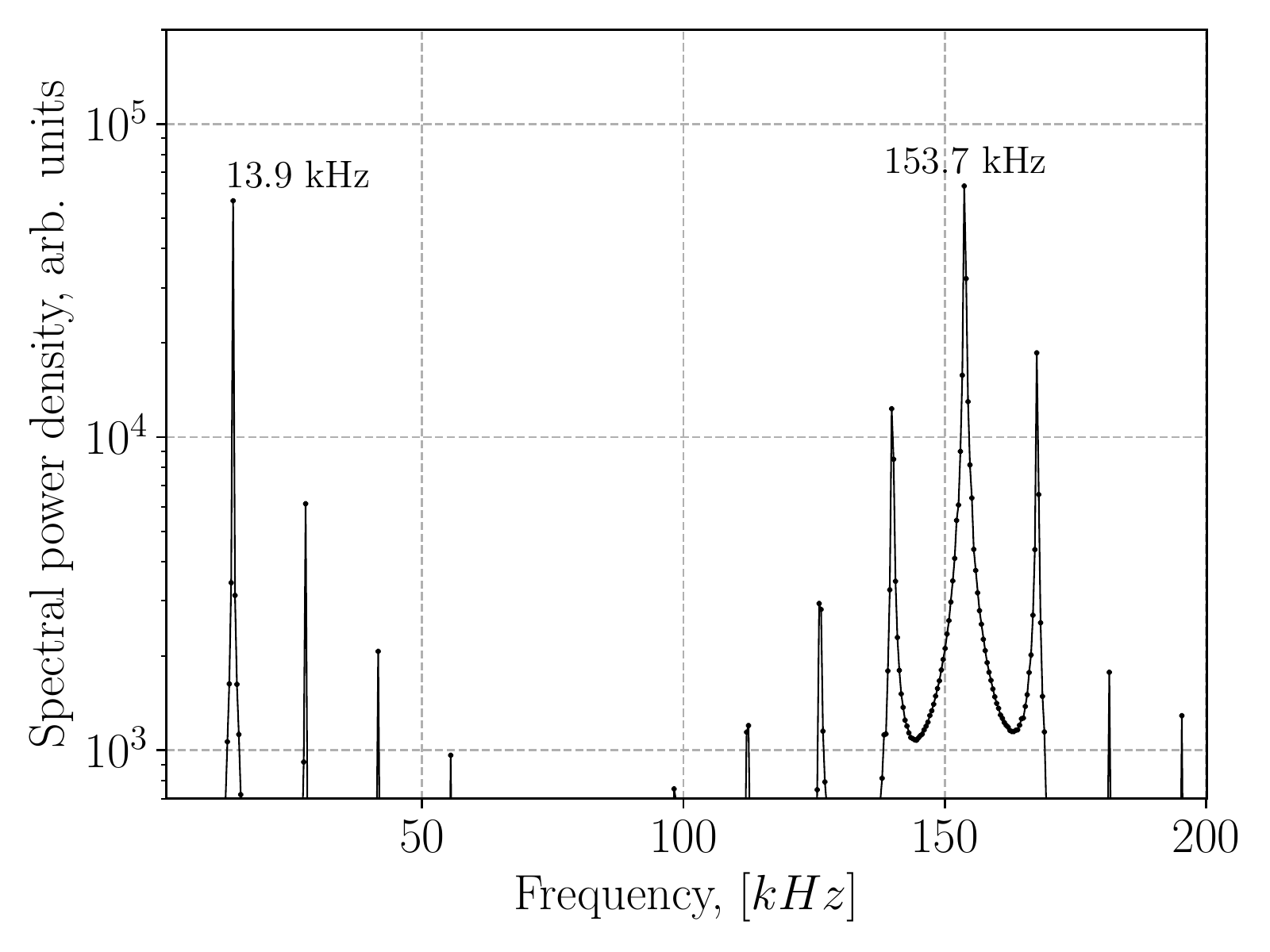}\label{ps_fluid_case1fixed}}
\caption{Comparison of total current in the fluid model and hybrid model (a) and the spectral power density of the total current in the fluid model (b). The result obtained is for the fluid model model with the ion pressure term included.}
\end{figure}

Based on the results above, the ion pressure force term was added to the fluid model with the temperature kept constant for simplicity. We understand that for the self-consistent treatment, the ion energy evolution shall be included or, at least, equation of state. We used $T_i = \SI{1.2}{eV}$ and the pressure $p_i = n T_i$ (average in Fig 6b is 1.7 eV, but fluid might result in stationary result). This results in a better agreement between two models, the total current amplitude decrease and the main low-frequency mode increase,  Figs.~\ref{cur_cmpr_case1fixed},\ref{ps_fluid_case1fixed}. The low-frequency mode increased to \SI{13.9}{kHz}, and the high-frequency peak is shifted to a lower value of \SI{153}{kHz}. The ratio of the ion current to the total current shows improved agreement, 45\%, the same as in the hybrid model. However, the time-averaged total current value of \SI{8.7}{A} became somehow larger than in the hybrid model (8.2 A). 
Also, the higher frequency component is closer to the hybrid result. 
As for the peak plasma density discrepancy (Fig.~\ref{case1_ni_cmpr1}), it remained similar.

\subsection{Case 2: High electron energy losses; the  solo regime  of the low frequency mode}

\begin{figure}[H]
\centering
\subfloat[]{\includegraphics[width=0.49\textwidth]{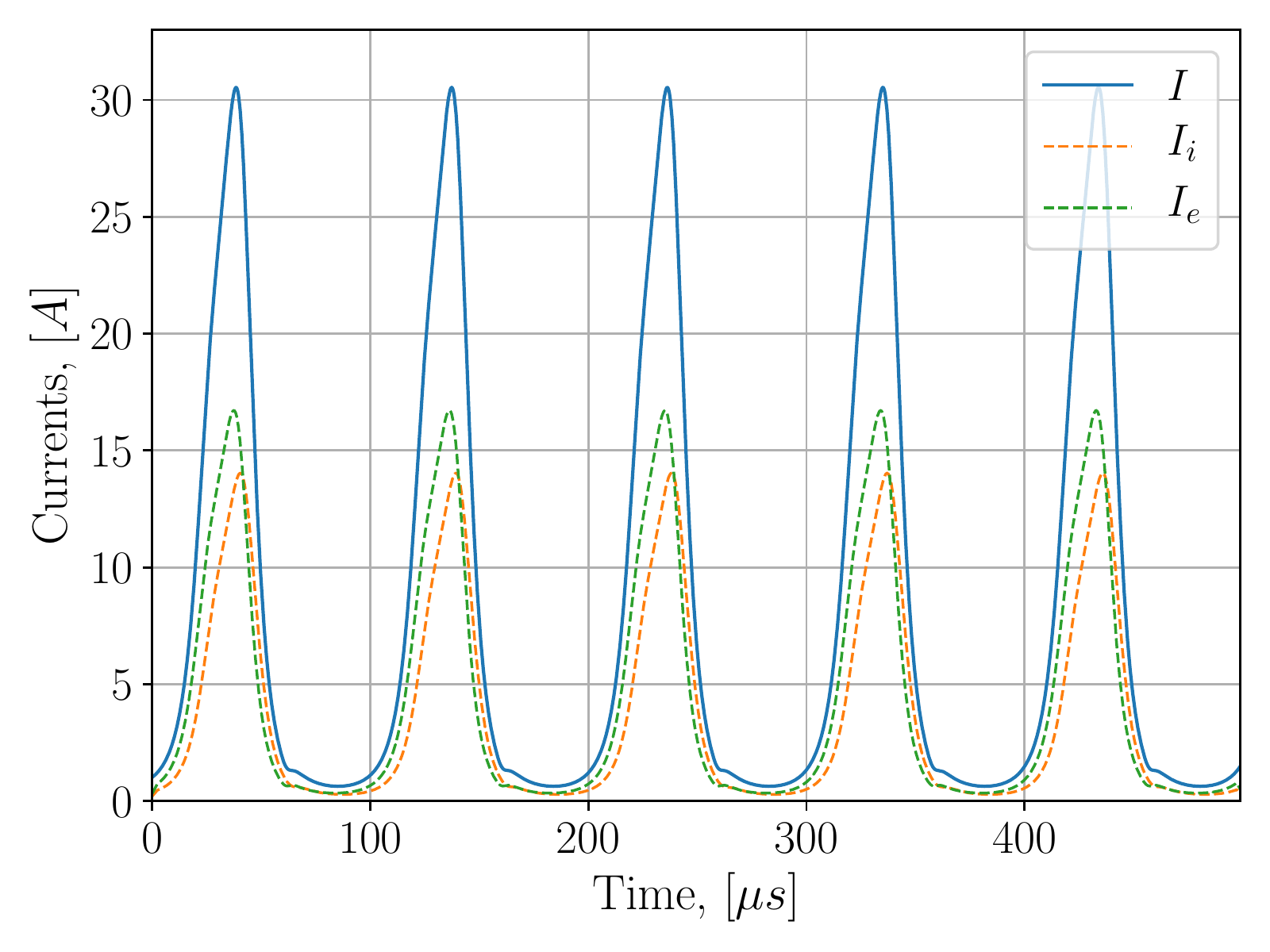}\label{case2_fluid_cur}}
\subfloat[]{\includegraphics[width=0.49\textwidth]{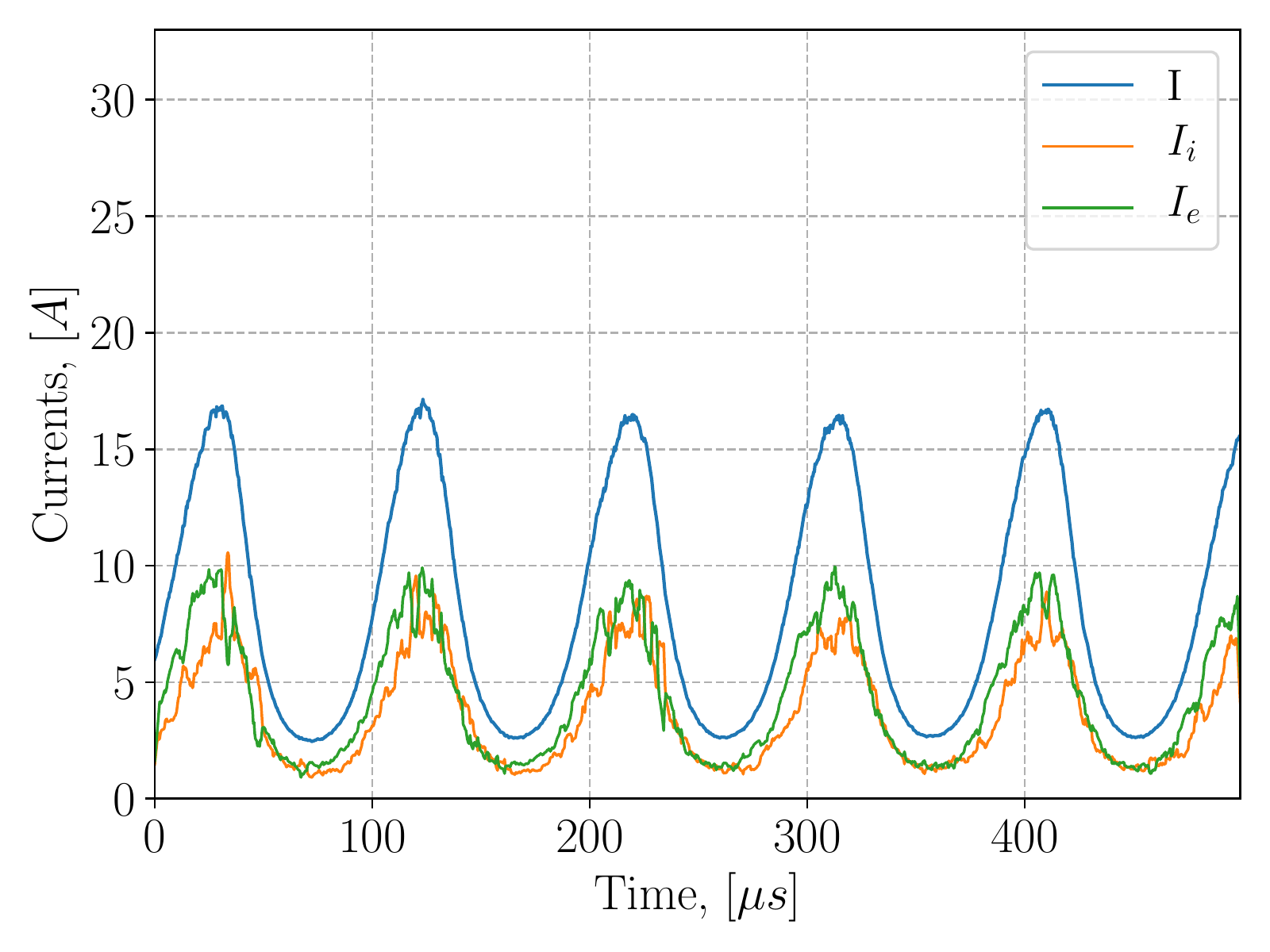}\label{case2_hybrid_cur}}
\caption{Amplitudes of total, ion, and electron currents in fluid model (a) and hybrid model (b) for Case 2. Ion and electron currents are evaluated at $x = \SI{5}{cm}$.}
\end{figure}

\begin{figure}[H]
\centering
\subfloat[]{\includegraphics[width=0.49\textwidth]{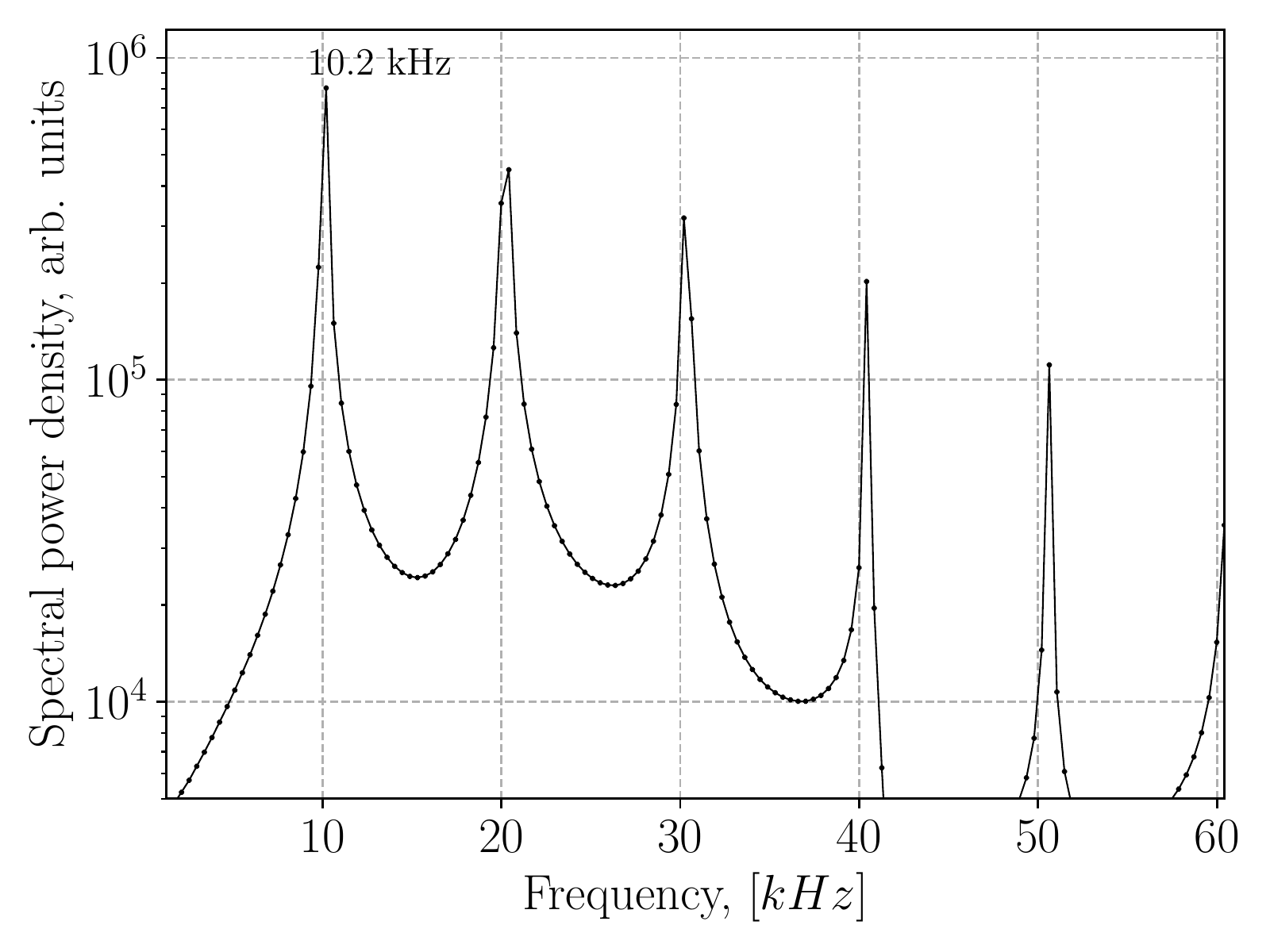}\label{ps_case2_fluid}}
\subfloat[]{\includegraphics[width=0.49\textwidth]{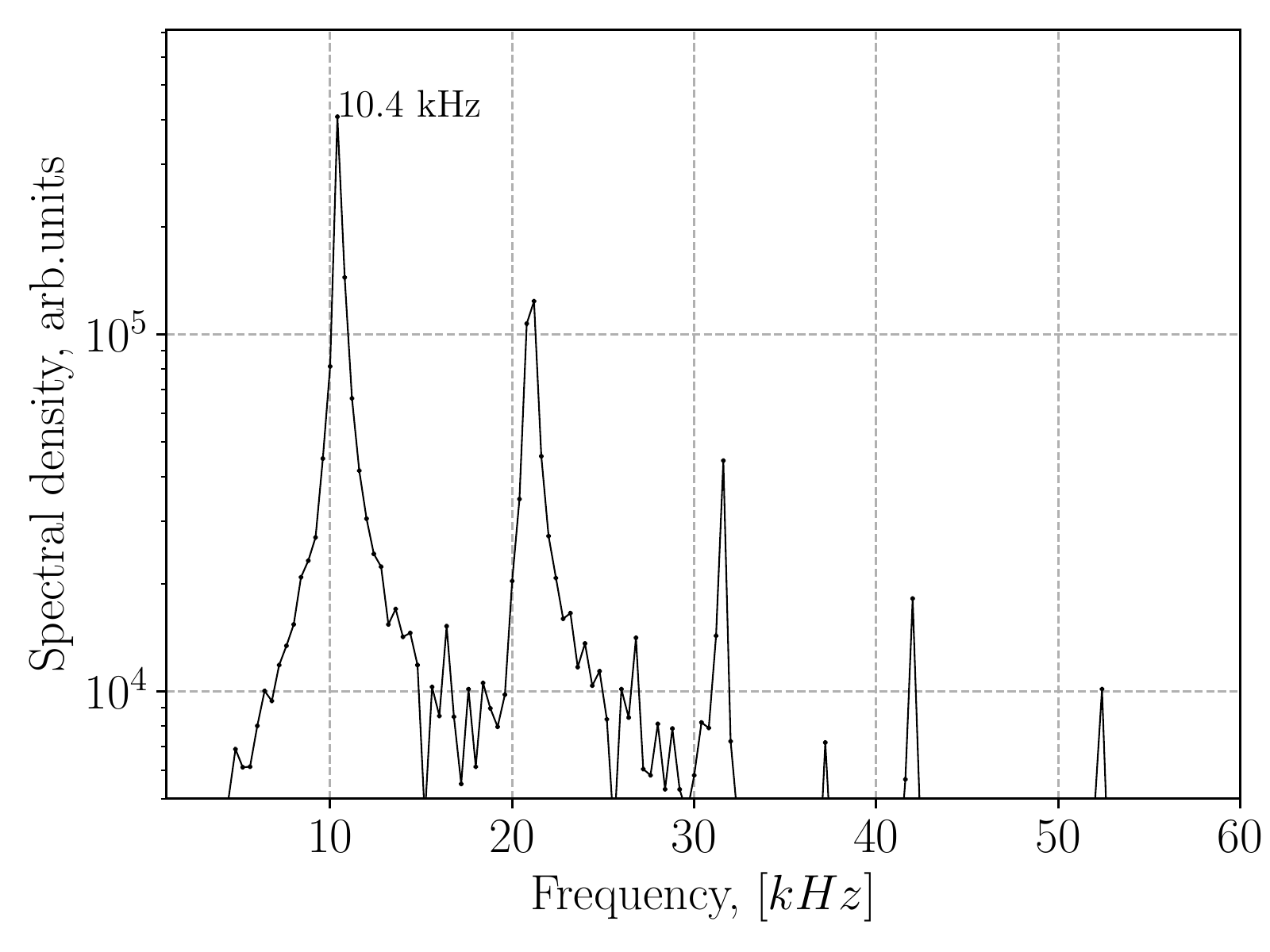}\label{ps_case2_hybrid}}
\caption{Spectral density of the total current for fluid model (a) and hybrid model (b).}
\end{figure}

This case is subject to the low-frequency oscillations only, with the only difference to Case 1 in the value of the anomalous electron energy loss coefficient, which is $\nu_{\varepsilon, \text{in}} = \SI{e7}{s^{-1}}$. As in the previous case, the total current amplitude is higher in the fluid model , Figs.~\ref{case2_fluid_cur}, \ref{case2_hybrid_cur}, but the main oscillation frequency in two models is similar, Figs.~\ref{ps_case2_fluid}, \ref{ps_case2_hybrid}.

\begin{figure}[ht]
\centering
\subfloat[]{\includegraphics[width=0.49\textwidth]{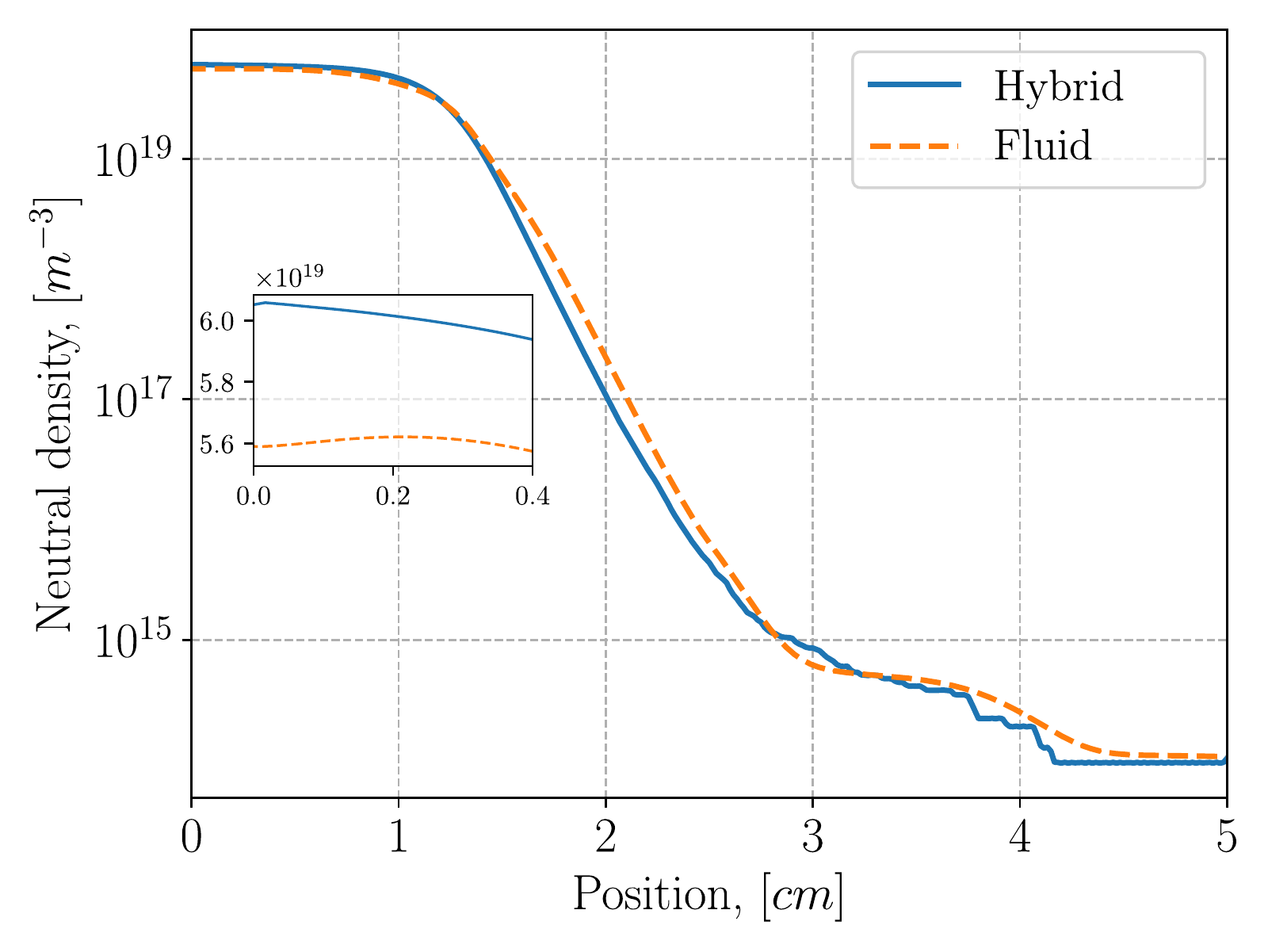}\label{case2_nn_cmpr}}
\subfloat[]{\includegraphics[width=0.49\textwidth]{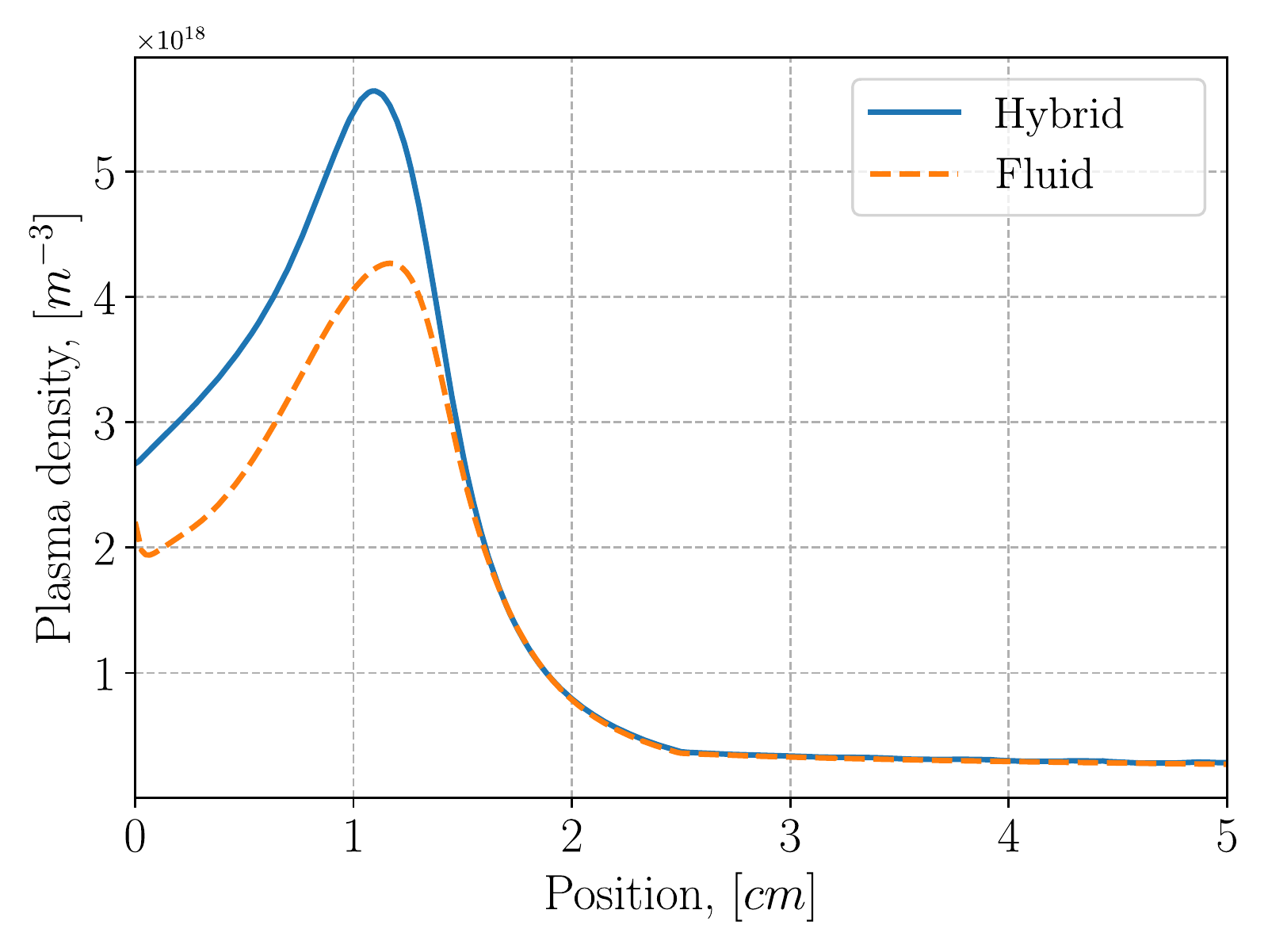}\label{case2_ni_cmpr}} \\
\subfloat[]{\includegraphics[width=0.49\textwidth]{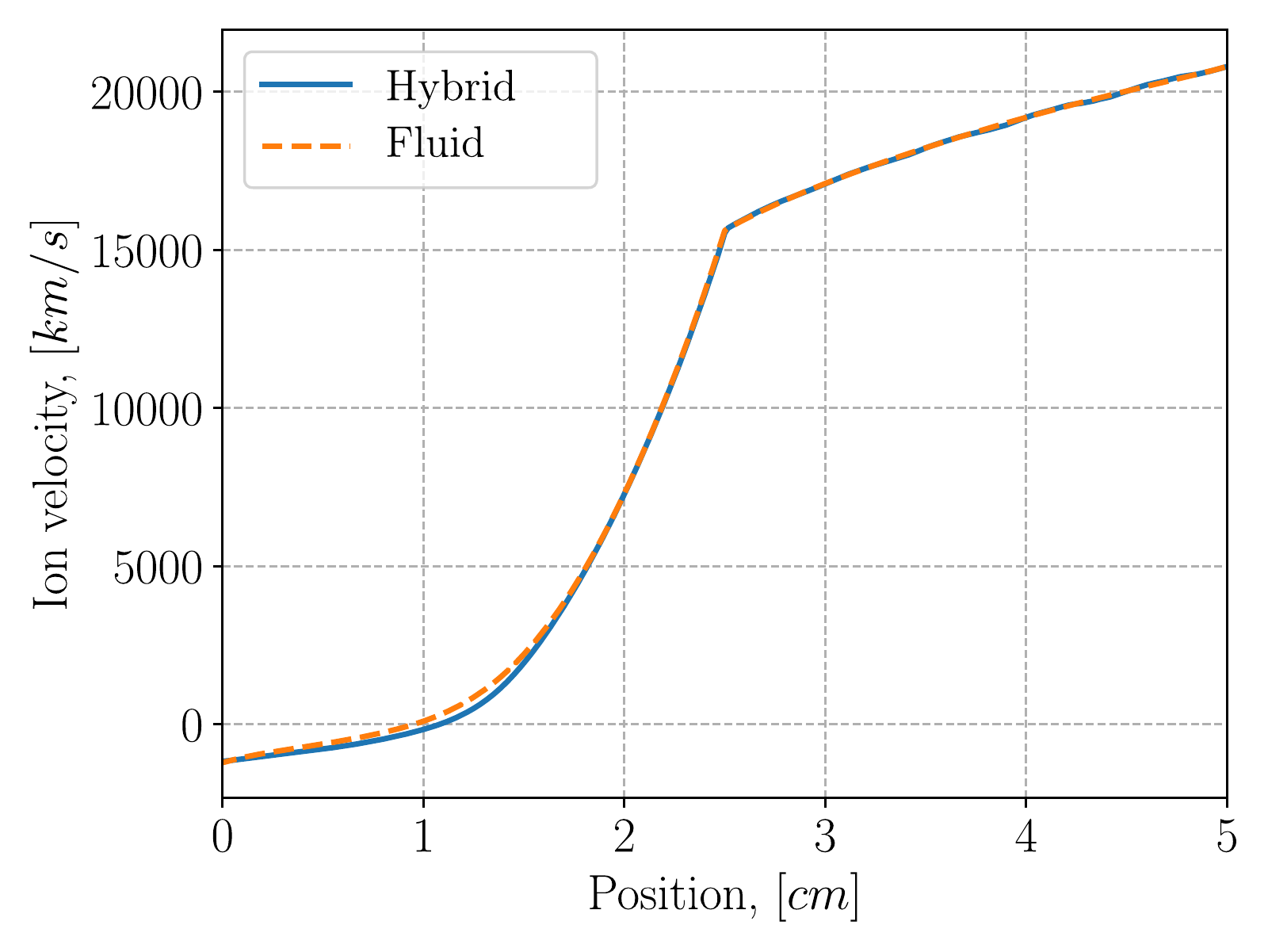}\label{case2_vi_cmpr}}
\subfloat[]{\includegraphics[width=0.49\textwidth]{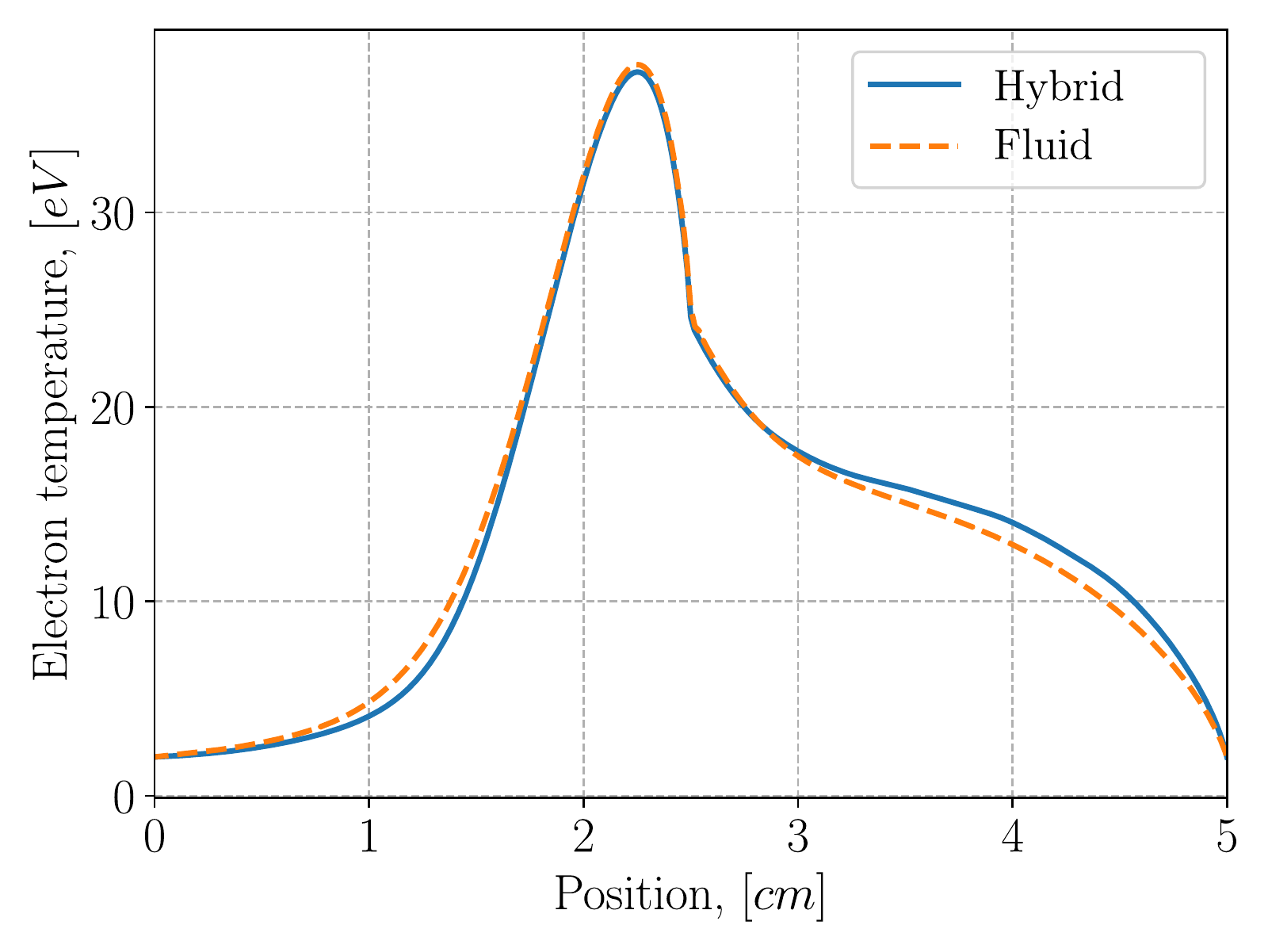}\label{case2_te_cmpr}}
\caption{Comparison of spatial distribution of time-averaged macroscopic profiles between fluid and hybrid models for Case 2.}
\end{figure}

Time-averaged profiles well agree in two models, with the only notable discrepancy in the plasma density profile, which is ${\sim} 30\%$ lower inside the channel region for the fluid model, see Figs.~\ref{case2_nn_cmpr}-\ref{case2_te_cmpr}.

\begin{figure}[H]
\centering
\subfloat[]{\includegraphics[width=0.5\textwidth]{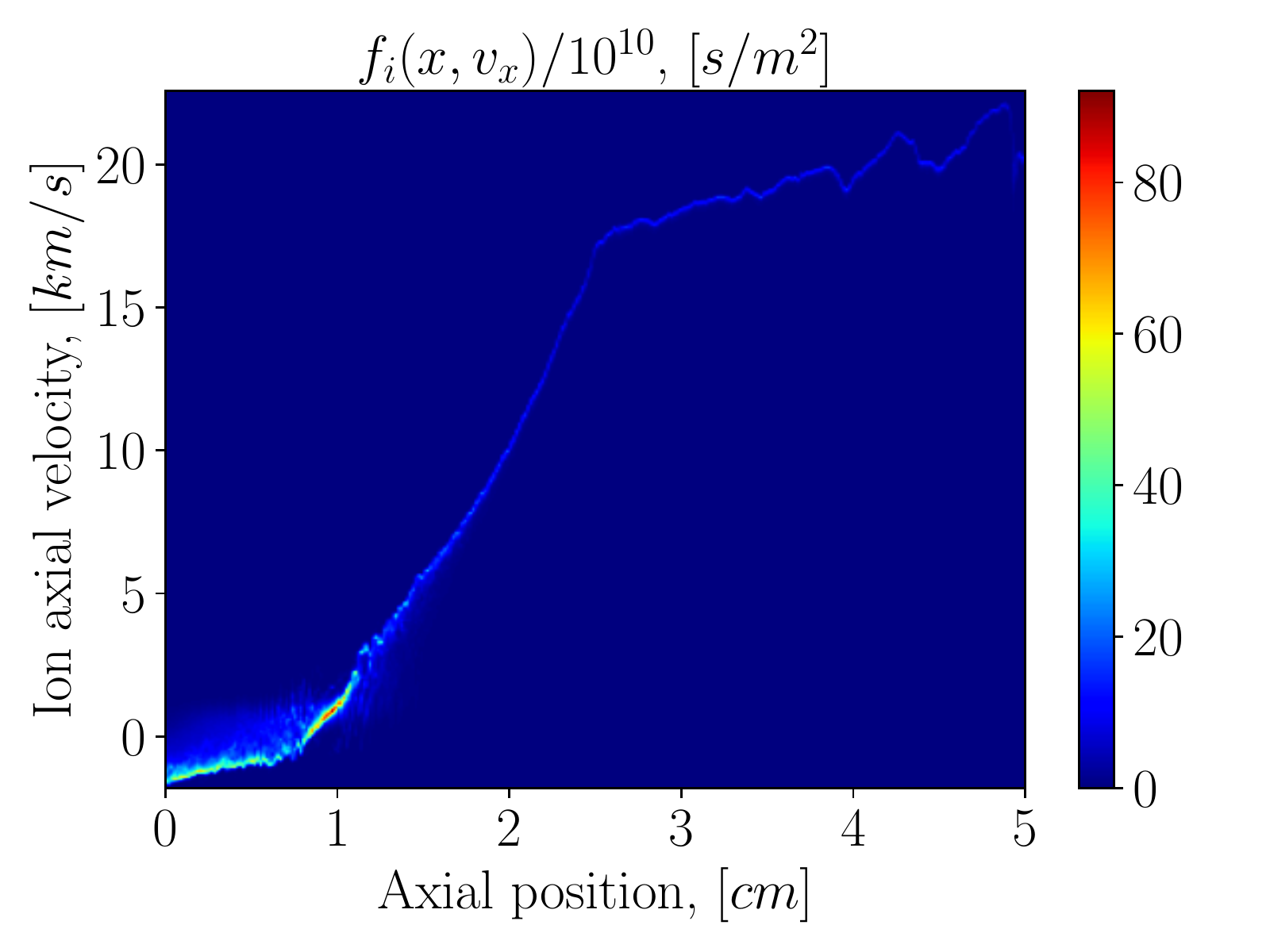}\label{phase_space_case2}}
\subfloat[]{\includegraphics[width=0.45\textwidth]{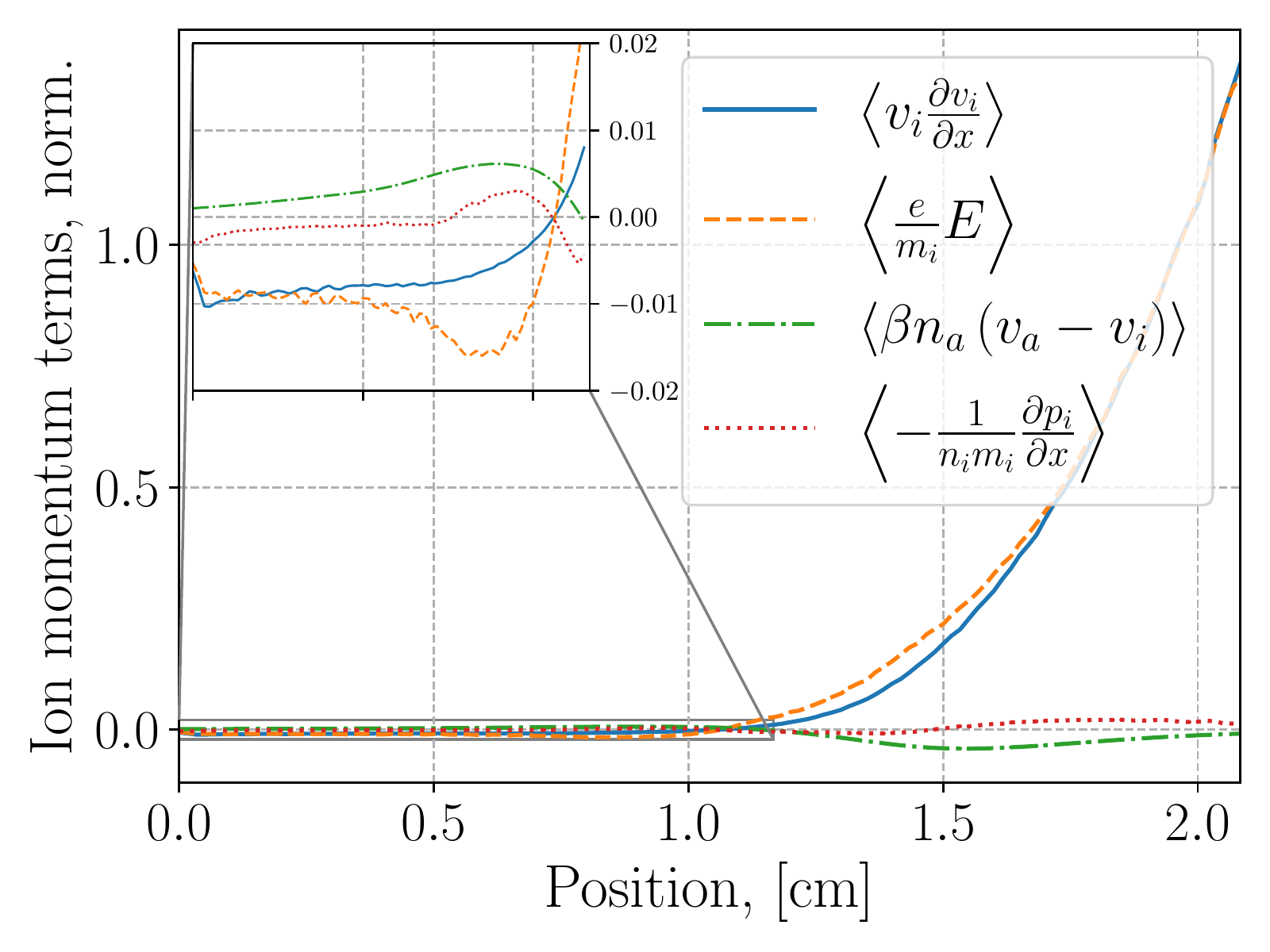}\label{ion_mom_balance_case2}}
\caption{Instantaneous image of ion distribution function (in space of axial coordinate and axial velocity) in the hybrid model (a). Separate terms of the ion momentum balance equation~(\ref{ion-mom}) evaluated from the ion kinetic description (hybrid model), averaged in time (denoted with angle brackets) over few periods of the low-frequency component (b).}
\end{figure}

Unlike in Case 1, the IVDF for this case (Fig.~\ref{phase_space_case2}) reveals that the ion population remains cold (everywhere except in the near-anode region), and the ion momentum balance must be well satisfied without the ion pressure term. This is seen in Fig.~\ref{ion_mom_balance_case2}, where the ion pressure force remains low everywhere in the channel. In fact, in this case, the ballistic ion acceleration is more pronounced, dominating everywhere except the stall point ($V_i \approx 0$). The average ion temperature, calculated similarly as in Case 1, is \SI{0.3}{eV} and not exceeding \SI{1}{eV} in the domain (about 5 times smaller than in Case 1).

\begin{figure}[ht]
\centering
\subfloat[]{\includegraphics[width=0.98\textwidth]{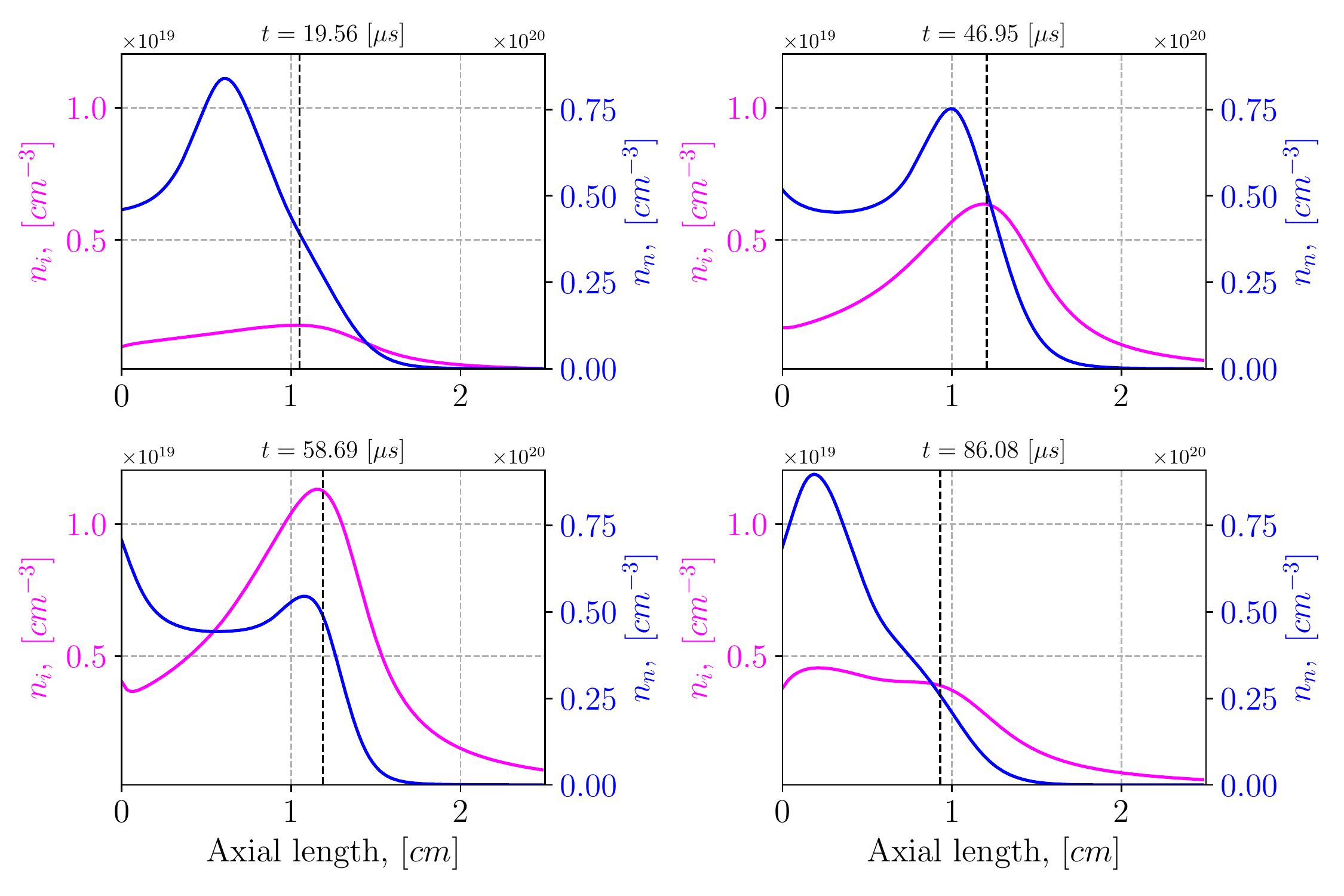}}
\caption{Neutral and plasma density evolution during one oscillation period. Dashed line separates the region with negative (to the left) and positive (to the right) ion velocity.}
\label{lf_dynamics}
\end{figure}

The main stages of the breathing mode dynamics are illustrated with plasma and atom densities, Fig.~\ref{lf_dynamics}. After atoms reach the ionization zone and undergo ionization, plasma density increases and quickly depleted  (${\sim}\SI{1}{km/s}$) to the left (due to the backflow region with negative velocity) and the right of ionization zone. Then all of the ion flux that reached the anode recombines and form the peak in neutral density at the anode, increasing the number of atoms advect to the ionization zone; this process repeats. In our previous work, Ref.~\onlinecite{chapurin2021mechanism}, this setup was studied in detail, and it was shown that the ion backflow region plays a crucial role in these oscillations and that they can exist without the recombination mechanism.

\begin{figure}[ht]
\centering
\includegraphics[width=0.6\textwidth]{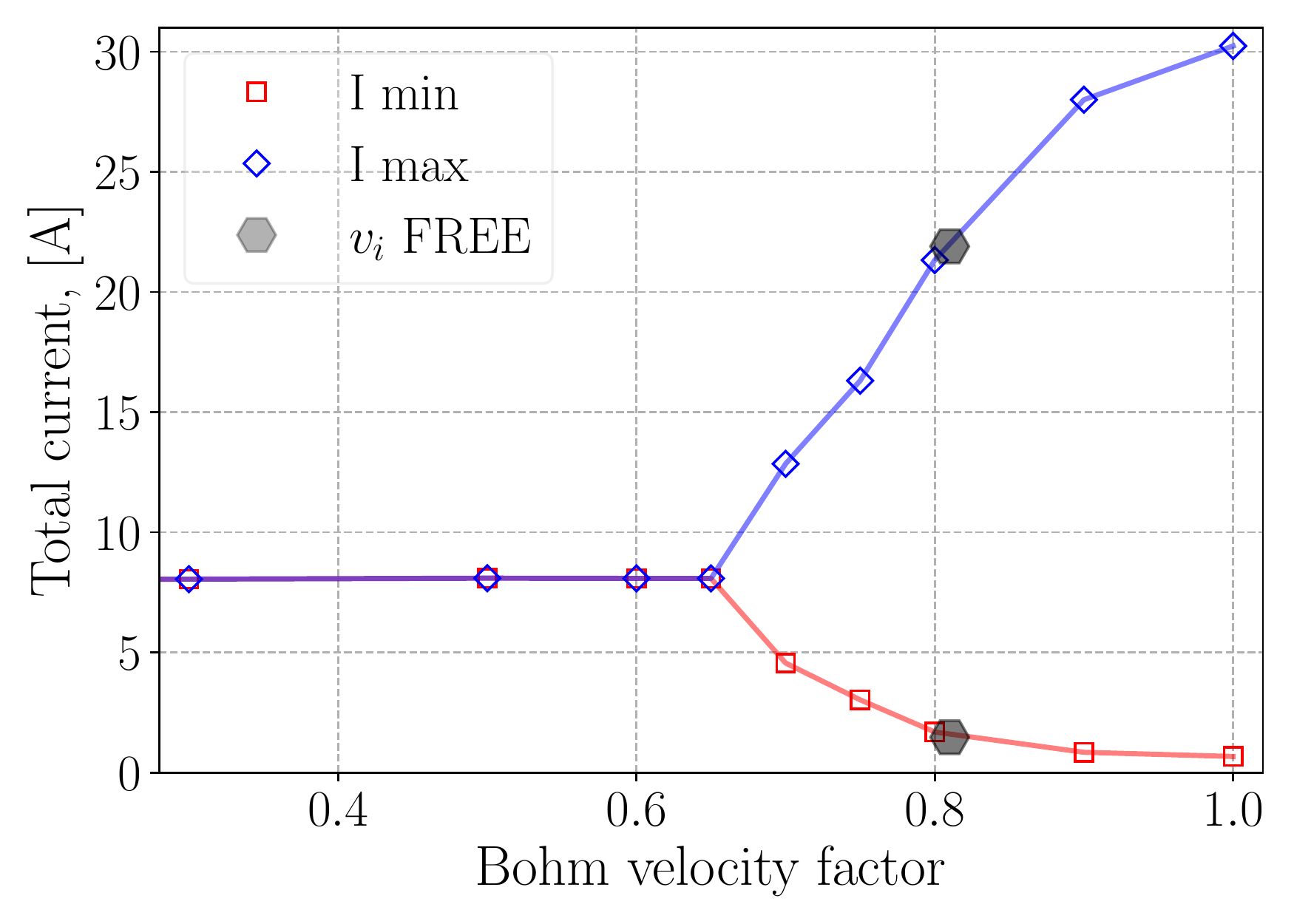}
\caption{Minimum and maximum total current values during oscillations for various ion velocities at the anode, expressed as fractions of the Bohm velocity. Note that oscillations are absent for $V_i < -0.65 c_s$. }
\label{bohm_fac}
\end{figure}

It is found that the oscillation dynamics in Case 2 highly depend on the recombination at the anode, modeled with Eq.~(\ref{nn_backflow}), and the number of recombined atoms proportional to the oscillation current amplitude. Note that in Case 1, recombination plays a minor role and does not affect the presented results. Turning off the recombination in Case 2 nullifies the oscillations and the stationary solutions obtained (in both models).
We suggest that the main difference in this case between the fluid and the hybrid model, in this case, lies in the differences with the ion velocity boundary condition. Normally, in the fluid quasineutral models, this boundary condition is fixed to the Bohm velocity, as in our fluid model.
When the Bohm boundary condition is scaled with the factor $b_v = 0 \textrm{--} 1$ (ion velocity effectively decreased at the anode) in the fluid model, the oscillations amplitude also decreases, see Fig.~\ref{bohm_fac}. It clearly shows that the plasma recombination provides the additional feedback in this configuration, and the main difference with the hybrid model lies in the ion boundary conditions causing larger oscillation amplitude in the fluid model. In the hybrid model (in our setup), ions are not forced to satisfy the Bohm condition, and the flow velocity is established self-consistently. We modeled this behaviour with the free boundary condition for the ion velocity at the anode in the fluid model (forcing $\partial_x^2 V_i(0) = 0$). In the presence of ion backflow (like we see in Case 1,2) in the quasineutral approximation, the ion velocity at the anode is defined self-consistently by the characteristic flowing from $v_i = 0$ inside the channel; thus, a fixed boundary is not required.

\begin{figure}[ht]
\centering
\includegraphics[width=0.65\textwidth]{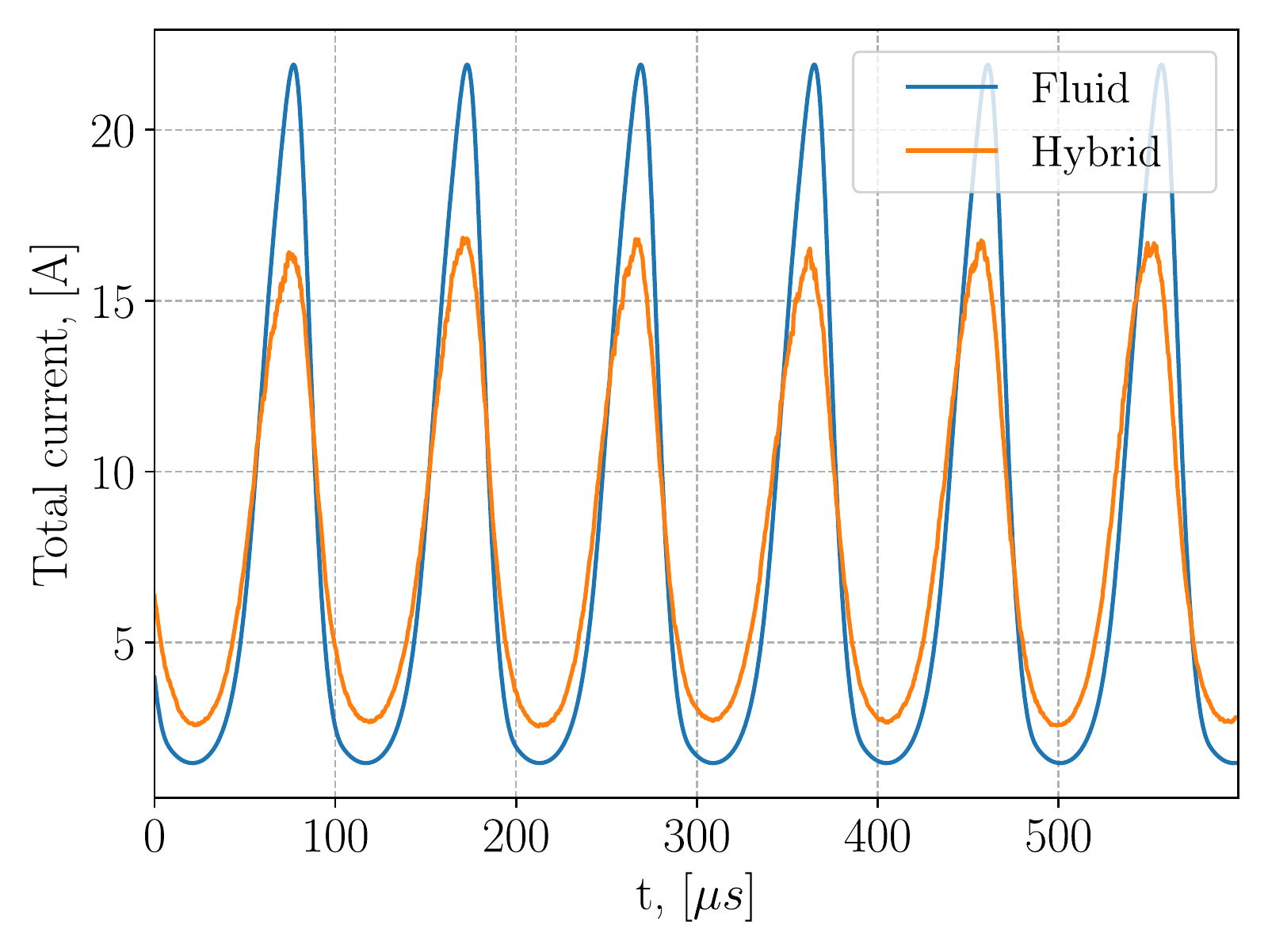}
\caption{Comparison of the total current obtained in both models where the free boundary condition used in the fluid model for ion velocity at the anode.}
\label{current_case2_vifree}
\end{figure}

\begin{figure}[H]
\centering
\subfloat[]{\includegraphics[width=0.49\textwidth]{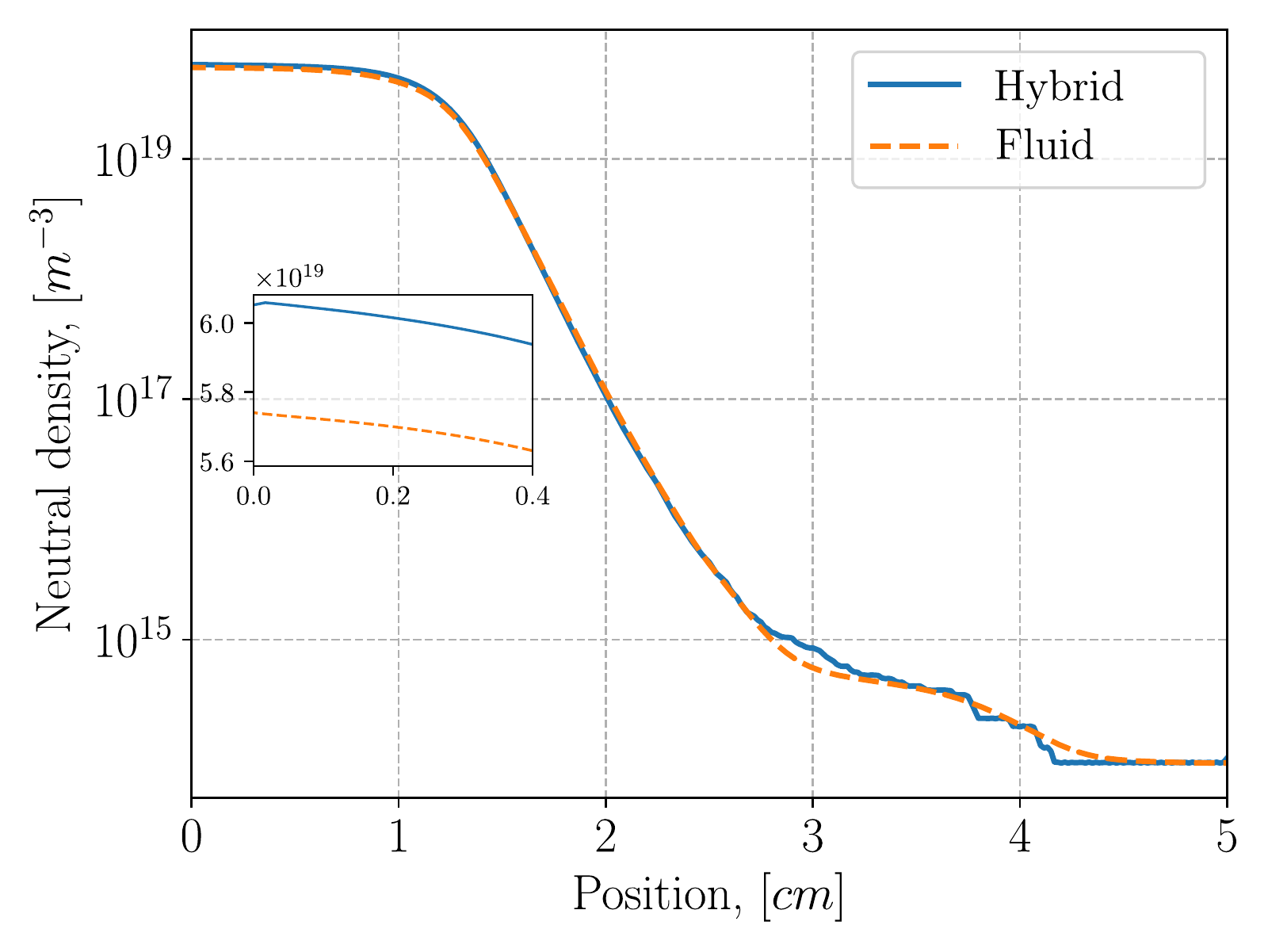}\label{nn_cmpr_vifree}}
\subfloat[]{\includegraphics[width=0.49\textwidth]{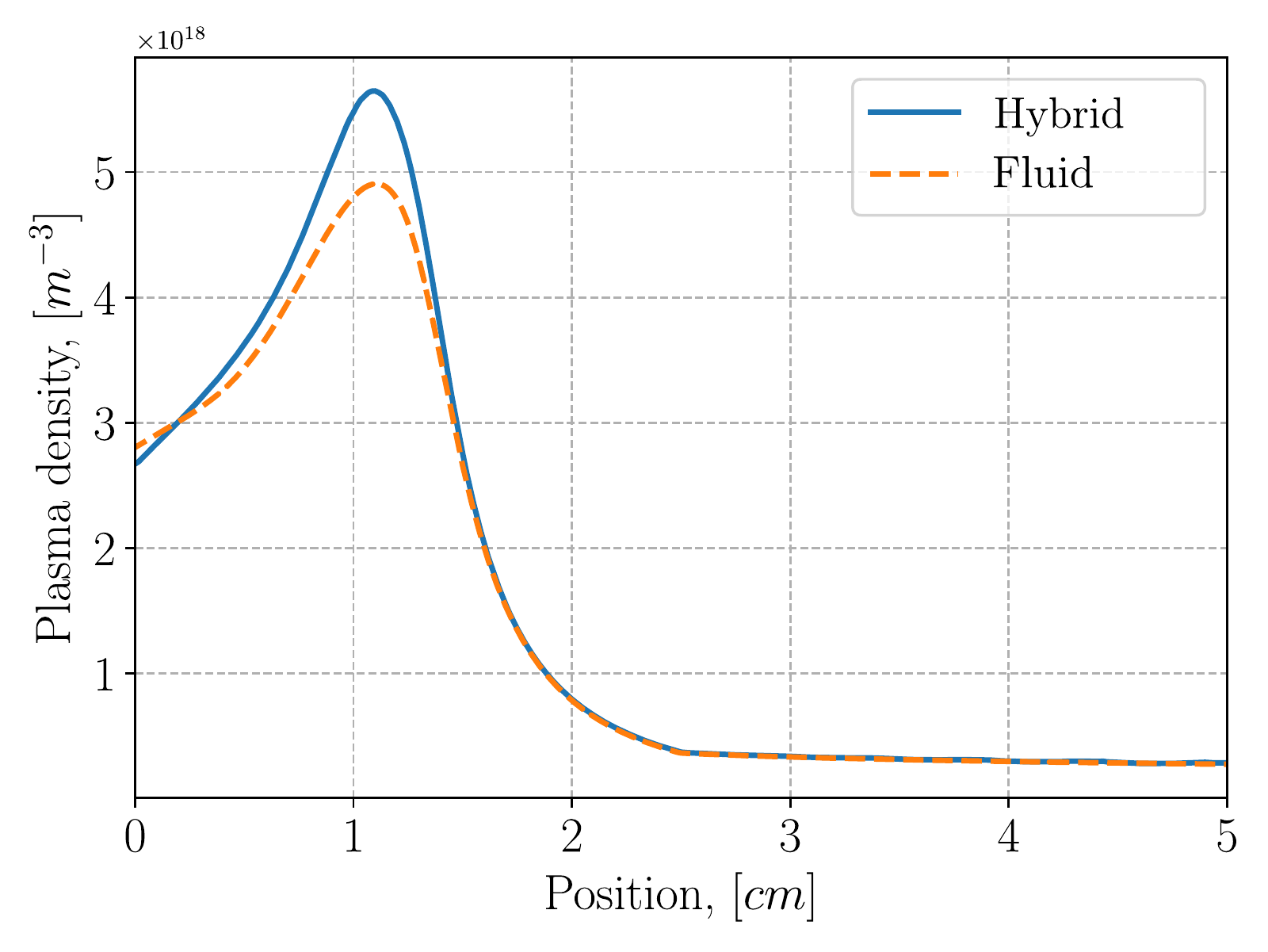}\label{ni_cmpr_vifree}} \\
\caption{Atom (a) and ion (b) axial profiles, averaged in time, compared in both models where the free boundary condition used in the fluid model.}
\end{figure}

The obtained result in the fluid model with the free boundary for the ion velocity at the anode reveals less violent oscillations and generally better agrees with the hybrid model results. The total current oscillation amplitude is lower, Fig.~\ref{current_case2_vifree}, closer to the hybrid mode result. 
Time-averaged atom and plasma density profiles agreement is improved, Figs.~\ref{nn_cmpr_vifree}, \ref{ni_cmpr_vifree}.



\subsection{Case 3: Effects of finite temperature of neutral atoms}

\begin{figure}[ht]
\centering
\subfloat[]{\includegraphics[width=0.65\textwidth]{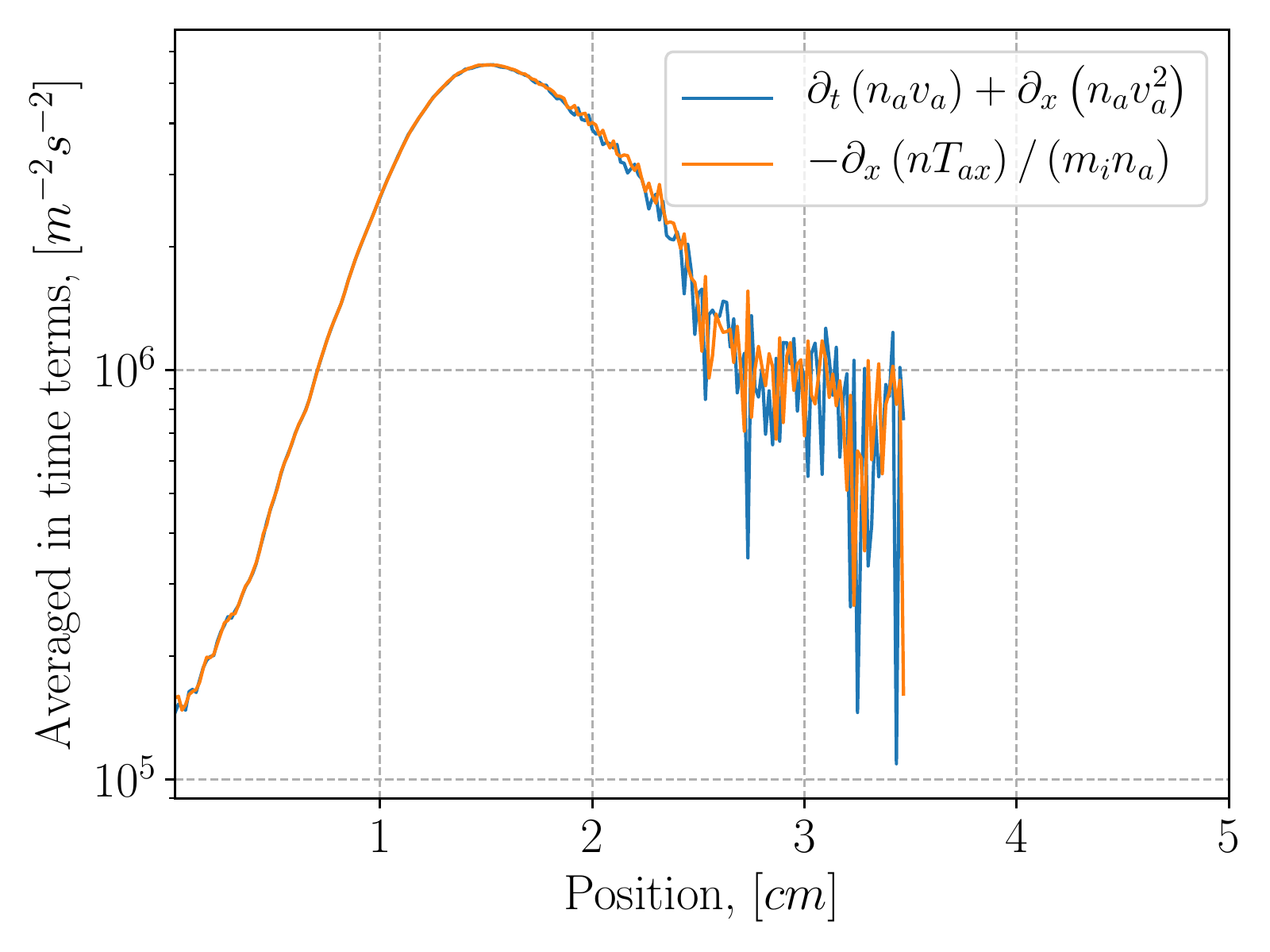}}
\caption{Atom momentum balance equation terms (a). Note that the number of macroparticles (and atom density) significantly decrease to the right along the channel elevating the noise.}
\label{atom_momentum}
\end{figure}

\begin{figure}[ht]
\centering
\subfloat[]{\includegraphics[width=0.49\textwidth]{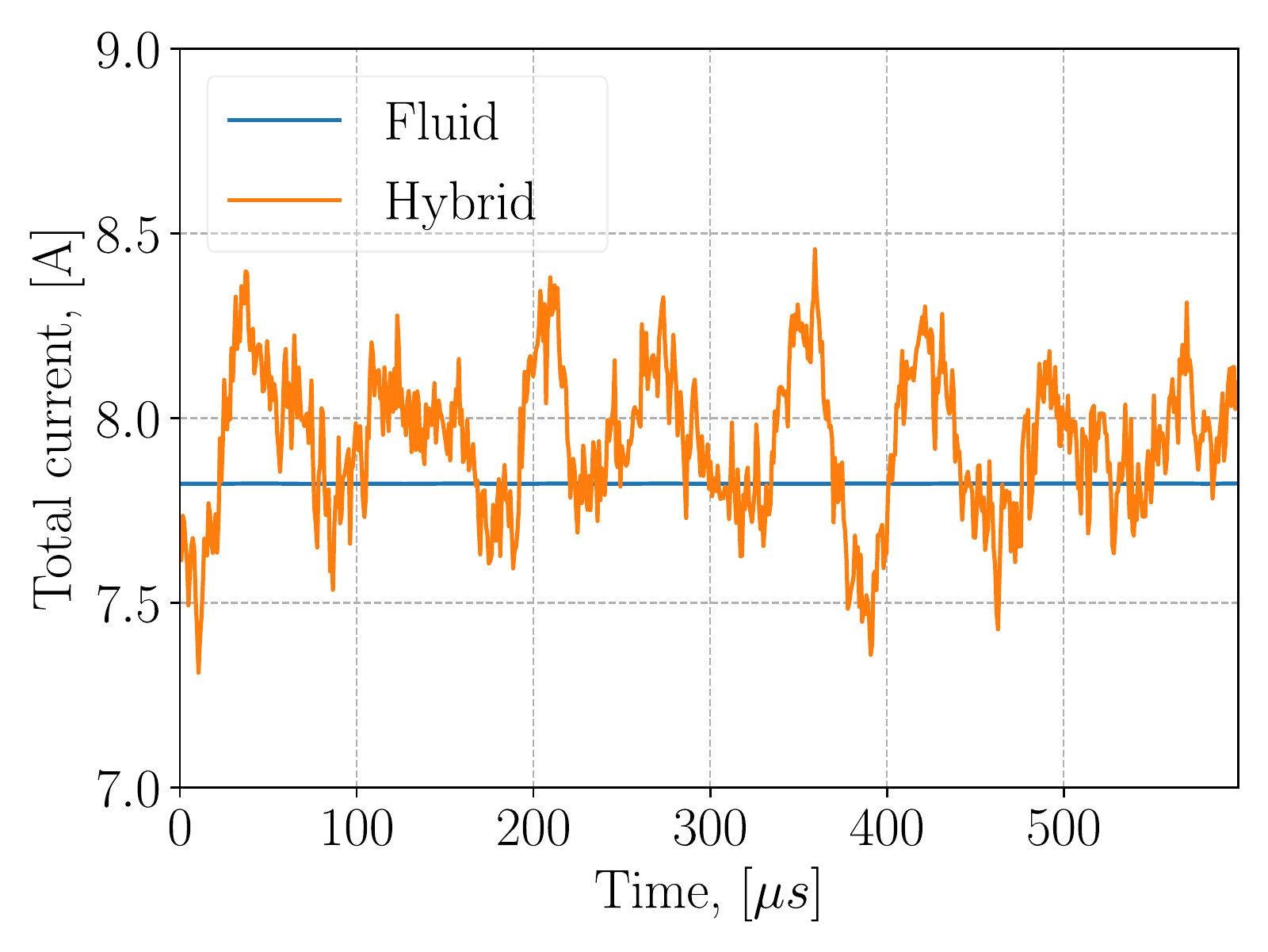}\label{c3:cur}}
\subfloat[]{\includegraphics[width=0.49\textwidth]{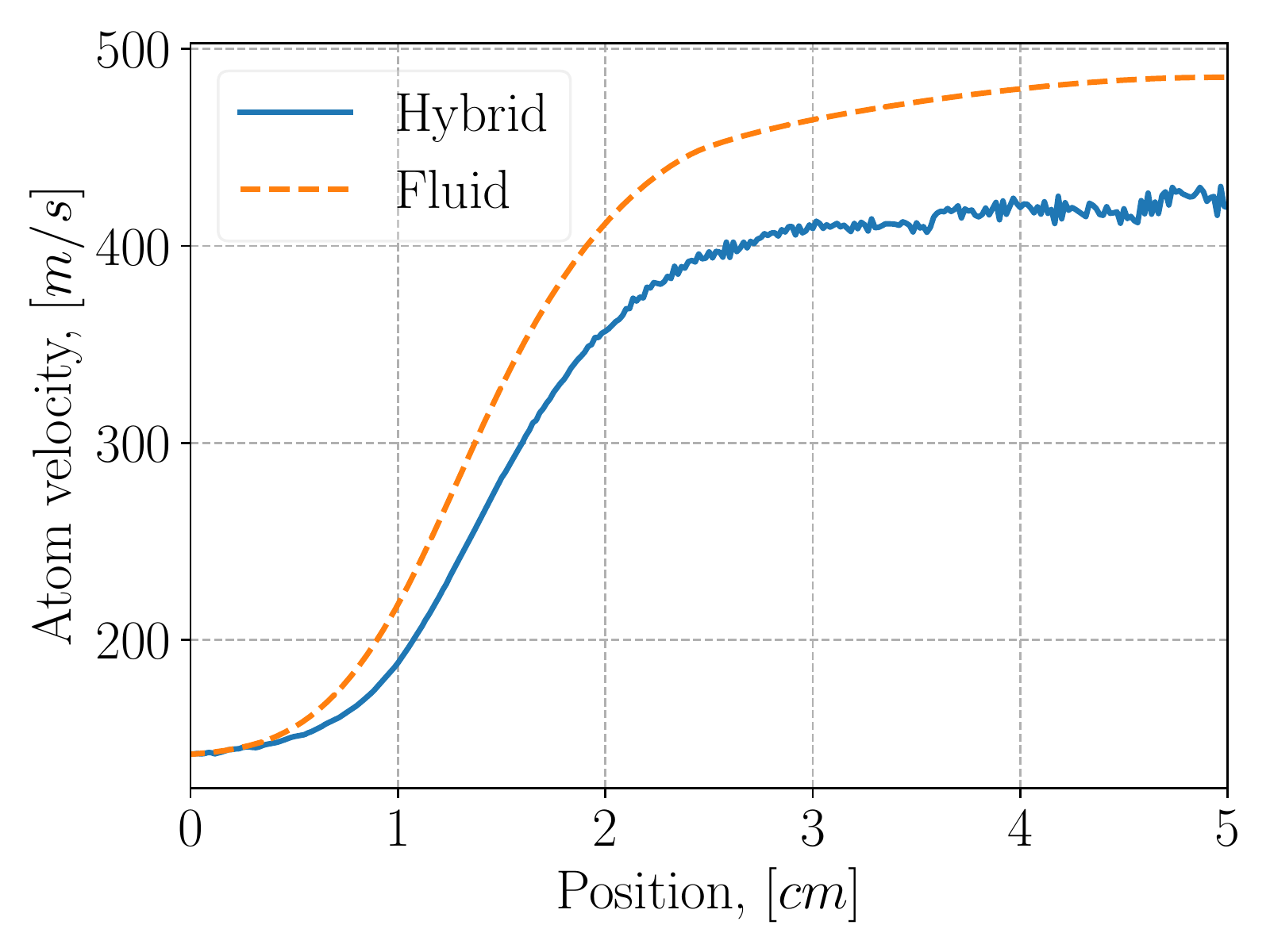}\label{case3_vels}}
\caption{The total current in fluid and hybrid simulations (a); comparison of time-averaged spatial profile of atom flow velocity for both models (b).}
\end{figure}

Unlike Case 2, the oscillation amplitude in this case is much smaller in the hybrid model, Fig.~\ref{c3:cur}, with the amplitude of ${\sim}\SI{1}{A}$ (in comparison to Case 2 with ${\sim}\SI{13}{A}$).
In the previous cases, injected atoms in the hybrid model were kept monokinetic (zero thermal energy). Here, the finite atom temperature is included while all other parameters are kept as in Case 2. Atoms are injected at the anode with the half-Maxwellian flux distribution, Eq.~(\ref{halfMaxw}) with a
temperature $T_a = \SI{500}{K}$. The average injection velocity for the half-Maxwellian is $V_0 = v_{Ta}/\sqrt{\pi} = \SI{142}{m/s}$, which is close to $\SI{150}{m/s}$ used in the previous monokinetic runs. It was noticed that the atom flow velocity in the hybrid model exhibits a clear spatial dependence, ``accelerating'' along the channel. This effect is known as selective ionization, observed both in experiments \cite{meezan2001anomalous, hargus2002interior} and simulations\cite{hara2012one}. It is clear that atoms in the fluid model with the simple advection equation~(\ref{na}) cannot capture this effect; hence the atom momentum balance equation was included in the fluid model:
\begin{equation}\label{am_c}
\frac{\partial \left(n_a V_a\right)} {\partial t} + \frac{\partial }{\partial x} \left( n_a V_a^2 \right) = -\beta n_a n_i V_a - \frac{T_a}{m_i}\frac{\partial \left(n_a\right)}{\partial x},
\end{equation}
where the closure is given with the constant temperature $T_a = \SI{500}{K}$. The illustration that the momentum balance given by Eq.~(\ref{am_c}) has a sufficient number of terms, the atom fluid moments were calculated from the kinetic representation of atoms in the hybrid model, and the balance is compared, see Fig.~\ref{atom_momentum}. Thus, Eq.~(\ref{am_c}) along with the continuity equation 
\begin{equation}\label{a_cont}
\frac{\partial n_a} {\partial t} + \frac{\partial \left(n_a V_a\right)} {\partial x} = -\beta n_a n_i
\end{equation}
solved in the fluid model, with $T_a = \SI{500}{K}$ and the fixed atom flow velocity at the anode $V_a(0) = \SI{142}{m/s}$ (Dirichlet condition), and the same recombination boundary condition given by Eq.~(\ref{nn_backflow}). This allowed to recover qualitatively the atom flow velocity behaviour, Fig.~\ref{case3_vels}, but in the same time the fluid model resulted in a completely stationary solution, Fig.~\ref{c3:cur}.

\begin{figure}[ht]
\centering
\includegraphics[width=0.65\textwidth]{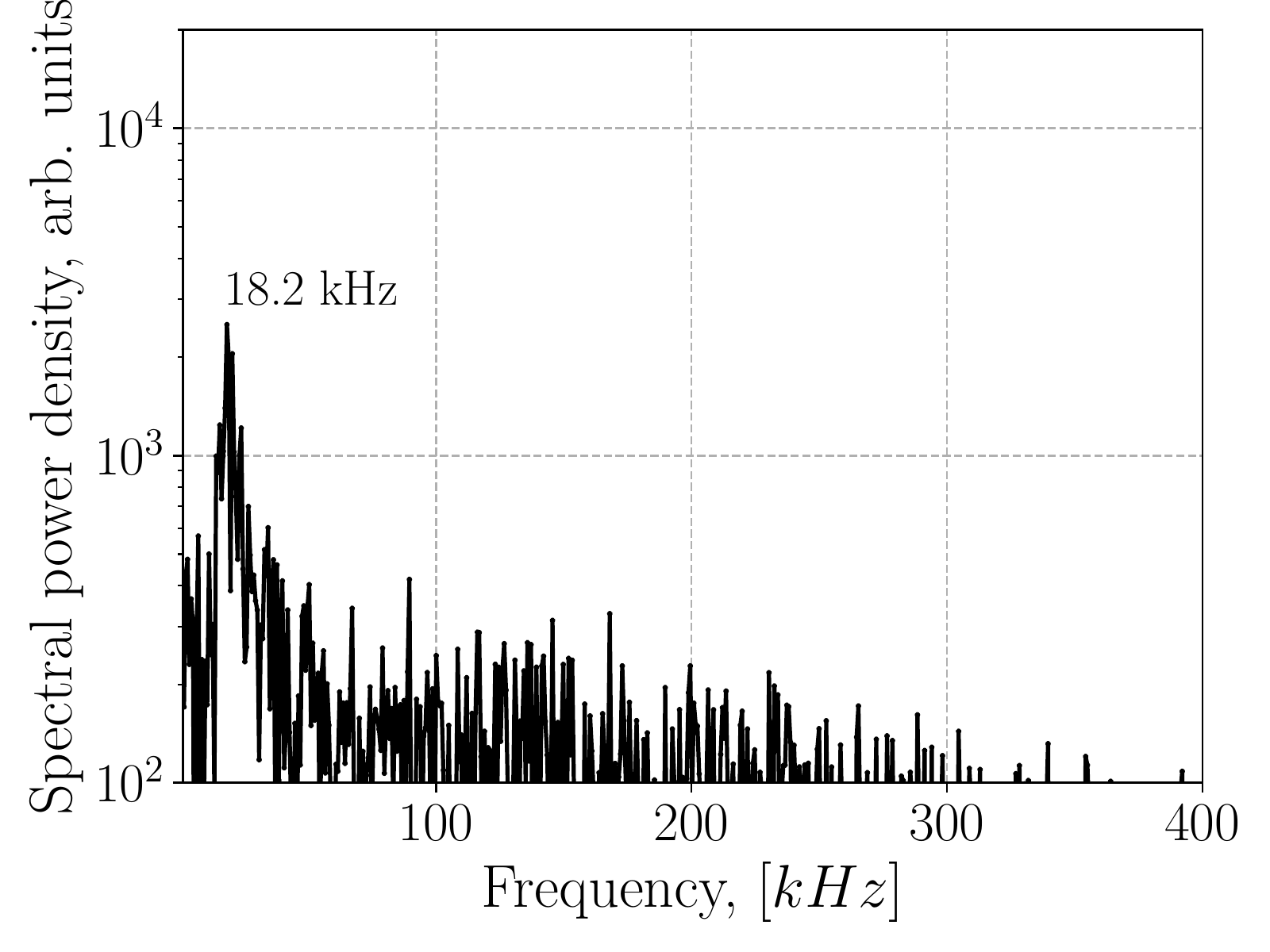}
\caption{Power spectrum of the total current in the hybrid simulation. Note that there is increase in frequency (from 10.2 kHz to 18.2 kHz) in compare to Case 2 (with monokinetic atoms).}
\label{c3:dft}
\end{figure}

To identify better the role of the atom distribution in the hybrid model, we configured the shifted Maxwellian distributions for atoms with various temperatures denoted as $T_{a,s}$, and the fixed shift $\SI{150}{m/s}$. Thus, the value of $T_{a,s} = \SI{0}{K}$ corresponds to Case 2. We found that even a low spread for atom velocities, e.g.\ corresponding to $T_{a,s} \approx \SI{50}{K}$, heavily damps the oscillation amplitude to the order $\SI{1}{A}$, similar to what we obtained with the half-Maxwellian and $T_a = \SI{500}{K}$. 

Interestingly, we were not able to find higher amplitude low-frequency oscillations for this case in the hybrid model (simply scanning parameters like $\nu_{\varepsilon}$ or $\beta_a$). However, slightly varying the position of the maximum magnetic field profile, higher amplitude breathing modes were observed even with thermal atoms. This important behaviour needs further attention which is out of the scope of this paper and is left for future studies.

\section{Summary}\label{sec_summary}

Generally, plasma models for Hall thruster configuration have to deal with long time scales (defined by slow neutral dynamics, requiring at least ${\sim}10^6$ electron plasma periods) and large spatial scales (above ${\sim}10^3~\lambda_{De}$). A scale separation by many orders is a typical problem in plasma modeling.
Full kinetic descriptions for the problem involving low-frequency dynamics in the axial direction may not be practical due to computational cost and potential numerical problems. Reduced plasma models, such as the fluid model formulated via the conservation laws derived from higher dimensional kinetic formulation, are less computationally expensive and capture main physical phenomena in many situations. Fluid formulations typically allow easier analysis and physical interpretation of the results. A good compromise is achieved with hybrid models. One species is modeled with fluid equations, and the other is kinetic (e.g.\, fluid electrons with neglected inertia are common for such low-frequency models). Typically, full kinetic simulations for plasma thrusters require some speed-up techniques, such as artificial increase of dielectric permittivity, an increase of electron to ion mass ratio, geometrical downscaling \cite{szabo2001fully, taccogna2005plasma,liu2010pic,charoy2020numerical}, all to overcome numerical constraints and resolve low-frequency dynamics. Full kinetic (particle-in-cell) 2D axial-azimuthal model (with the artificial increase of dielectric permittivity) reproduces the low-frequency ionization oscillations with no assumptions made on anomalous electron transport coefficients (electron current develops self-consistently via azimuthal drift instabilities), presented in Ref.~\onlinecite{coche2014two}.
Note that the radial coordinate may introduce important modifications to the azimuthal drift mode (excitation of large scale MTSI modes) with an influence on the anomalous electron transport\cite{janhunen2018evolution, villafana2021rz, jimenez2021twod}. Thus, ideally, the detailed model of electron transport will be reproduced in a 3D simulation.

Low-frequency plasma dynamics in the axial direction of a Hall thruster are studied with full fluid and hybrid (kinetic/fluid) models. The model parameters are taken close to those of the LANDMARK benchmarking project\cite{landmark}. We identified and distinguished two low-frequency oscillations regimes: one is subject to the low-frequency oscillations only, and another with both low and higher (time of ion flyby oscillations) frequency components. Another distinct feature is the extent of the ion backflow region: in the case with the pure low-frequency component, the ion backflow on average is about half of the channel length, while for the regime where two modes coexist, it is much shorter, Fig.~\ref{vi_pros_nueps}. These regimes are observed in fluid and hybrid models and presented as Cases 1,2 in the numerical experiments. Finally, Case 3 illustrates the effects of finite atom temperature.

The main difference between the full fluid and hybrid model results, as expected, is caused by the different approaches in heavy species modeling. The kinetic method for heavy particles in the hybrid model automatically includes effects of the non-equilibrium distribution function (i.e.\ higher fluid moments effects, only limited by statistical of the noise of the PIC approach). However, as seen from the results, a simple fluid model for both ion and atom components (first two moments) is well sufficient to reproduce the main results and capture breathing mode frequency and the resistive mode presence.

It is found that in Case 1 (regime with co-existence of low and higher frequency modes), the finite value of ion pressure played a notable role in oscillatory behaviour. Ion temperature is often assumed negligible in Hall thrusters operation (similar to a neutral temperature, up to about $\SI{1000}{K}$ or $\SI{0.1}{eV}$), thus it was not included in the primary fluid model. Using kinetic representation from the hybrid model, we show that the ion momentum balance without the ion pressure was notably violated. Thus the ion pressure (with constant temperature) term was added into the fluid model. It resulted in better conformity between the two models. Nevertheless, both fluid and hybrid models predicted the existence of two modes in this configuration even without the ion pressure in the fluid model, suggesting that these oscillations are the so-called resistive modes \cite{chable2005numerical, koshkarov2018current}. We believe these modes are observed in the models without electron pressure, many of which were done in the primary studies and modeling of axial dynamics of Hall thrusters\cite{morozov2000fundamentals, chable2005numerical}. We conjecture that the resistive modes of higher frequencies play an important role in the excitation of the low-frequency modes. 

For Case 2, which shows only the low-frequency oscillations, we already performed extensive studies, Ref.~\onlinecite{chapurin2021mechanism}, where we identified that the mechanism of these oscillations lies in the ion backflow region (presheath) and that they can be additionally reinforced with the plasma recombination. This led to the observation that the difference in the ion boundary conditions at the anode played a crucial role in our benchmark. Bohm velocity for ions in the fluid model generated a larger atom yield (due to recombination, Eq.~(\ref{nn_backflow})) from the anode in comparison to the hybrid model where this velocity was unconstrained and found to be smaller on average (thus leading to smaller atom yield). Modifying the ion velocity boundary condition in the fluid model (replacing the Bohm condition with a free boundary condition) resulted in better conformity between the two models. Similar to the previous case, we conclude that the main physical behaviour was identified in both models even without this modification.

Finally, in Case 3, with finite atom temperature in the hybrid model but the same values of other parameters as in Case 2, it is shown that the advection equation in the fluid model, Eq.~(\ref{na}), with the constant flow velocity is not sufficient to describe atom dynamics. An effect of selective ionization of neutral particles is observed in this case, where average macroscopic velocity increased more than twice along the channel. Thus, the fluid model was supplemented with the atom momentum balance equation~(\ref{am_c}), which resulted in a better agreement between the two models, reproducing the selective ionization effect. Though it resulted in a completely stationary solution, the hybrid model preserved the breathing oscillations (with similar frequency to Case 2) but with much smaller amplitude.

In summary,  benchmarking of the fluid and hybrid models show very close agreement for averaged plasma parameters profiles. Both models reveal the existence of the two different regimes of the low-frequency oscillations in Hall thrusters. 
While qualitatively, the two regimes are identifiable in both models, there are some quantitative differences in the frequencies and the amplitude of the oscillations. These differences are attributed to the ion finite thermal (pressure) effects which were not originally included in the fluid model. 
Account of the finite ion pressure improves the agreement. The finite temperature (energy) spread of the neutral atoms provides a strong stabilizing effect on the oscillations. 
These results highlight limitations of the fluid models that have to be considered in future modeling of practical devices.

\appendix

\section{Resistive modes}\label{apdx_resistive}

\begin{figure}[H]
\centering
\includegraphics[width=0.65\textwidth]{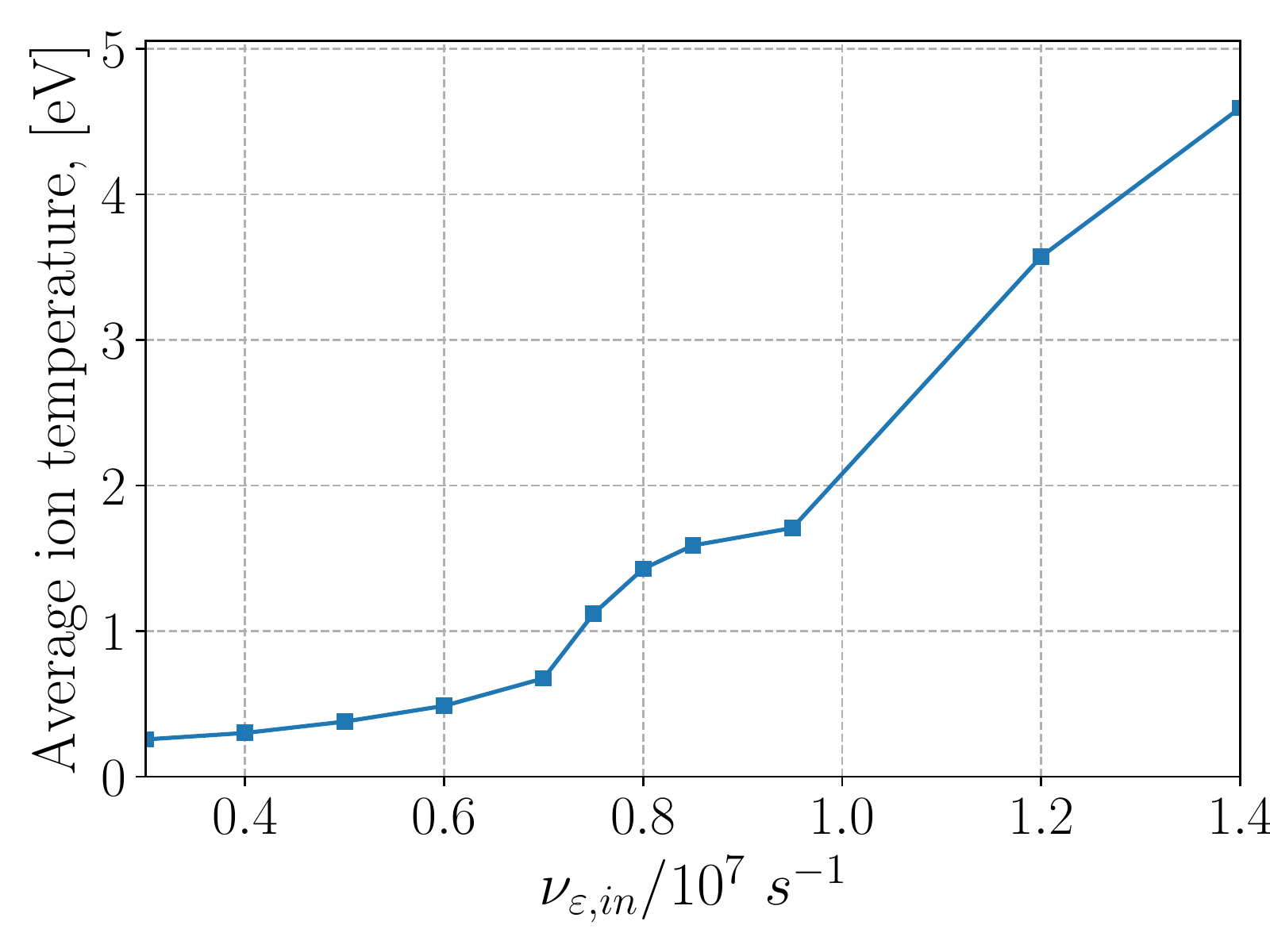}
\caption{Ion average thermal energy as a function of the anomalous electron energy loss coefficient. As we move into regime with resistive modes present (studied in Case 1), ion heating increases and ion pressure effects may become important.}
\label{ti_nueps}
\end{figure}

\begin{figure}[H]
\centering
\subfloat[]{\includegraphics[width=0.5\textwidth]{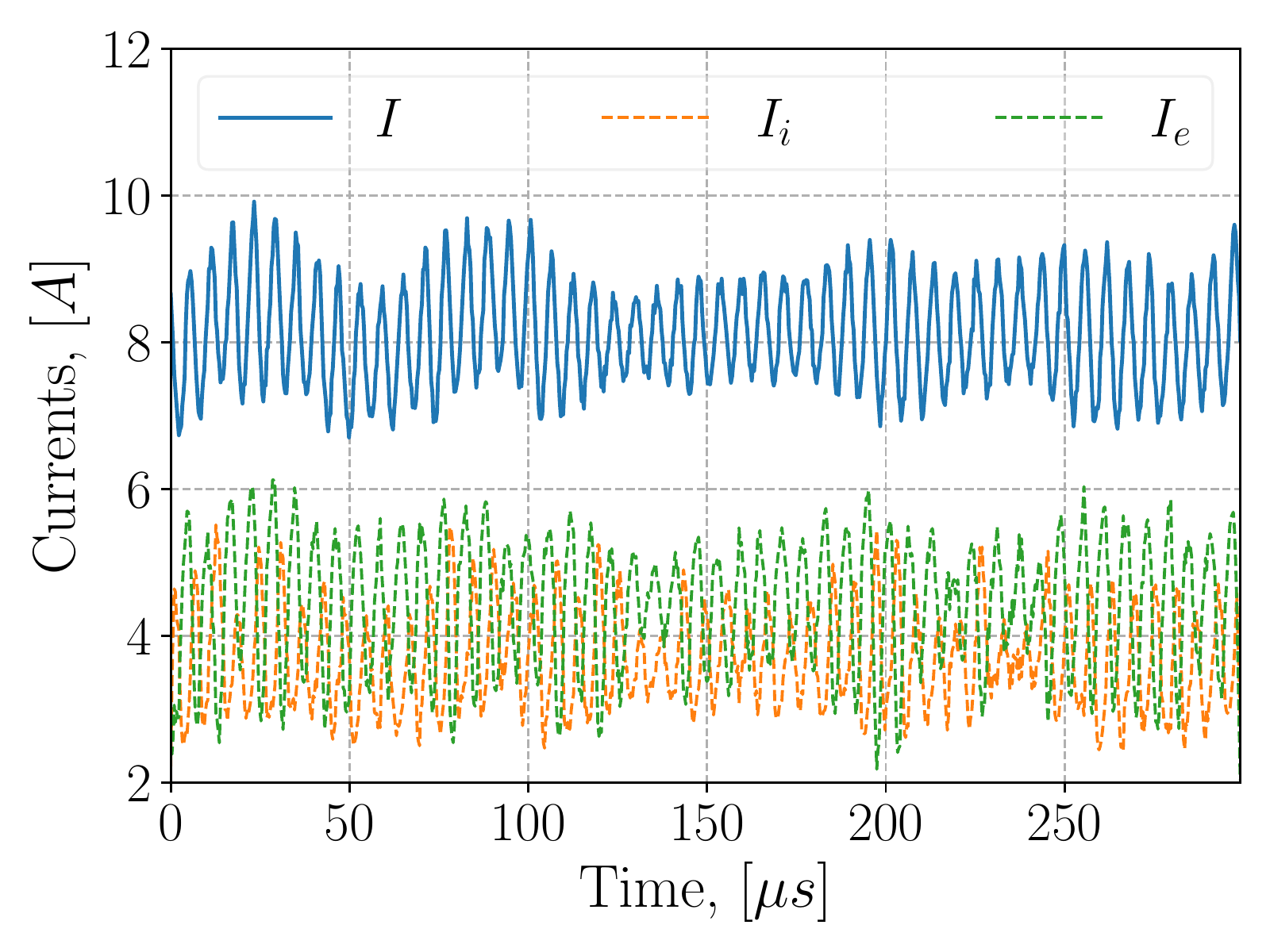}\label{cur_case1_v2}}
\subfloat[]{\includegraphics[width=0.5\textwidth]{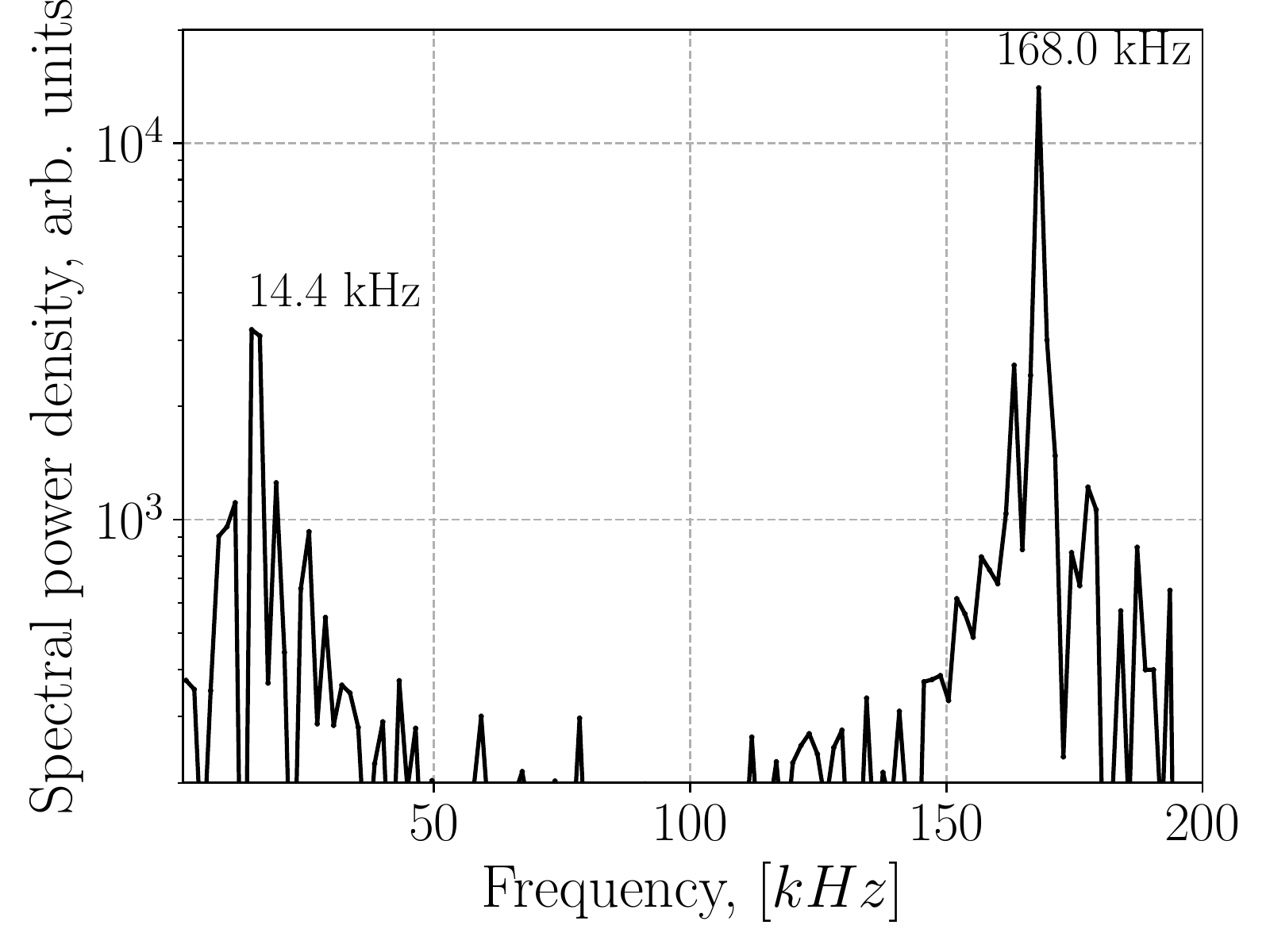}\label{fft_case1_v2}}
\caption{Currents (a) and the total current spectra (b).}
\end{figure}

Additionally to Case 1 with the coexistence of both breathing and resistive modes with relatively small amplitudes, we present cases with more prominent resistive modes and show their effect on ion heating. Larger value of electron anomalous energy losses, $\nu_{\varepsilon,\text{in}} = \SI{1.2e7}{s^{-1}}$, leads to a solution where restive mode dominate, see Fig.~\ref{cur_case1_v2}. It reveals larger amplitude and clearly distinct higher-frequency oscillations (168 kHz), with a small low-frequency modulation (14.4 kHz), Fig.~\ref{fft_case1_v2} its spectrum. As it was demonstrated for Case 1, ion temperature effects were not negligible, and ions were heated to a few electron volts. We noticed that the average ion temperature is higher with presence of the resistive modes as shown in Fig.~\ref{ti_nueps}, where the resistive modes appear for values $\nu_{\varepsilon,\text{in}} = \SI{0.9}{s^{-1}}$ and higher.

\begin{figure}[H]
\centering
\subfloat[]{\includegraphics[width=0.5\textwidth]{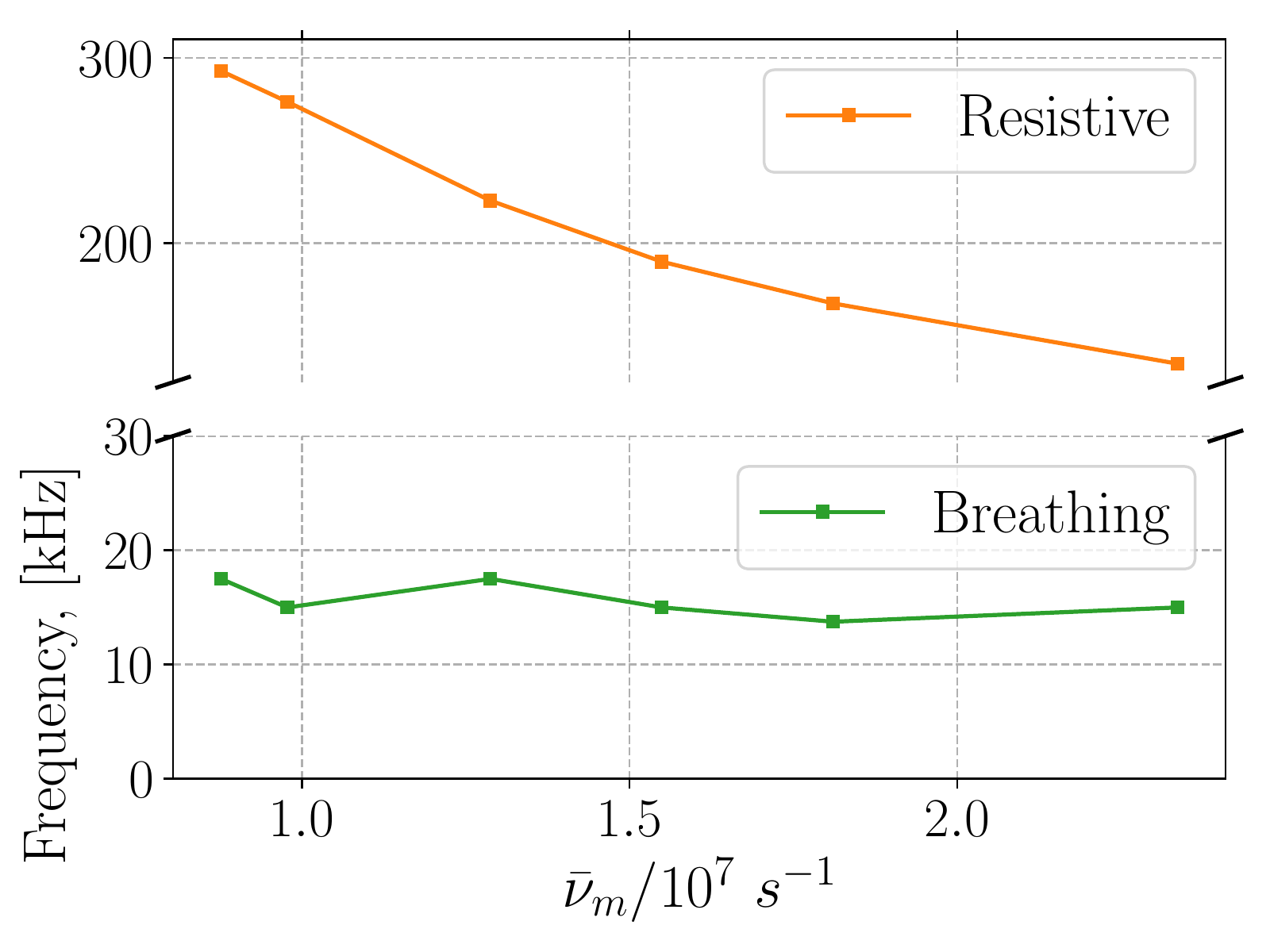}\label{hf_betain}}
\subfloat[]{\includegraphics[width=0.5\textwidth]{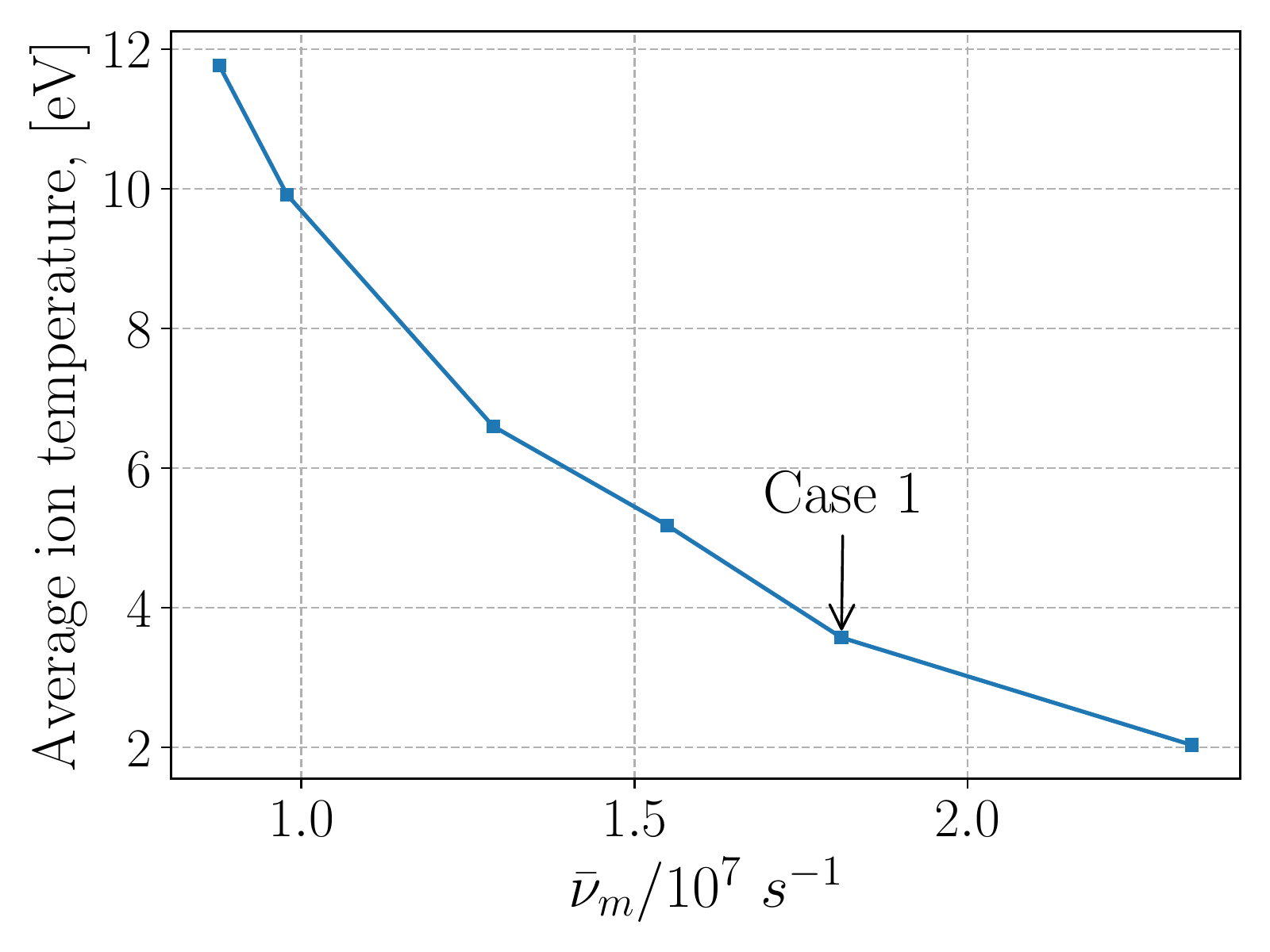}\label{tiavg_betain}}
\caption{Frequency of breathing and resistive modes as a function of averaged (over time and space inside the channel) total electron momentum exchange frequency $\bar{\nu}_m$ (a). Average ion temperature (in time and whole domain) as function of $\bar{\nu}_m$ (b). The parameter varied in this study was the anomalous Bohm coefficient $\beta_{a, in}$ inside the channel, directly affecting the total electron frequency, Eq.~(\ref{nue}).}
\end{figure}

Now, we show that the frequency of the resistive mode scales with the average electron collision frequency $\bar{\nu}_m$, Fig.~\ref{hf_betain}, and thus higher electron mobility ${\sim}1/\bar{\nu}_m$ leads to higher frequencies. At the same time, the ion heating effect is stronger for the higher frequency of resistive modes, Fig.~\ref{tiavg_betain}.  
It is interesting to note that the breathing mode frequency stays approximately the same (Fig.~\ref{hf_betain}), along with the size of the presheath region.

\acknowledgments{This work is supported in part by NSERC Canada, and US Air Force Office of
Scientific Research FA9550-15-1-0226.}

\section*{Data availability}
The data that support the findings of this study are available from the corresponding author upon reasonable request.


\bibliographystyle{unsrt}
\bibliography{ref}

\begin{thebibliography}{10}

\bibitem{morozov2000fundamentals}
A.I. Morozov and V.V. Savelyev.
\newblock Fundamentals of stationary plasma thruster theory.
\newblock In {\em Reviews of plasma physics}, pages 203--391. Springer, 2000.

\bibitem{smolyakov2016fluid}
A.I. Smolyakov, O.~Chapurin, W.~Frias, O.~Koshkarov, I.~Romadanov, T.~Tang,
  M.~Umansky, Y.~Raitses, I.D. Kaganovich, and V.P. Lakhin.
\newblock Fluid theory and simulations of instabilities, turbulent transport
  and coherent structures in partially-magnetized plasmas of discharges.
\newblock {\em Plasma Physics and Controlled Fusion}, 59(1):014041, 2016.

\bibitem{lafleur2016theory}
T.~Lafleur, S.~D. Baalrud, and P.~Chabert.
\newblock Theory for the anomalous electron transport in {H}all effect
  thrusters. i. insights from particle-in-cell simulations.
\newblock {\em Physics of Plasmas}, 23(5):053502, 2016.

\bibitem{janhunen2018nonlinear}
S.~Janhunen, A.~Smolyakov, O.~Chapurin, D.~Sydorenko, I.~Kaganovich, and
  Y.~Raitses.
\newblock Nonlinear structures and anomalous transport in partially magnetized
  {E}{$\times$}{B} plasmas.
\newblock {\em Physics of Plasmas}, 25(1):011608, 2018.

\bibitem{janhunen2018evolution}
S.~Janhunen, A.~Smolyakov, D.~Sydorenko, M.~Jimenez, I.~Kaganovich, and
  Y.~Raitses.
\newblock Evolution of the electron cyclotron drift instability in
  two-dimensions.
\newblock {\em Physics of Plasmas}, 25(8):082308, 2018.

\bibitem{villafana2021rz}
W.~Villafana, F.~Petronio, A.~C. Denig, M.~J. Jimenez, D.~Eremin, L.~Garrigues,
  F.~Taccogna, A.~Alvarez-Laguna, J.~P. Boeuf, A.~Bourdon, P.~Chabert,
  T.~Charoy, B.~Cuenot, K.~Hara, F.~Pechereau, A.~Smolyakov, D.~Sydorenko,
  A.~Tavant, and O.~Vermorel.
\newblock 2{D} radial-azimuthal particle-in-cell benchmark for {E}{$\times$}{B}
  discharges.
\newblock {\em Plasma Sources Science and Technology}, 30(7):075002, jul 2021.

\bibitem{jimenez2021twod}
M.~Jimenez.
\newblock {2D3V} {P}article-in-cell simulations of {E}lectron {C}yclotron
  {D}rift {I}nstability and anomalous electron transport in {E}{$\times$}{B}
  plasmas.
\newblock Master's thesis, University of Saskatchewan, 2021.

\bibitem{kaganovich2020physics}
I.D. Kaganovich, A.~Smolyakov, Y.~Raitses, E.~Ahedo, I.G. Mikellides, B.~Jorns,
  F.~Taccogna, R.~Gueroult, S.~Tsikata, A.~Bourdon, J.-P. Boeuf, M.~Keidar,
  A.~T. Powis, M.~Merino, M.~Cappelli, K.~Hara, J.~A. Carlsson, N.~J. Fisch,
  P.~Chabert, I.~Schweigert, T.~Lafleur, K.~Matyash, A.V. Khrabrov, R.~W.
  Boswell, and A.~Fruchtman.
\newblock Physics of {E}{$\times$}{B} discharges relevant to plasma propulsion
  and similar technologies.
\newblock {\em Physics of Plasmas}, 27(12):120601, 2020.

\bibitem{esipchuk1974plasma}
Y.B. Esipchuk, A.I. Morozov, G.N. Tilinin, and A.V. Trofimov.
\newblock Plasma oscillations in closed-drift accelerators with an extended
  acceleration zone.
\newblock {\em Soviet Physics Technical Physics}, 18:928, 1974.

\bibitem{fife1997numerical}
J.~Fife, M.~Martinez-Sanchez, J.~Szabo, J.~Fife, M.~Martinez-Sanchez, and
  J.~Szabo.
\newblock {\em A numerical study of low-frequency discharge oscillations in
  {H}all thrusters}, page 3052.
\newblock 2012.

\bibitem{barral2008origin}
S.~Barral and E.~Ahedo.
\newblock On the origin of low frequency oscillations in {H}all thrusters.
\newblock {\em AIP Conference Proceedings}, 993(1):439--442, 2008.

\bibitem{hara2014perturbation}
Kentaro Hara, Michael~J Sekerak, Iain~D Boyd, and Alec~D Gallimore.
\newblock Perturbation analysis of ionization oscillations in {H}all effect
  thrusters.
\newblock {\em Physics of Plasmas}, 21(12):122103, 2014.

\bibitem{chapurin2021mechanism}
O.~Chapurin, A.I. Smolyakov, G.~Hagelaar, and Y.~Raitses.
\newblock On the mechanism of ionization oscillations in {H}all thrusters.
\newblock {\em Journal of Applied Physics}, 129(23):233307, 2021.

\bibitem{LafleurJAP2021}
T.~Lafleur, P.~Chabert, and A.~Bourdon.
\newblock The origin of the breathing mode in hall thrusters and its
  stabilization.
\newblock {\em Journal of Applied Physics}, 130(5):053305, 2021.

\bibitem{chable2005numerical}
S.~Chable and F.~Rogier.
\newblock Numerical investigation and modeling of stationary plasma thruster
  low frequency oscillations.
\newblock {\em Physics of plasmas}, 12(3):033504, 2005.

\bibitem{makowski2001review}
K.~Makowski, Z.~Peradzynski, S.~Barral, and M.A. Dudeck.
\newblock Review of the plasma fluid models in stationary plasma thrusters.
\newblock {\em High Temperature Material Processes: An International Quarterly
  of High-Technology Plasma Processes}, 5(2), 2001.

\bibitem{hagelaar2004modelling}
G.J.M. Hagelaar, J.~Bareilles, L.~Garrigues, and J.-P. Boeuf.
\newblock Modelling of stationary plasma thrusters.
\newblock {\em Contributions to Plasma Physics}, 44(5-6):529--535, 2004.

\bibitem{boeuf1998low}
J.-P. Boeuf and L.~Garrigues.
\newblock Low frequency oscillations in a stationary plasma thruster.
\newblock {\em Journal of Applied Physics}, 84(7):3541{H}all3554, 1998.

\bibitem{barral2001fluid}
S.~Barral, Z.~Peradzynski, K.~Makowski, and M.A. Dudeck.
\newblock Fluid model of {H}all thruster -- comparison with hybrid model.
\newblock {\em High Temperature Material Processes: An International Quarterly
  of High-Technology Plasma Processes}, 5(2), 2001.

\bibitem{barral2009low}
S.~Barral and E.~Ahedo.
\newblock Low-frequency model of breathing oscillations in {H}all discharges.
\newblock {\em Physical Review E}, 79(4):046401, 2009.

\bibitem{giannetti2021numerical}
V.~Giannetti, M.M. Saravia, L.~Leporini, S.~Camarri, and T.~Andreussi.
\newblock Numerical and experimental investigation of longitudinal oscillations
  in {H}all thrusters.
\newblock {\em Aerospace}, 8(6):148, 2021.

\bibitem{MorozovPPR2000h}
A.~I. Morozov and V.~V. Savel'ev.
\newblock One-dimensional hybrid model of a stationary plasma thruster.
\newblock {\em Plasma Physics Reports}, 26(10):875--880, 2000.

\bibitem{ShashkovPoP2017}
Andrey Shashkov, Alexander Lovtsov, and Dmitry Tomilin.
\newblock A one-dimensional with three-dimensional velocity space hybrid-pic
  model of the discharge plasma in a hall thruster.
\newblock {\em Physics of Plasmas}, 24(4):043501, 2017.

\bibitem{gavrikov2021hybrid}
Gavrikov M.B. and Taiurskii A.A.
\newblock Hybrid model of a stationary plasma thruster (in {R}ussian).
\newblock {\em Preprints Keldysh Institute of Applied Mathematics}, (0):35--47,
  2021.

\bibitem{dudson2009bout++}
B.D. Dudson, M.V. Umansky, X.Q. Xu, P.B. Snyder, and H.R. Wilson.
\newblock Bout++: A framework for parallel plasma fluid simulations.
\newblock {\em Computer Physics Communications}, 180(9):1467{H}all1480, 2009.

\bibitem{hagelaar2002two}
G.J.M. Hagelaar, J.~Bareilles, L.B.J.P. Garrigues, and J.-P. Boeuf.
\newblock Two-dimensional model of a stationary plasma thruster.
\newblock {\em Journal of Applied Physics}, 91(9):5592--5598, 2002.

\bibitem{hagelaar2003role}
G.J.M. Hagelaar, J.~Bareilles, L.~Garrigues, and J.-P. Boeuf.
\newblock Role of anomalous electron transport in a stationary plasma thruster
  simulation.
\newblock {\em Journal of Applied Physics}, 93(1):67--75, 2003.

\bibitem{landmark}
Low temper{A}ture mag{N}etize{D} plas{MA} benchma{RK}s.
\newblock \url{https://www.landmark-plasma.com/}.

\bibitem{litvak2001resistive}
A.~A. Litvak and Nathaniel~J. Fisch.
\newblock Resistive instabilities in {H}all current plasma discharge.
\newblock {\em Physics of Plasmas}, 8(2):648--651, 2001.

\bibitem{fernandez2008growth}
E.~Fernandez, M.~K. Scharfe, C.~A. Thomas, N.~Gascon, and M.~A. Cappelli.
\newblock Growth of resistive instabilities in {E}{$\times$}{B} plasma
  discharge simulations.
\newblock {\em Physics of Plasmas}, 15(1):012102, 2008.

\bibitem{koshkarov2018current}
O.~Koshkarov, A.I. Smolyakov, I.V. Romadanov, O.~Chapurin, M.V. Umansky,
  Y.~Raitses, and I.D. Kaganovich.
\newblock Current flow instability and nonlinear structures in dissipative
  two-fluid plasmas.
\newblock {\em Physics of Plasmas}, 25(1):011604, 2018.

\bibitem{MorozovPPR2000f}
A.~I. Morozov and V.~V. Savel’ev.
\newblock One-dimensional hydrodynamic model of the atom and ion dynamics in a
  stationary plasma thruster.
\newblock {\em Plasma Physics Reports}, 26(3):219--224, 2000.

\bibitem{gavrikov2021some}
Gavrikov M.B. and Taiurskii A.A.
\newblock Some mathematical questions of plasma ionization (in {R}ussian).
\newblock {\em Preprints Keldysh Institute of Applied Mathematics}, (94):1--48,
  2021.

\bibitem{hagelaar2005solving}
G.J.M. Hagelaar and L.C. Pitchford.
\newblock Solving the {B}oltzmann equation to obtain electron transport
  coefficients and rate coefficients for fluid models.
\newblock {\em Plasma Sources Science and Technology}, 14(4):722, 2005.

\bibitem{siglo}
{SIGLO} database.
\newblock \url{www.lxcat.net/SIGLO}.
\newblock [Retrived: June-2013].

\bibitem{CohenZurPoP2002}
A.~Cohen-Zur, A.~Fruchtman, J.~Ashkenazy, and A.~Gany.
\newblock Analysis of the steady-state axial flow in the {H}all thruster.
\newblock {\em Physics of Plasmas}, 9(10):4363--4374, 2002.

\bibitem{AhedoPoP2005}
E.~Ahedo and J.~Rus.
\newblock Vanishing of the negative anode sheath in a {H}all thruster.
\newblock {\em Journal of Applied Physics}, 98(4):043306, 2005.

\bibitem{dorf2003anode}
L.~Dorf, V.~Semenov, and Y.~Raitses.
\newblock Anode sheath in {H}all thrusters.
\newblock {\em Applied Physics Letters}, 83(13):2551--2553, 2003.

\bibitem{dorf2005experimental}
L.~Dorf, Y.~Raitses, and N.~J. Fisch.
\newblock Experimental studies of anode sheath phenomena in a {H}all thruster
  discharge.
\newblock {\em Journal of Applied Physics}, 97(10):103309, 2005.

\bibitem{smolyakov2019stationary}
A.~Smolyakov, O.~Chapurin, I.~Romadanov, Y.~Raitses, I.~Kaganovich,
  G.~Hagelaar, and J.-P. Boeuf.
\newblock Stationary profiles and axial mode oscillations in {H}all thrusters.
\newblock In {\em AIAA Propulsion and Energy 2019 Forum}, pages 4080--4080,
  2019.

\bibitem{birdsall1991particle}
C.K. Birdsall.
\newblock Particle-in-cell charged-particle simulations, plus {M}onte {C}arlo
  collisions with neutral atoms, {PIC-MCC}.
\newblock {\em IEEE Transactions on plasma science}, 19(2):65--85, 1991.

\bibitem{hagelaar2008modelling}
G.J.M. Hagelaar.
\newblock {\em Modelling methods for low-temperature plasmas}.
\newblock PhD thesis, Universit{\'e} Toulouse III Paul Sabatier (UT3 Paul
  Sabatier), 2008.

\bibitem{cartwright2000loading}
K.L. Cartwright, J.P. Verboncoeur, and C.K. Birdsall.
\newblock Loading and injection of {M}axwellian distributions in particle
  simulations.
\newblock {\em Journal of Computational Physics}, 162(2):483--513, 2000.

\bibitem{meezan2001anomalous}
N.B. Meezan, W.A. Hargus~Jr, and M.A. Cappelli.
\newblock Anomalous electron mobility in a coaxial {H}all discharge plasma.
\newblock {\em Physical Review E}, 63(2):026410, 2001.

\bibitem{hargus2002interior}
W.A. Hargus and M.A. Cappelli.
\newblock Interior and exterior laser-induced fluorescence and plasma
  measurements within a {H}all thruster.
\newblock {\em Journal of Propulsion and Power}, 18(1):159--168, 2002.

\bibitem{hara2012one}
K.~Hara, I.D. Boyd, and V.I. Kolobov.
\newblock One-dimensional hybrid-direct kinetic simulation of the discharge
  plasma in a {H}all thruster.
\newblock {\em Physics of Plasmas}, 19(11):113508, 2012.

\bibitem{szabo2001fully}
J.J. Szabo.
\newblock {\em Fully kinetic numerical modeling of a plasma thruster}.
\newblock PhD thesis, Massachusetts Institute of Technology, 2001.

\bibitem{taccogna2005plasma}
F.~Taccogna, S.~Longo, M.~Capitelli, and R.~Schneider.
\newblock Plasma flow in a {H}all thruster.
\newblock {\em Physics of Plasmas}, 12(4):043502, 2005.

\bibitem{liu2010pic}
H.~Liu, B.~Wu, D.~Yu, Y.~Cao, and P.~Duan.
\newblock Particle-in-cell simulation of a {H}all thruster.
\newblock {\em Journal of Physics D: Applied Physics}, 43(16):165202, apr 2010.

\bibitem{charoy2020numerical}
T.~Charoy.
\newblock {\em Numerical study of electron transport in {H}all thrusters}.
\newblock PhD thesis, Institut polytechnique de Paris, 2020.

\bibitem{coche2014two}
P.~Coche and L.~Garrigues.
\newblock A two-dimensional (azimuthal-axial) particle-in-cell model of a
  {H}all thruster.
\newblock {\em Physics of Plasmas}, 21(2):023503, 2014.

\end{thebibliography}

\end{document}